\documentclass[aps, prb, reprint, superscriptaddress]{revtex4-1}

\usepackage{amsmath, amssymb, times, graphicx, mathrsfs, hyperref, array, bbm,ulem}
\usepackage[usenames,dvipsnames]{xcolor}
\usepackage{times}
\usepackage{braket}
\usepackage{afterpage}

\hypersetup{colorlinks=false}

\graphicspath{{./}{./figures/}}

\begin{document}
\title{Thermodynamic classification of three-dimensional Kitaev spin liquids}

\author{Tim Eschmann}
\email[E-mail: ]{eschmann@thp.uni-koeln.de}
\affiliation{Institute for Theoretical Physics, University of Cologne, 50937 Cologne, Germany}
\author{Petr A. Mishchenko}
\affiliation{Department of Applied Physics, The University of Tokyo, Tokyo 113-8656, Japan}
\affiliation{NTT Secure Platform Laboratories, Tokyo 180-8585, Japan}
\author{Kevin O'Brien}
\affiliation{Institute for Theoretical Physics, University of Cologne, 50937 Cologne, Germany}
\author{Troels A. Bojesen}
\affiliation{Department of Applied Physics, The University of Tokyo, Tokyo 113-8656, Japan}
\author{Yasuyuki Kato}
\affiliation{Department of Applied Physics, The University of Tokyo, Tokyo 113-8656, Japan}
\author{Maria Hermanns}
\affiliation{Department of Physics, Stockholm University, AlbaNova University Center, SE-106 91 Stockholm, Sweden}
\affiliation{Nordita, KTH Royal Institute of Technology and Stockholm University, SE-106 91 Stockholm, Sweden}
\author{Yukitoshi Motome}
\affiliation{Department of Applied Physics, The University of Tokyo, Tokyo 113-8656, Japan}
\author{Simon Trebst}
\affiliation{Institute for Theoretical Physics, University of Cologne, 50937 Cologne, Germany}


\begin{abstract}

In the field of frustrated magnetism, Kitaev models provide a unique framework  
to study the phenomena of spin fractionalization and emergent lattice gauge theories in two and three spatial dimensions.
Their ground states are quantum spin liquids, which can typically be described in terms of a Majorana band structure and 
an ordering of the underlying $\mathbb{Z}_2$ gauge structure. Here we provide a comprehensive classification of the ``gauge physics" of a family of elementary three-dimensional Kitaev models, discussing how their thermodynamics and ground state order depends on the underlying lattice geometry.
Using large-scale, sign-free quantum Monte Carlo simulations we show that 
the ground-state gauge order can generally be understood in terms of the length of elementary plaquettes -- 
 a result which extends the applicability of Lieb's theorem to lattice geometries beyond its original scope.
At finite temperatures, the proliferation of (gapped) vison excitations destroys the gauge order at a critical temperature
scale, which we show to correlate with the size of vison gap for the family of three-dimensional Kitaev models.
We also discuss two notable exceptions where the lattice structure gives rise to ``gauge frustration'' or intertwines the
gauge ordering with time-reversal symmetry breaking. 
In a more general context, the thermodynamic gauge transitions in such 3D Kitaev models are one of the most
natural settings for phase transitions beyond the standard Landau-Ginzburg-Wilson paradigm.

\end{abstract}

\maketitle


\section{Introduction}

Quantum spin liquids occur in strongly correlated magnetic quantum systems, when frustration effects prevent the local magnetic moments from ordering and spin dynamics persist even at zero temperature\cite{Balents2010spin}.  While there is no long-range magnetic order and corresponding order parameter, such systems are nevertheless highly special in that they form long-range entanglement \cite{KitaevPreskill,levinWen} and possibly harbor exotic excitations, such as, e.g.,~Majorana fermions.
Despite this conceptual understanding, it remains notoriously hard to study the formation of quantum spin liquids in a 
microscopic context. This is particularly true for three-dimensional (3D) settings. As such, the recent introduction of 3D generalizations \cite{Mandal2009exactly,Hermanns2014quantum,Hermanns2015weyl,Obrien2016classification,Yamada2017} of the Kitaev honeycomb model \cite{Kitaev2006anyons} are of special importance. These 3D Kitaev models not only remain largely tractable by analytical and numerical tools, they also harbor a broad variety of different spin liquid ground states \cite{Obrien2016classification,Yamada2017}. The underlying physics very much resembles what is well known from the two-dimensional (2D) Kitaev model: The original spin degrees of freedom fractionalize into itinerant Majorana fermions and a static $\mathbb{Z}_2$ gauge field.
As the latter is generically gapped, the zero temperature physics can be understood in terms of (non-interacting) Majorana fermions moving in a static, fixed flux background. The resulting Majorana band structure describes the low-energy physics
of the quantum spin liquid, typically as a gapless ``Majorana metal'' if the three different Kitaev couplings are of more or less equal strength
\cite{Hermanns2017physics}. 
The precise nature of these Majorana metals, however, depends on the underlying lattice geometries with the nodal manifold being described as Majorana Fermi surfaces, nodal lines, or Weyl points for different lattices \cite{Obrien2016classification}. 

A necessary ingredient for the classification of these 3D Kitaev spin liquids is a correct assignment of the gauge order, i.e., the
flux background in which the Majorana fermions move. Determining the appropriate flux background is, in general, non-trivial. 
Analytically, one can resort to a theorem by Lieb \cite{Lieb1994}, which makes a connection between the length of the elementary plaquettes of a lattice and the ground state flux pattern. However, the applicability of this theorem is restricted to lattices that exhibit certain mirror symmetries. Examples include the honeycomb lattice in two spatial dimensions, for which the ground state is ``flux free", and only one of the 3D lattices which we consider in the following. For all other lattices, the strict requirements of Lieb's theorem are not met. For those lattices, an unambiguous identification of the ground-state gauge order
and corresponding flux assignment can alternatively be obtained through (much more demanding) numerical calculations. 
Since the Kitaev model in its parton description (i.e., in the language of Majorana fermions coupled to a $\mathbb{Z}_2$ gauge field) does not exhibit a sign problem \cite{Nasu2014vaporization}, one can perform quantum Monte Carlo (QMC) simulations to 
track the finite-temperature ordering of the gauge field and infer its low-temperature order.

It is the purpose of this manuscript, to elucidate and comprehensively classify the thermodynamics of 3D Kitaev spin liquids via extensive, sign-free quantum Monte Carlo simulations. Going to finite temperatures, it is generally expected that the flux or ``vison'' excitations of the $\mathbb{Z}_2$ gauge field become important, as their proliferation destroys all ground state order \cite{Read1991,Senthil2000}. In three spatial dimensions, this proliferation is generally expected to occur at a {\sl finite} temperature since the visons are loop-like excitations (whereas in two spatial dimensions they are point-like excitations allowing for an instant proliferation at any non-zero temperature). This is indeed what previous numerical simulations of 3D Kitaev models on certain lattice structures have found \cite{PhysRevB.89.115125, Nasu2014vaporization, mishchenko_prb_96_2017}, indicating that the thermal gauge ordering transition is suppressed by about two orders of magnitude with regard to the bare Kitaev couplings. Here, we perform a systematic study of a large set of 3D Kitaev spin liquids considered first in  Refs.~\onlinecite{Obrien2016classification,Yamada2017} in classifying the aforementioned Majorana metals. We show that the finite-temperature behavior is very systematic, except for two cases where additional frustration and/or spontaneous symmetry breaking effects become important \cite{kato_prb_96_2017, 2019EschmannGaugeFrustration, 2020MishchenkoChiralSpinLiquids}. In particular, we find a strong correlation between the transition temperature and the local flux/vison gap, whereas there is no correlation to the minimal loop length
\footnote{
It has previously been argued \cite{Kimchi2014three} that the loop {\sl tension} should play a crucial role in setting the transition temperature.
Technically, it is a tedious task to sharply define the loop tension consistently due to the strong anisotropies for the different lattice geometries
that we consider in this study. However, one might argue that the loop tension itself is also correlated with the vison gap, explaining the 
consistency of these arguments with the numerically observed correlation.}.
Our study also shows that Lieb's theorem  predicts the correct flux ground state even in cases where its requirements are not fulfilled. This suggests that there should exist a more general version of the theorem. 

Due to the substantial amount of specific results for the various lattice geometries under consideration and the considerable length of the manuscript, we start by giving an overview of our results, together with the necessary background. A more in-depth discussion of our findings can be found in the subsequent sections. In  particular, in section \ref{3DKitaevModels} we review the definition and solution approach for the Kitaev model, as well as a  brief review on Lieb's theorem. 
In section \ref{SignFreeQMC}, we outline the quantum Monte Carlo method that was used in the numerical studies,  focusing on the modifications we made in contrast to earlier works in the field. Following that, in section \ref{Thermodynamics} we give a detailed presentation of our numerical results on the bipartite ``families'' of lattice systems (8,3)$x$ and (10,3)$x$. In this context, we discuss the different behaviors of the relevant thermodynamic observables and relate them to elementary geometric properties of the underlying lattices (e.g., its fundamental symmetries), and to the different Majorana (semi-)metal ground states that these systems exhibit. We also review results from former works of our collaboration\cite{2019EschmannGaugeFrustration, 2020MishchenkoChiralSpinLiquids} on the more exotic systems (8,3)c (section \ref{(8,3)c}), which exhibits ``gauge frustration", and the non-bipartite lattice (9,3)a (section \ref{(9,3)a}), which intertwines gauge ordering and time-reversal symmetry breaking (due to the odd length of its elementary plaquettes).
Finally, we sum up the conclusions following from our studies and give an outlook on further research directions in the field of 3D Kitaev systems in section \ref{Conclusions}. In addition to the main manuscript, there is an extensive appendix which explains a variety of technical details that we chose not to cover in the main text. 

\begin{figure}[t]
   \centering
    \includegraphics[width=\columnwidth]{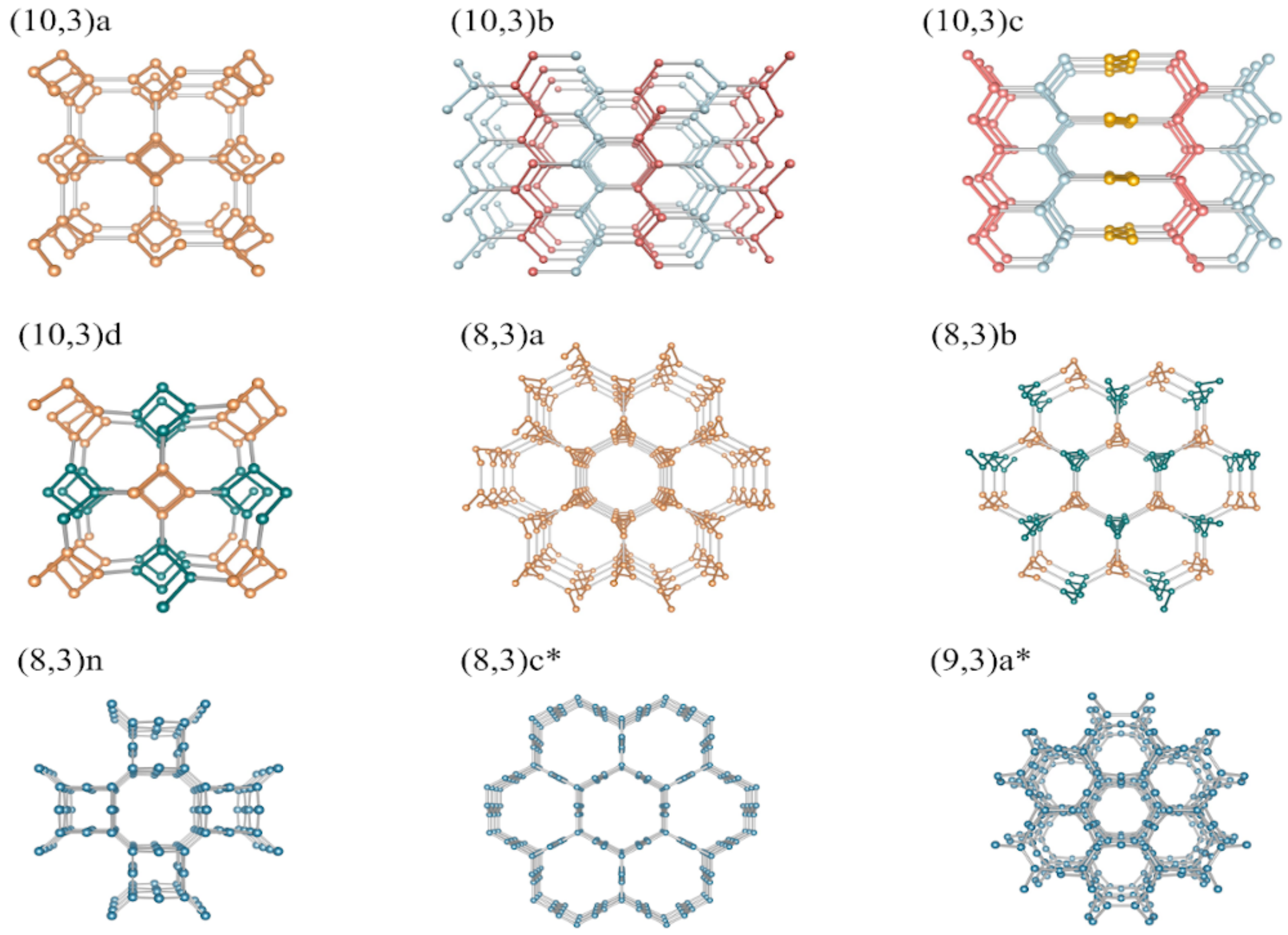}  
   \caption{{\bf Elementary tricoordinated 3D lattices} considered in this manuscript \cite{Wells1977, Obrien2016classification}. The lattices are named and ordered according to the Schl\"afli notation $(p,c)x$, specifying the elementary plaquette length $p$ and the coordination number $c$ (followed by an index letter $x$). Spirals colored blue (orange)  rotate clockwise (anti-clockwise). The colors red/lightblue/yellow highlight different directions of `zigzag' chains. For the lattices (8,3)c and (9,3)a, marked with an asterisk, the gauge sectors behave differently from the other tricoordinated lattices. These systems were regarded separately.}
    \label{fig:Lattices}
\end{figure}

\subsection*{Overview of results}

We start with a summary of our main results. The main ingredient of our study is a family of elementary 3D tricoordinated lattice geometries, illustrated in Fig.~\ref{fig:Lattices}, which allow to define 3D generalizations of the Kitaev honeycomb model. It is the same family of lattices that has been considered in  earlier classification work discussing the ground states of 3D Kitaev models as Majorana metals \cite{Obrien2016classification}, augmented by one of the lattices considered in a follow-up study~\cite{Yamada2017}. 

One central result of our study concentrating on the ``gauge physics" of these 3D Kitaev models is the numerical observation that the predictions of Lieb's theorem \cite{Lieb1994} on the ground state flux sectors holds for {\sl all} 3D lattice geometries that we considered, including those that do not  possess  the mirror symmetry requirements for its rigorous applicability. 

\begin{figure}[t]
   \centering
    \includegraphics[width=\columnwidth]{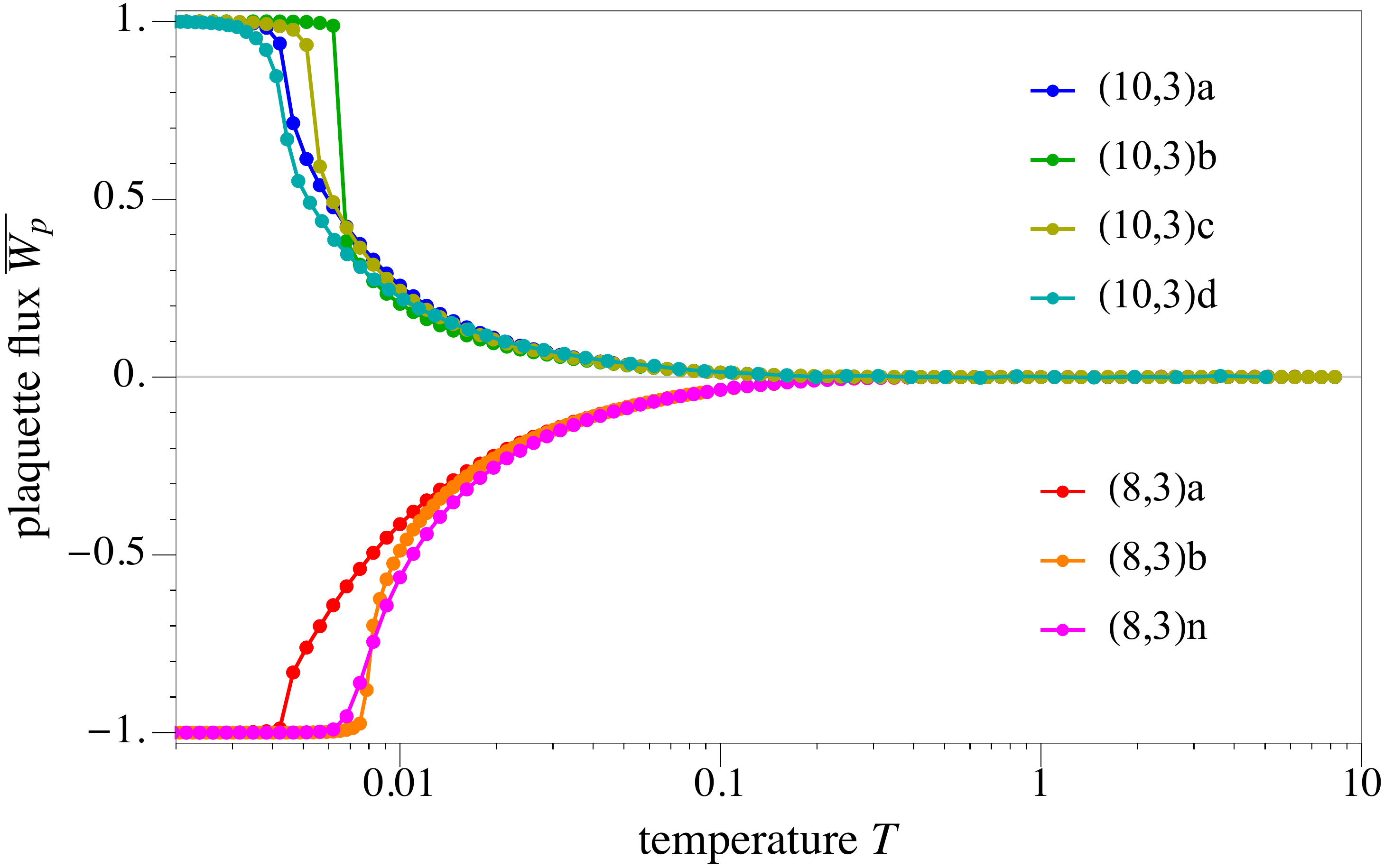}  
   \caption{{\bf Flux per plaquette.} The numerical data shows that the ground state flux sector $0 / \pi$, corresponding to the average flux operator eigenvalue $\overline{W_p} = \pm 1$ (see Eq. \eqref{eq:AveragePlaquetteFlux}) is determined by the elementary plaquette length $p$ for each lattice. For $p \mod 4 = 0$, the ground state has a $\pi$-flux per plaquette, while the flux for systems with $p \mod 4 = 2$ is 0. This prediction made by Lieb's theorem is valid for all bipartite 3D Kitaev systems, even for those that do not fulfill the mirror symmetry conditions that are required for its strict proof (which are all except (8,3)b). Data shown is for the linear system size $L=7$, except for (8,3)n and (10,3)d (here: $L = 5$).}
    \label{fig:Wp_all}
\end{figure}

According to this theorem, it is the (even) length of the elementary plaquette $p$ which determines the ground state flux: If 
$p \mod 4 = 2$, the energy is minimized if all plaquettes remain flux-free, while for $p \mod 4 = 0$, it is a $\pi$-flux per plaquette which produces the flux ground state. First used to predict the flux-free ground state of the 2D Kitaev honeycomb model, this theorem is, in principal, applicable to lattice systems of arbitrary spatial dimensionality, and thus, also to 3D Kitaev systems. There is a caveat, however, since the statement on ground state flux sectors provided by Lieb is strictly proven only for systems and plaquettes which fulfill certain mirror symmetries. Among the lattice systems considered in our studies, these symmetries are only a feature of the lattice denoted (8,3)b. As a consequence, there exists no mathematically rigorous prediction on the energy-minimizing flux sector for all other systems.
In our former (mostly analytical) study on exact ground states\cite{Obrien2016classification}, it was nonetheless assumed, backed by symmetry considerations and benchmark calculations {on periodic flux configurations with small unit cells}, that Lieb's theorem provides the correct guideline to find the ground state flux sector, also for the other 3D lattices. Here, based on large-scale, numerically exact quantum Monte Carlo studies, we  confirm that this assumption was indeed well justified. Our results unambiguously show that, despite the lack of the particular mirror symmetries, all bipartite systems with a plaquette length $p$ dividable by 4 possess a $\pi$-flux ground state, while all lattice systems with $p \mod 4 = 2$ have a ground state where all plaquettes are flux-free. This is illustrated in Fig.~\ref{fig:Wp_all}, where we plot a central results of our quantum Monte Carlo simulations: the flux per elementary plaquette versus temperature for all lattices (with an even length elementary plaquette).

Based on this result, the elementary 3D Kitaev systems can be systematically categorized into different families, which are characterized by their plaquette length and the corresponding ground state flux sector. For this purpose, it is convenient to use the {\it Schl\"afli notation} for the naming of the lattices: A lattice is denominated by the expression $(p,c)x$, where the number $p$ is the elementary plaquette length, $c$ the coordination number, and both are followed by an index letter $x$. Among the bipartite lattices, we find the family (10,3)$x$ with flux-free ground states, and the family (8,3)$x$, where the plaquettes generally carry $\pi$-flux  in the ground state. In addition, we briefly discuss two exceptional lattice systems that have already been presented in earlier works, namely (8,3)c, which possesses a {\it frustrated} gauge sector\cite{2019EschmannGaugeFrustration}, and the non-bipartite lattice (9,3)a, where the ground state of the Kitaev system carries  $\pm \pi/2$-flux per plaquette, thus  spontaneously breaking time-reversal symmetry\cite{2020MishchenkoChiralSpinLiquids}.

\begin{figure}[t]
    \centering
    \includegraphics[width=\columnwidth]{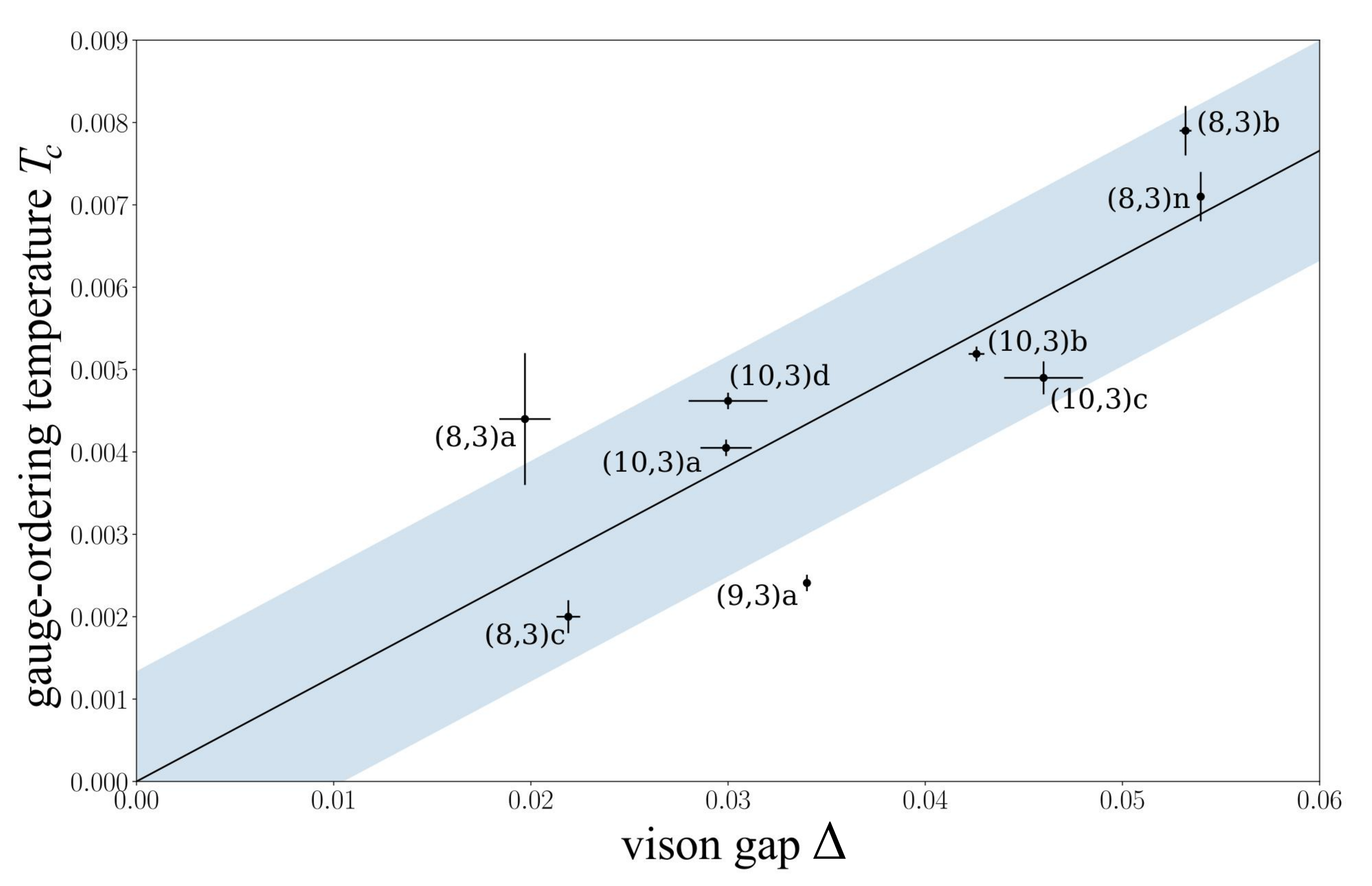}  
      \caption{{\bf Correlation of ``gauge ordering" temperature $T_c$ and the vison gap $\Delta$}. Both quantities are clearly correlated, which is indicated by a linear fit (black line), which does not include the origin. Here, the shaded area indicates the error of the $y$-intersect. The system size dependence of the vison gap $\Delta$ is shown in Fig. \ref{fig:VisonGapsPlot}. The values of $T_c$ were extrapolated from the positions of the low-temperature peaks in the specific heat $C_v$ (see Fig. \ref{fig:CriticalTemperaturesPlot}). At this phase transition temperature, the loop-like vison excitations of the 3D Kitaev systems proliferate and break open into line-like objects.}
    \label{fig:CorrelationPlot}
\end{figure}

Extending earlier numerical studies on the thermodynamics of Kitaev systems\cite{Nasu2014vaporization, Nasu2015thermal}, we could verify that the main characteristic thermodynamic signatures -- namely a double-peak structure in the specific heat -- that were so far identified for (10,3)b (hyperhoneycomb)\cite{Nasu2014vaporization} and (10,3)a (hyperoctagon)\cite{mishchenko_prb_96_2017}, are also observed for the other (regular) systems of plaquette length 10 and 8. This validates the general expectation that 3D Kitaev systems indeed exhibit separate energy scales for the fractionalization of the spin degrees of freedom -- resulting in a cross-over feature in the specific heat around the strength of the bare Kitaev coupling   and a consecutive thermal gauge ordering transition. The precise energy/temperature scale of the latter is a priori not known, but its existence is expected  to generically occur \cite{Read1991,Senthil2000} in 3D $\mathbb{Z}_2$ lattice gauge theories:
The elementary gapped flux/vison excitations of the ground-state flux sector of 3D Kitaev systems have a {\sl loop-like} form (illustrated for an example lattice in Fig.~\ref{fig:VisonLoop} below), caused by the geometric constraint on plaquettes forming the boundary of closed volumes, and therefore differ from the 2D vison excitations, which are {\sl point-like} objects. These loop-like visons are first created as local excitations at low temperatures and break up and form system-spanning objects only at a critical temperature $T_c$. For systems with semi-open boundary conditions this effect can be measured in terms of a non-local order parameter, a Wilson loop. This phase transition between two topologically distinct ``loop regimes" in temperature space can be recast in terms of domain boundaries in a 3D Ising model, with the roles of high and low temperature phases inverted \cite{PhysRevB.89.115125, Nasu2014vaporization}. As such, we generically expect this transition to be in the (inverted) 3D Ising universality class if it is a continuous transition. Our numerical data is indeed consistent with such a continuous phase transition for most systems of plaquette length 10 and 8, but our limited system sizes do not allow to solidly establish its universality class. 
This general argument for the existence of a finite-temperature phase transition in 3D Kitaev systems, however, does not provide any quantitative estimate for the transition temperature of the gauge ordering transition. But since the transition depends on the proliferation of visons, which are gapped excitations, one might expect to see a correlation between the transition temperature and the vison gap. This is indeed what we find for our family of 3D Kitaev models -- an almost perfect linear correlation between the two energy scales, as illustrated in Fig.~\ref{fig:CorrelationPlot}.

We mention only in passing that also the gapless excitations in the Majorana sector are generally expected to leave a fingerprint on thermodynamic observables. For instance, the low-temperature specific heat is expected to exhibit an algebraic temperature dependence $C_v(T) \propto T^{d_c}$, where the co-dimension $d_c$ is the difference between spatial dimension and Fermi surface dimension, i.e., $d_c = 1,2,3$ for a Majorana (semi-)metal with a Majorana Fermi surface, nodal line, and Weyl points in three spatial dimensions, respectively. Searching for such algebraic signatures in the specific heat has not been the focus of the current study
\footnote{In passing, we mention that algebraic signatures in the specific heat have been the subject of a recent study of the {\sl intermediate} temperature regime of a 2D Kitaev spin liquid \cite{Nasu2015thermal}.}.  

Methodologically, we have generalized the sign-free QMC simulation approach for Kitaev models. In particular, we expanded the approach from a purely Jordan-Wigner (JW) transformation-based, i.e., {\it non-local} approach, to an ansatz which makes use of Kitaev's original, local transformation from spins to Majorana fermions. We could show that the exactness of the local approach is guaranteed on lattice geometries where  a JW transformation is also applicable, and that it can be extended to larger gauge-sectors and systems with periodic boundary conditions. Under these conditions, the effect of non-physical states, which might arise in the local approach, is negligible already for moderately large system sizes (like the ones in the current study). 


\section{3D Kitaev models}  
\label{3DKitaevModels}

The Kitaev model is the prototypical example of an exactly solvable quantum spin liquid model \cite{Kitaev2006anyons}. Its essential ingredient are bond-directional Ising-type exchanges of the form
\begin{equation}
	\mathcal{H}_\text{Kitaev} = \sum_{\langle i,j\rangle, \gamma} J_{\gamma} \, \sigma^\gamma_i \sigma^{\gamma}_j \,,
	\label{eq:KitaevModel}
\end{equation}
which connect spin-1/2 local moments, represented by the Pauli matrices $\sigma_i$, along bonds labeled by $\gamma =x,y,z$, indicating three subclasses of bonds. Any given spin, subject to these three types of bond-directional interactions, cannot  simultaneously satisfy all couplings.  The resulting ``exchange frustration'' manifests itself already 
on the classical level \cite{Chandra2010,Sela2014} and is the origin for the emergence of spin liquid physics also in the quantum model. 

With three principle bond-directional exchange types it might be most natural to define the Kitaev model on tricoordinated lattice structures in two \cite{PhysRevB.76.180404, Kells_2011, yao-kivelson} and three spatial dimensions \cite{Mandal2009exactly,Hermanns2014quantum,Hermanns2015weyl,Obrien2016classification,Yamada2017}.
Generalizations to lattice geometries with higher coordination numbers have also been considered for lattice systems with odd \cite{WuHungGammaMatrix2009, ortiz_bond_algebra, DwivediCornerModes2018} and also even \cite{yao_algebraic_spin_liquid, ryu_kitaev_diamond_lattice, ryu_kitaev_sqr_lattice} coordination numbers. Typically, these generalization also consider higher-dimensional spin operators, which can be captured, e.g., via representations of $\Gamma$-matrices \cite{WuHungGammaMatrix2009, ortiz_bond_algebra, DwivediCornerModes2018, yao_algebraic_spin_liquid, ryu_kitaev_diamond_lattice, ryu_kitaev_sqr_lattice}. 

To solve the spin-1/2 Kitaev model for a tricoordinated lattice structure, a local transformation from spins to Majorana fermions can be applied. In doing so, each spin operator is replaced by four Majorana fermion operators via
\begin{equation}
\sigma_i^\gamma = ib_i^\gamma c_i \,.
\label{eq:LocalTransformation}
\end{equation}
The Majorana operators $b_i^\gamma$, $c_i$ fulfill the canonical commutation relations
\begin{align}
\{b_i^\alpha, b_j^\beta\} &= 2 \delta_{ij}\delta_{\alpha \beta},\nonumber\\
\{c_i, c_j\} &= 2 \delta_{ij},\nonumber\\
\{c_i, b_j\} &= 0,
\end{align}
and reproduce two of the four relations which define the algebra of spins: $(\sigma^\gamma)^2 = 1$ and $(\sigma^\gamma)^\dagger = \sigma^\gamma$. There remains a subtlety, however, since the other two relations $[\sigma^\alpha, \sigma^\beta] = 2i\epsilon_{\alpha \beta \gamma} \sigma^\gamma$ and $\{\sigma^\alpha, \sigma^\beta\} = 2\delta_{\alpha \beta}$ require the introduction of an additional condition. The reason for this is that the Majorana operators act on a 4-dimensional Fock space, while the Fock space of the spin operators has only dimension 2. The transformation \eqref{eq:LocalTransformation} thus artificially increases the local Hilbert space of each spin. Only a subspace reproduces the whole spin algebra, though, and can therefore be considered as physical. This physical subspace is obtained by introducing the local gauge transformation $D_i = b_i^x b_i^y b_i^z c_i$ and consists of all states $\ket{\xi}$ which are gauge-invariant, i.e., for which $D_i \ket{\xi} = \ket{\xi}$.

The transformation \eqref{eq:LocalTransformation} replaces the Ising terms of the Hamiltonian according to $\sigma_i^\gamma \sigma_j^\gamma = -i(ib_i^\gamma b_j^\gamma) c_i c_j$, i.e., it reveals an interaction that is quadratic in the Majorana fermions $c_i$. The remaining quantity in the interaction term are the bond operators $\hat{u}_{ij}^\gamma = ib_i^\gamma b_j^\gamma$, with $\hat{u}_{ji} = - \hat{u}_{ij}$. They have eigenvalues $\pm 1$ and commute with each other as well as with the Hamiltonian. Therefore, they can be replaced by their eigenvalues in the Hamiltonian, which then assumes the form
\begin{align}
	\mathcal{H} &= \frac{i}{4} \sum_{i,j} A_{ij} c_i c_j, \nonumber\\
	A_{ij} &= 2 J^\gamma u_{ij}^\gamma,
	\label{eq:MajoranaHamiltonian}	
\end{align}
for connected sites $i$,$j$ (otherwise $A_{ij}$ = 0).

Physically, this parton construction leads to a system where the spins are {\it fractionalized}. The new degrees of freedom are non-interacting Majorana fermions $c_i$, which are coupled to a static $\mathbb{Z}_2$ gauge field $\{u_{ij}\}$ on the lattice bonds. Given a $\mathbb{Z}_2$ gauge field configuration $\{u_{ij}\}$, the model is exactly solvable by diagonalizing the complex, Hermitian tight-binding matrix $iA$. The eigenvalues of $iA$ come in pairs $\pm \epsilon_\lambda$ and are the single-particle energy levels that can be occupied by $N/2$ spinless fermionic modes ($N$ being the number of lattice sites). The latter are obtained from the Majorana operators by applying a basis transformation to normal modes $b_\lambda', b_\lambda''$, and introducing fermionic operators $a_\lambda^\dagger = (b_\lambda' - ib_\lambda'')/2, \hspace{0.2cm} a_\lambda = (b_\lambda' + ib_\lambda'')/2$ (see Appendix \ref{sec:LocalTransformation} for the details). The diagonal representation of the Kitaev Hamiltonian then reads
\begin{equation}
\mathcal{H} = \sum_{\lambda = 1}^{N/2} \epsilon_\lambda \left(a_\lambda^\dagger a_\lambda - \frac{1}{2}\right). 
\label{eq:DiagHamiltonian}
\end{equation}
Note that the ground-state energy of this system  
\begin{equation}
E = -\frac{1}{2} \sum_{\lambda = 1}^{N/2} \epsilon_\lambda 
\label{eq:GSEnergy} 
\end{equation}
is simply the sum over the lower half of the energy levels, divided by two.

\subsection*{Lieb's Theorem}

As a result of the parton construction described above, the problem of obtaining the ground state is reduced to finding the $\mathbb{Z}_2$ gauge field configuration $\{u_{ij}\}$ which minimizes the energy \eqref{eq:GSEnergy}. However, the bond operators $\hat{u}_{ij}^\gamma$ are not gauge-invariant, since they change sign under the gauge transformation $D_i$. Therefore, underlying the $\mathbb{Z}_2$ gauge field $\{u_{ij}\}$, there has to be another, gauge-invariant quantity, which is expected to be associated with some physical observable. 

As it turns out, this quantity is already well known. It has been long established (see  Appendix \ref{sec:LiebTheorem}) that the energy spectrum of a tight-binding Hamiltonian with nearest-neighbor hopping  $t_{ij} = |t_{ij}|e^{i\phi_{ij}}$  only depends on the flux $\Phi$ around the elementary plaquettes $p$ 
of the underlying lattice\cite{Lieb1993, Lieb1994, Macris1996}, which is defined by

\begin{equation}
\exp \left( i \Phi \right) =  \frac{\prod_{\langle i,j\rangle \in p} t_{ij}}{\prod_{\langle i,j\rangle \in p} |t_{ij}|}.
\label{eq:PhaseFactor}
\end{equation}

In Kitaev systems, the plaquette flux operator defined in Eq.~\eqref{eq:PhaseFactor} is usually denoted by $\hat{W}_p$ and calculated by taking the product of the Kitaev bond terms around the plaquettes
\begin{equation}
	\hat{W}_p = \prod_{\langle i,j\rangle, \gamma \in p} \sigma^\gamma_i \sigma^{\gamma}_j
	        = \prod_{\langle i,j\rangle, \gamma \in p} (-i \hat{u}_{ij}^\gamma) \,. 
	\label{eq:PlaquetteFlux}
\end{equation}
The eigenvalues of this plaquette flux operator are $\pm 1$ for plaquettes with even and $\pm i$ for those with odd length. In accordance with Eq.~\eqref{eq:PhaseFactor}, we use the convention that an eigenvalue $W_p = +1 = e^{i0}$ is identified with the {\it absence} of a $\mathbb{Z}_2 $ plaquette flux ($\hat{=}$ $0$-flux), while an eigenvalue $-1 = e^{i\pi}$ signifies the {\it presence} of a $\mathbb{Z}_2 $ plaquette flux ($\hat{=}$ $\pi$-flux). We also define the average 
\begin{equation}
\overline{W_p} = \frac{1}{N_p} \sum_i^{N_p} W_{p_i}, 
\label{eq:AveragePlaquetteFlux}
\end{equation}
with respect to all $N_p$ elementary plaquettes in the lattice, for which numerical results are shown in Fig. \ref{fig:Wp_all}. $\hat{W}_p$ is a conserved quantity of the Kitaev model, as it commutes with the Hamiltonian. The Hilbert space is therefore divided into eigenspaces of $\hat{W}_p$, which are also referred to as flux sectors. In the Majorana basis, the flux sectors $\{W_p\}$ are realized by choosing an adequate $\mathbb{Z}_2$ gauge field configuration $\{u_{ij}\}$.

The remaining problem is to find the ground state plaquette flux for a tight-binding Hamiltonian with a half-filled band. This has been the subject of numerous studies in mathematical physics \cite{Lieb1993, Lieb1994, Macris1996}. The most striking result is a theorem by Lieb\cite{Lieb1994}, which states that the flux ground state of a lattice with (at least) semi-periodicity is determined by its elementary plaquette length $p$. Specifically, the plaquette flux  $\Phi$ takes the values
\begin{equation}
	\Phi = \pi(p-2)/2 \quad \quad \quad (\text{mod  } 2\pi) \,.
	\label{eq:LiebTheorem}
\end{equation}
That is, if $p\mod 4 = 0$ (e.g., for the square lattice), the energy of the half-filled system is minimized by a plaquette flux $\Phi = \pi$. Strikingly, this means that the presence of a magnetic flux {\it lowers} the energy of the system, a phenomenon which can only occur for systems where the electron density is high. On the other hand, if $p\mod 4 = 2$ (e.g., for the honeycomb lattice), the ground state plaquette flux is $\Phi = 0$. 

\begin{figure}[t]
    \centering
    \includegraphics[width=\columnwidth]{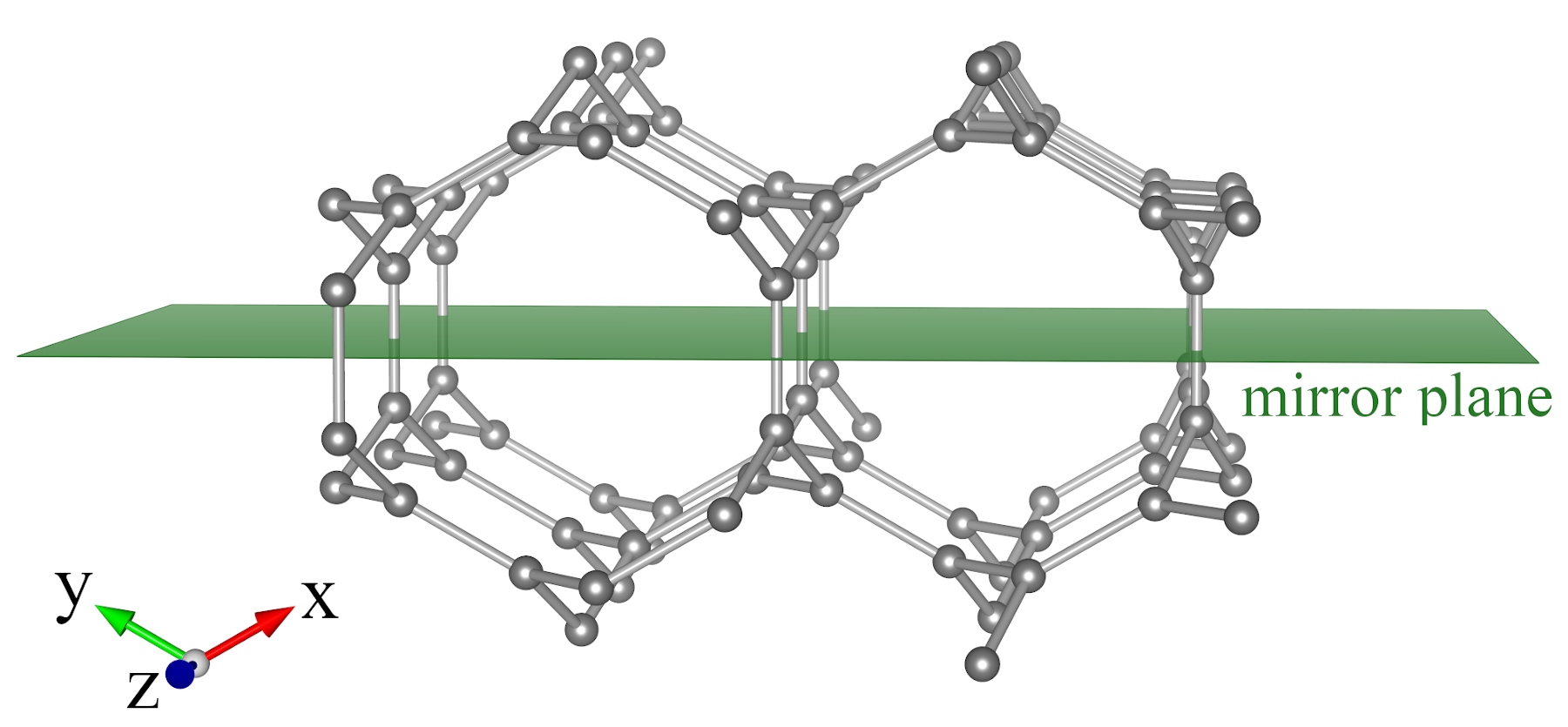}  
      \caption{{\bf Geometric condition for Lieb's theorem.} (8,3)b is the only elementary, tricoordinated 3D lattice that possesses a global mirror symmetry for which the mirror plane (green) does not cut through any lattice vertices. Two other such mirror planes are generated by 120-degree rotations around the $z$-axis (blue).}
    \label{fig:MirrorPlane}
\end{figure}

This theorem, however, is only rigorously proven for systems with a specific geometric condition: The whole lattice (i.e., the sites and bonds along with the configuration of coupling constants $J_\gamma$), and the individual plaquettes, for which the flux is minimized, have to be invariant under reflection symmetry. Also, the corresponding mirror plane may not cut through any vertices of the lattice. This condition is fulfilled for the honeycomb lattice
\footnote{Strictly speaking, Lieb's theorem only applies as long as at least two of the coupling constants are equal. But the presence of a finite vison gap ensures that the zero flux state must be the stable ground state configuration for a finite region in the phase diagram.}
, on which the Kitaev model was first defined:
Here, the elementary plaquettes are hexagonal and mirror-symmetric, and the ground state is flux-free ($W_p = 1$). On the other hand, from the 3D lattices studied in this manuscript, only the (8,3)b lattice possess mirror planes which fulfill this geometric condition -- it is therefore the only 3D lattice for which we can rigorously predict the ground-state flux assignment using Lieb's theorem
\footnote{
For completeness, we note that the (8,3)n lattice also possesses a number of mirror planes which fulfill the geometric condition of Lieb's theorem. However, these symmetries allow to strictly assign only seven of the eight elementary plaquettes in this lattice (each with a) $\pi$-flux in the ground state. 
}.

\begin{table}[b]
\footnotesize
\begin{tabular*}{\columnwidth}{@{\extracolsep{\fill} } l c c c}
\hline 
\hline
Lattice & Alternative names & Ground state flux sector & Lieb theorem \\ 
\hline 
(10,3)a & hyperoctagon\cite{Hermanns2014quantum} & 0 &  \\ 
(10,3)b & hyperhoneycomb \cite{Takayama2015hyperhoneycomb} & 0 &  \\ 
(10,3)c &   & 0 &  \\ 
(10,3)d &   & 0 &  \\ 
\noalign{\smallskip}
(8,3)a &   & $\pi$ &  \\ 
(8,3)b & hyperhexagon  & $\pi$ & Yes \\  
(8,3)n &   & $\pi$ &  \\ 
\noalign{\smallskip}
(8,3)c* &   & frustrated &  \\ 
(9,3)a* & hypernonagon & $\pm \pi/2$ &  \\ 
\hline
(6,3) & honeycomb & 0 & Yes \\
\hline 
\hline
\end{tabular*} 
\caption{{\bf Ground-state flux assignments} for all elementary 3D tricoordinated lattices. The plaquette length $p$ is the quantity which determines the ground state flux sector for all bipartite lattice systems (see Fig. \ref{fig:Wp_all}), while Lieb's Theorem is only strictly applicable for (8,3)b. The lattices (8,3)c and (9,3)a, indicated by an asterisk, do not possess the conventional ground state flux sector $0 / \pi$: For (8,3)c, additional geometric conditions determine a frustrated flux ground state, while the flux sector of the non-bipartite system (9,3)a is characterized by spontaneous breaking of time-reversal symmetry and plaquette fluxes $\pm \pi/2$.}
\label{TableLattices}
\end{table}

For all other lattice geometries, it is only through the numerically exact quantum Monte Carlo simulations discussed in the following that we can unambiguously assign the ground-state flux assignments -- see the numerical data in Fig.~\ref{fig:Wp_all} and the summary in Table \ref{TableLattices}. Notably, all lattice geometries with an even plaquette length (except the lattice (8,3)c where the fluxes are geometrically frustrated, see below) turn out to follow the general guidance of Lieb's theorem \eqref{eq:LiebTheorem}. This might suggest that the symmetry requirements for Lieb's theorem to hold might be relaxed at least for some lattice geometries.

If the elementary plaquette has an odd length -- i.e., the lattice is non-bipartite, as it is the case for (9,3)a -- Lieb's theorem does not apply at all. Here, the nature of the flux ground state is fundamentally different from the bipartite systems, since there is no energetic selection of one distinguished flux state, which all plaquettes assume. Instead, {\sl two} possible flux states $W_p = +i$ ($\hat{=} \pi/2$-flux) and $W_p = -i$ ($\hat{=} -\pi/2$-flux) are connected by time-reversal symmetry, and the ground state spontaneously breaks time-reversal symmetry by selecting either of the two states for all plaquettes. 

\subsection*{Vison excitations}

The finite-temperature thermodynamic signatures of the $\mathbb{Z}_2$ gauge theory underlying Kitaev spin liquids are closely linked to the elementary flux or ``vison" excitations. For all 3D Kitaev models considered in this manuscript, these vison excitations are gapped excitations, just like in the original Kitaev honeycomb model. The magnitude of the vison gap, however, varies for the various lattice geometries under consideration, as summarized in Table \ref{TableVisonGaps}.

\begin{table}[b]
\footnotesize
\begin{tabular*}{\columnwidth}{@{\extracolsep{\fill} } l c c c c}
\hline 
\hline
Lattice & Vison loop length & Vison gap $\Delta$ & Transition temperature $T_c$ & $T_c / \Delta$\\ 
\hline 
(10,3)a & 10 & 0.0299(13) & 0.00405(9)\cite{mishchenko_prb_96_2017} & 0.135(7) \\ 
(10,3)b & 6 & 0.0426(4) & 0.00519(9) \cite{Nasu2014vaporization} &  0.121(2) \\ 
(10,3)c & 3 & 0.046(2) & 0.0049(2) & 0.107(6) \\ 
(10,3)d & 6 & 0.030(2) & 0.00462(1) & 0.154(10) \\ 
\noalign{\smallskip}
(8,3)a & 2 & 0.0197(13) & 0.0044(8) & 0.22(4) \\ 
(8,3)b & 2 & 0.0532(3) & 0.0079(3) & 0.148(6) \\  
(8,3)n & 2 & 0.05397(10) & 0.0071(3) & 0.132(6) \\ 
\noalign{\smallskip}
(8,3)c* & 4 & 0.0219(6) & 0.0020(2) & 0.091(9) \\ 
(9,3)a* & 4 & 0.034(1)\cite{2020MishchenkoChiralSpinLiquids} & 0.00244(4)\cite{2020MishchenkoChiralSpinLiquids} & 0.072(2) \\ 
\hline
(6,3) & 0 & 0.09\cite{Obrien2016classification} & - & - \\
\hline 
\hline
\end{tabular*} 
\caption{Overview of the {\bf vison gaps $\Delta$ and critical temperatures $T_c$ of the 3D Kitaev systems}. Quantum Monte Carlo simulations show that the temperature at which the Kitaev systems undergo a thermal phase transition from a disordered to an ordered $\mathbb{Z}_2$ spin liquid is correlated with the size of the vison gap (see Fig. \ref{fig:CorrelationPlot}). For the lattices (8,3)c and (9,3)a, indicated by an asterisk, the physical mechanism underlying the low-temperature phase transition is different from the rest due to special characteristics of the gauge sectors in these systems.}
\label{TableVisonGaps}
\end{table}

It was already mentioned above that the nature of the vison excitations depends on the spatial dimensionality of the underlying 
lattice structure -- while vison in two spatial dimensions are point-like objects, they are loop-like objects in 3D settings. To understand the formation of a loop-like excitation for 3D lattice geometries, note that the elementary plaquettes form boundaries of {\sl closed volumes} in three dimensions. This gives rise to a geometric constraint on the fluxes, as one can show that the product of $W_p$ for plaquettes forming any closed volume is necessarily restricted to $1$. One may think of this geometric constraint as a divergence-free condition: Whenever a flux enters a closed volume through one plaquette, it also has to leave through another one. That is, there are no ``flux monopoles" allowed in the $\mathbb{Z}_2$ gauge sector of the Kitaev systems. As a consequence, the plaquette flux operators $W_p$ are linearly dependent. For the 3D lattice systems considered in this manuscript, we can find $M/2$ linearly independent plaquette flux operators per unit cell (with $M$ being the number of sites per unit cell). 

\begin{figure}[t]
   \centering
    \includegraphics[width=\columnwidth]{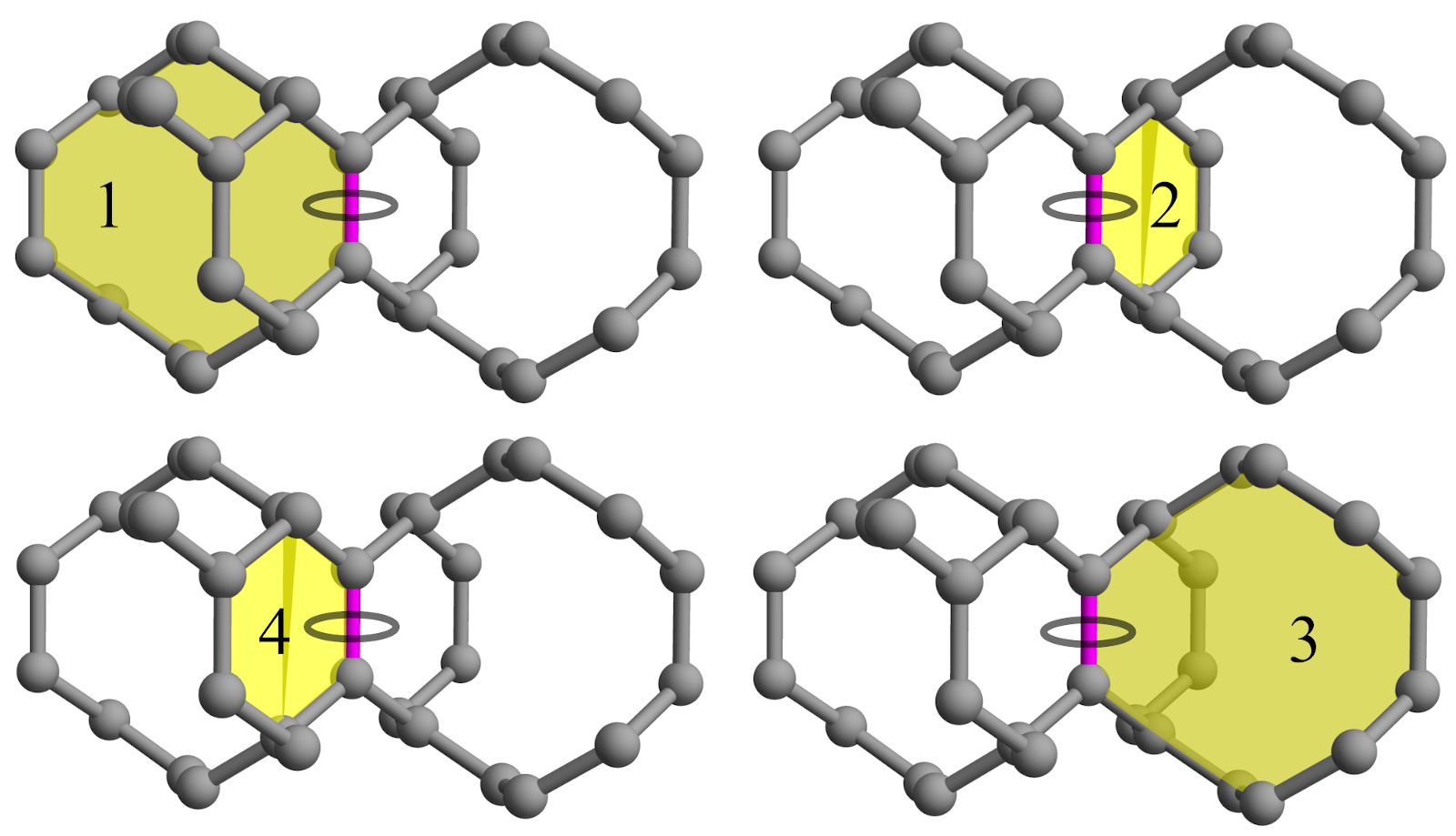}  
        \caption{{\bf Loop-like vison excitation on (8,3)b.} The excitation of a single $\mathbb{Z}_2$ gauge variable on an $x$-bond (red) creates a vison loop, which is generated by four linearly independent lattice plaquettes (yellow): Two plaquettes of length 8 (2,4) and two of length 12 (1,3). At the critical temperature $T_c$, the so-created vison loops proliferate and break open into system spanning objects. 
        }
    \label{fig:VisonLoop}
\end{figure}

The divergence-free condition has a remarkable effect on the visons: In 2D systems, the local change of a $\mathbb{Z}_2$ variable $u_{ij}^\gamma$ excites a pair of plaquette fluxes (resp. one plaquette flux, if the gauge variable is located at a boundary of the system). The visons are thus point-like. In addition, they can be arbitrarily far separated with a finite energy cost. In contrast to this, due to the linear dependence of plaquettes, the flip of a local gauge variable in three spatial dimensions always excites all plaquettes that surround the respective bond, i.e., it creates a loop-like excitation \footnote{We define the minimal vison loop length as the number of \emph{elementary} plaquettes whose eigenvalue is flipped by flipping a single bond operator. Thus, for the example of (8,3)b in Fig. \ref{fig:VisonLoop}, we count the two plaquettes of length 8, but not the ones of length 12. Note that if the bonds are not symmetry-equivalent (i.e., for all lattices except (10,3)a) the length depends on the bond type. In Table \ref{TableVisonGaps}, we state the smallest such value, usually obtained by flipping the bond operator on an $x$- or $y$-bond, for more details see \cite{Obrien2016classification} or \cite{Wells1977}.}. Enlarging the loop requires an energy cost that grows unbounded with the loop length. This fundamental difference in the nature of visons results in a no less fundamental dissimilarity of the thermodynamics: While in 2D, the $\mathbb{Z}_2$ plaquette fluxes {\it locally} freeze into their ordered ground state configuration at sufficiently low temperatures, realizing a {\it thermal crossover}, things are entirely different in 3D: Here, it is a {\it thermal phase transition} that separates the ordered from a disordered $\mathbb{Z}_2$ spin liquid regime: When the system temperature is increased away from zero, loop-like visons will start to form, extend and proliferate in the system. At a critical temperature $T_c$, these vison loops  transform into system-spanning excitations, a topological phenomenon which is highly {\it non-local}. 

\begin{figure}[t]
    \centering
    \includegraphics[width=\columnwidth]{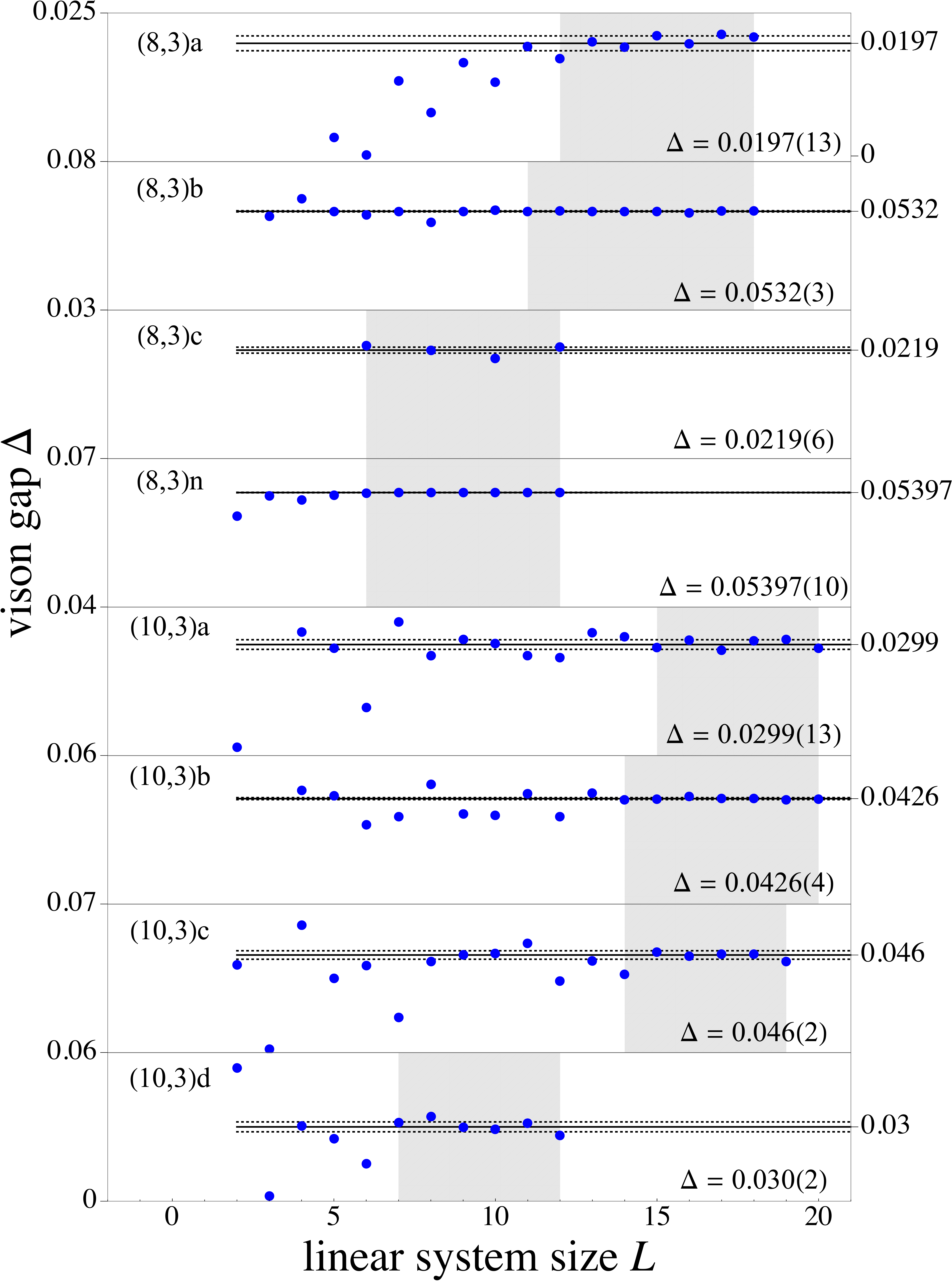}  
      \caption{{\bf Vison gap $\Delta$ of the smallest vison loop as a function of linear system size $L$.} For all lattices except (8,3)c and (10,3)d, we rescaled and refitted the data which was first presented in Ref.~\onlinecite{Obrien2016classification}. The shaded region indicates the data points included in the least-squares fit. The dashed horizontal lines indicate the uncertainty range of the fit.}
    \label{fig:VisonGapsPlot}
\end{figure}


\section{Sign-free quantum Monte Carlo} 
\label{SignFreeQMC} 
While the analytical approach to find exact solutions of Kitaev systems at zero temperature has been known since the original introduction of the model, its numerical investigation (at finite temperature) has initially been a challenge. While exact diagonalization (ED) of the spin system is always an option, its limitation to small system sizes makes it impractical for the 3D systems of interest here. Because of the non-commuting, frustrating nature of the bond-directional spin exchanges, conventional QMC approaches have also been discarded due to their intrinsic sign problem (in this spin basis). 
Instead, the key insight here is to actually use the change of perspective provided in the analytical approach. In its parton construction, i.e., in the basis of Majorana fermions and $\mathbb{Z}_2$ gauge fields, the QMC approach turns out to be completely free of the sign problem \cite{Nasu2014vaporization}. Such an approach, which is closely related to Monte Carlo studies of double-exchange models \cite{motome_MC_1999}, in fact makes use of the exact solubility of the Kitaev model in a fixed $\mathbb{Z}_2$ gauge field configuration $\{u_{ij}\}$. The key idea then is to do a Monte Carlo sampling of the $\mathbb{Z}_2$ gauge field as a local Ising variable, accompanied by an exact diagonalization of the respective tight-binding Majorana Hamiltonian \eqref{eq:MajoranaHamiltonian}. Since the latter is quadratic in the $c_i$-operators, an analytic expression for the partial Majorana partition function in a given $\mathbb{Z}_2$ gauge field background $\{u_{ij}\}$ can be found to be 
\begin{equation}
\mathcal{Z}_{\rm Maj}(\{u_{ij}\}) = \prod_{\lambda = 1}^{N/2} \left\{2 \cosh \left(\frac{\beta \epsilon_\lambda}{2}\right)\right\}.
\label{eq:MajoranaPartitionFunction}
\end{equation}
(see  Appendix \ref{sec:Observables} for the details of the derivation). From this quantity, all thermodynamic observables can be directly derived, particularly the free energy 
\begin{align}
F(\{u_{ij}\}) &= -T  \ln \mathcal{Z}_{\rm Maj}(\{u_{ij}\})\nonumber\\
&=  -T \sum_{\lambda = 1}^{N/2} \ln \left\{ 2 \cosh\left(\frac{\beta \epsilon_\lambda}{2} \right) \right\},
\label{eq:MajoranaFreeEnergy}
\end{align}
which is used to determine the Boltzmann weights $e^{-\beta \Delta F}$ in a Metropolis algorithm, giving the probability with which an update of the $\mathbb{Z}_2$ gauge field,  $\{u_{ij}\} \rightarrow \{u_{ij}'\}$, is accepted.  

\subsection*{Jordan-Wigner and local transformation}

There are, in fact, several different ways to transform the spin degrees of freedom of the system into Majorana fermions and a $\mathbb{Z}_2$ gauge field \cite{FuKnollePerkins2018}, two of which are considered in this manuscript: The first one, which was used in the first Kitaev-QMC simulations \cite{Nasu2014vaporization}, is built on a JW transformation \cite{chen_prb_76_2007, feng_prl_98_2007, chen_j_phys_a_41_2008, Nasu2014vaporization}. A second approach, on the other hand, is the one introduced above, which follows more closely Kitaev's original, local transformation approach \cite{Kitaev2006anyons} -- at the expense of an enlarged, local Hilbert space (whose unphysical parts are avoided in the JW approach).  

The JW transformation is well known as an exact solution method for one-dimensional spin models like the Heisenberg chain. Instead of locally transforming a spin according to \eqref{eq:LocalTransformation}, here it is a chainwise transformation that the spins undergo to turn into their Majorana representation. The JW chains have to be chosen to consist of two of the three subclasses of $\gamma$-bonds, while only the third class of bonds will carry the $\mathbb{Z}_2$ gauge degrees of freedom, here usually denoted by $\{\eta\}$. The major advantage of this approach is its faithfulness to the Hilbert space dimensionality of the spin model: In contrast to the local transformation, the JW ansatz makes no use on an artificial Hilbert space extension. This makes sure that one does not integrate over unphysical states when calculating the Majorana partition function and, subsequently, any thermodynamic observables. However, there remains a weak spot in this approach: In order to avoid the introduction of nonlocal parity terms that have to be considered when closing the JW strings, boundary conditions with broken JW strings (such as open boundary conditions in at least one spatial direction) have to be imposed \footnote{We can further reduce the number of broken bonds by imposing twisted boundary conditions\cite{Nasu2014vaporization}.}. 

In contrast, the local transformation approach imposes no restrictions on any boundary conditions, but at the expense of a Hilbert space that is artificially increased by a factor 2 for each spin. Thus, in order to avoid any unphysical contributions which might blur the numerical results, it might seem that a projection to the physical subspace needs to be included in each step of the QMC simulation. Such a projection operator was indeed introduced by Kitaev as 
\[
\mathcal{P}= \prod_i (1 + D_i)/2 \,,
\]
i.e., a symmetrization over all gauge transformations $D_i = b_i^xb_i^yb_i^zc_i$. A detailed analysis of this operator shows that on a given lattice, a fixed $\mathbb{Z}_2$ gauge field configuration will allow for fermionic states with only either even or odd parity\cite{Pedrocchi2011}. Thus, a summation over all fermionic states, regardless of the parity, is, in a strict sense, unphysical, and, in the derivation of the Majorana partition function, Eq.~\eqref{eq:MajoranaPartitionFunction}, parity weights have to be considered for each $\mathbb{Z}_2$ gauge field configuration\cite{Udagawa2018, udagawa2019spectroscopy}, which increases the computational cost of every QMC step and puts a restriction on the accessible system sizes.

However, it can be shown that for lattice systems whose principal geometry allows for a JW transformation, both approaches will lead to the same Hamiltonian, if, within the local approach, the $\mathbb{Z}_2$ gauge field on two subclasses of bonds is fixed to a specific configuration. In this case, the remaining $\mathbb{Z}_2$ gauge degrees of freedom $u_{ij}^\gamma$ on the $\gamma$-bonds are equivalent to the $\mathbb{Z}_2$ gauge variables $\eta$ from the JW approach (see also Appendices \ref{sec:LocalTransformation} and \ref{sec:JWTransformation}). Since we know that the JW transformation leads to an exact analytic expression for the Majorana partition function \eqref{eq:MajoranaPartitionFunction} and, consequently, all quantities derived from it, we are, in this case, guaranteed, that the QMC results are exact (see the Appendices \ref{sec:LocalTransformation} and \ref{sec:JWTransformation} for the technical details). It turns out that this is in fact the case for {\sl all} 3D Kitaev systems considered in this manuscript, if suitable (open) boundary conditions are applied in the direction of the JW strings. However,  benchmark calculations have shown us that even if we move away from this exact equivalence and sample over {\sl all} the $u_{ij}$ as $\mathbb{Z}_2$ gauge variables, our results still remain within the error margins of the exact results. The effect of sampling over ``too many bonds'' can therefore be interpreted as a mere overcounting of physical states, which does not affect the measurement results of the physical observables.

In addition, from the perspective of both transformation approaches, it is a scaling argument which justifies the extension of the QMC simulation also to systems with periodic boundary conditions in all spatial directions. From the perspective of the local transformation approach,  
the effect of adding a single additional fermion to a system with $N$ sites and a $\mathbb{Z}_2$ gauge field configuration $\{u_{ij}^\gamma\}$ (strictly allowing only for even / odd fermionic parity) would scale \cite{Zschocke2015} as $1/N$, such that the effect of the ``false fermion" can be neglected in the thermodynamic limit. Exactly the same argument holds for the parity term which appears in the non-local approach whenever a JW string is closed.

Based on these arguments, we performed the QMC simulations presented in this manuscript with the local transformation approach (treating the $\mathbb{Z}_2$ gauge field $\{u_{ij}\}$ on all bonds as free Ising variables), but assured that a JW transformation is possible on all underlying lattice geometries (see Appendix \ref{sec:JWTransformation}). The results for the (9,3)a Kitaev system that have formerly been presented in Ref. \onlinecite{2020MishchenkoChiralSpinLiquids} were obtained with the JW approach. With one exception \footnote{The (10,3)d system is the only system, where we performed the QMC simulations on systems with open boundary conditions along two of the spatial dimensions (due to a failure of the KPM-approach).
}, we performed the QMC simulations on systems with periodic boundary conditions in all spatial directions and system sizes up to $\sim 2000$ lattice sites, which not only justifies the neglect of parity terms but also allows for an extrapolation of infinite-size estimates for some relevant quantities of interest (e.g., the critical temperature $T_c$).

Finally, we note that empirically we find in our numerical simulations that the behavior of the thermodynamic observables becomes more systematic if we work with lattice geometries/boundary conditions where all JW strings have approximately equal length.

\subsection*{Green-function-based kernel polynomial method}

One bottleneck in the numerical simulation is the exact diagonalization of the Majorana Hamiltonian, i.e., the tight-binding matrix $\tilde{A} := iA$ of Eq.~\eqref{eq:MajoranaHamiltonian}: For every single Monte Carlo update between two $\mathbb{Z}_2$ gauge field configurations $\{u_{ij}\}$ and $\{u_{ij}'\}$, the calculation of transition probability via the Boltzmann weights  $e^{-\beta (F' - F)}$ requires a calculation of the Majorana free energy 
\[
   F = -T \sum_{\lambda = 1}^{N/2} \ln \left\{ 2 \cosh\left(\frac{\beta \epsilon_\lambda}{2} \right) \right\} \,,
   \label{eq:MajoranaFreeEnergy2}
\]  
based on the full sets of eigenvalues, $\{\epsilon_\lambda\}$ and $\{\epsilon_\lambda'\}$. Doing a full-fledge calculation of these eigenvalues via an exact diagonalization step, we denote the resulting quantum Monte Carlo approach as QMC-ED. For a lattice with $N$ sites, this requires that in every step a matrix of size $N \times N$ has to be diagonalized - a calculation that scales as $\mathcal{O} (N^3)$, using the conventional divide-and-conquer algorithms. Even on state-of-the-art high-performance compute clusters, this limits the accessible system sizes to about $N = 1000$ sites. While this has proved sufficient to extract most of the physics for 2D systems, an exhaustive study of their 3D counterparts asks for a significant speed-up of the algorithm. 

Computationally, a local update of the $\mathbb{Z}_2$ gauge variable on a particular bond $\langle i,j \rangle$ means changing the signs of a single pair of entries $\tilde{A}_{ij}, \tilde{A}_{ji}$ of the Hermitian matrix $\tilde{A}$. This update can be written as $\tilde{A} \longrightarrow \tilde{A}' = \tilde{A} + \Delta$, and the question arises if the information stored in $\tilde{A}$ can be further used, in order to avoid the recalculation of the whole spectrum. This is indeed the case, as shown for the  "Green-function-based kernel polynomial method" (GF-KPM)\cite{Weisse2006, Weisse2009}  -- an efficient algorithm that enables us to calculate the Majorana free energy \eqref{eq:MajoranaFreeEnergy2} change under a local bond update without explicit exact diagonalization \cite{mishchenko_prb_96_2017}. Its rationale is the use of Green functions, which are approximated in terms of Chebyshev polynomials.

Defining the Green function belonging to the matrix $\tilde{A}$ as $G(E) = (\tilde{A} - E\cdot \mathbb{I})^{-1}$, the spectrum of the updated matrix $\tilde{A}'$ is given by the roots of the function $d(E) = \det \{\mathbb{I} + G(E)\Delta\}$. Since $\Delta$ is a rank-2 matrix, the expression for $d(E)$ contains only four Green functions\cite{Weisse2009}: 
\begin{align}
d(E) = \{1 &+ \Delta_{ij}G_{ji}(E)\}\{1 - \Delta_{ji}G_{ij}(E)\} \nonumber\\ &+ \Delta_{ij}\Delta_{ji} G_{jj}G_{ii}\,.
\end{align} 
The off-diagonal Green functions $G_{ij}(E)$ can be further expressed in terms of the diagonal Green functions $G_{ii}(E)$ (see Appendix \ref{sec:KPM} for the details).

We extend the domain of the function $d(E)$ to the complex plane by the analytic continuation $E \rightarrow z:= E + i \epsilon$. It can now be related to the change in the density of states (DOS) during the MC update by 
\begin{equation}
N\left\{\rho(E) - \rho(E')\right\} = \frac{1}{\pi} {\rm Im} \left\{ \lim_{\epsilon \rightarrow 0} \frac{d \ln d(z)}{d z} \right\}\,.
\label{eq:deltaDOS}
\end{equation}
The key step in the MC update is the calculation of the diagonal Green functions $G_{ii}(z)$ by the Chebyshev approximation

\begin{equation}
G_{ii}(E + i \epsilon) = i\frac{\mu_0 + 2 \sum_{m=1}^{M-1}\mu_m \exp \left\{-i m \arccos(E/s) \right\}}{\sqrt{s^2 - E^2}} \,,
\end{equation}
where $s$ is the bandwidth of the system. It has to be calculated in the beginning of the simulation by a sufficient number, e.g., 1000, of ED calculations of $\tilde{A}$ with random $\mathbb{Z}_2$ gauge field configurations. The $\mu_m$ are the Chebyshev moments: 

\begin{equation}
\mu_{m} = g_m \bra{i} T_m(H/s) \ket{i}. 
\end{equation}

The moments $\bra{i} T_m(H/s) \ket{i}$ are iterated by the recursion $T_m(x) = 2x T_{m-1}(x) - T_{m-2}(x)$, and $g_m$ denotes the Jackson kernel factor (see Appendix \ref{sec:KPM}). In our simulations, we usually used 256-512 Chebyshev moments. Having the Green functions and $d(z)$, the expression for the free energy change during the step $\tilde{A} \rightarrow \tilde{A}'$ follows easily from the change in the DOS by partial integration

\begin{equation}
\Delta F = -\frac{N}{\pi} \int_0^\infty \lim_{\epsilon \rightarrow 0} d(E + i\epsilon) \frac{1}{2} \tanh \left( \frac{\beta E}{2} \right) dE,
\end{equation}
which enables us to calculate the Boltzmann weights without performing another exact diagonalization. In practice, we here perform a standard numerical integration that is restricted to the half-open interval $[0,s)$. In order to optimize the convergence of the integration, the number of abscissas should correspond to the number of Chebyshev moments. 

\begin{figure}[t]
\centering
\includegraphics[width=0.49\columnwidth]{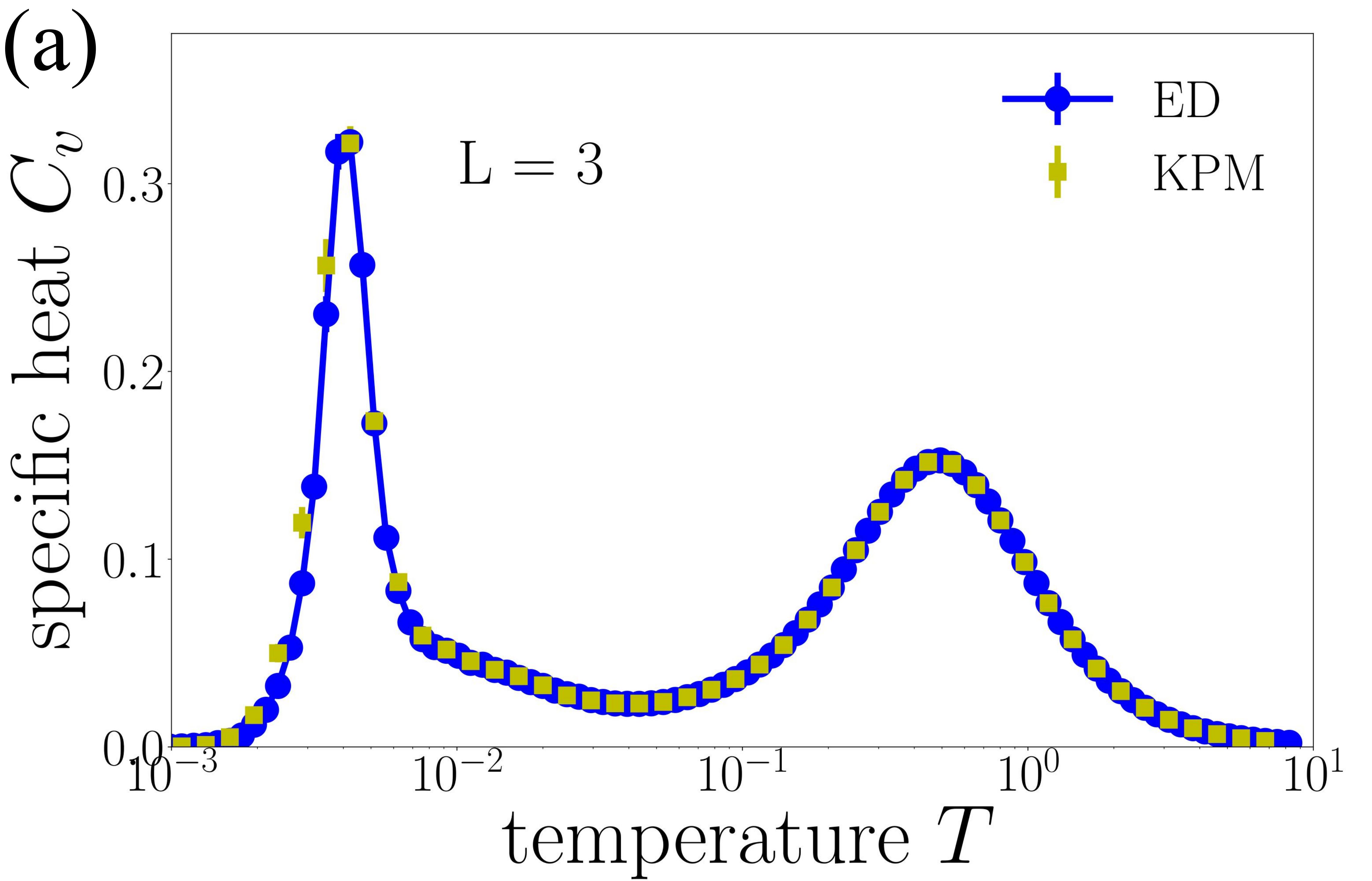}  
\includegraphics[width=0.49\columnwidth]{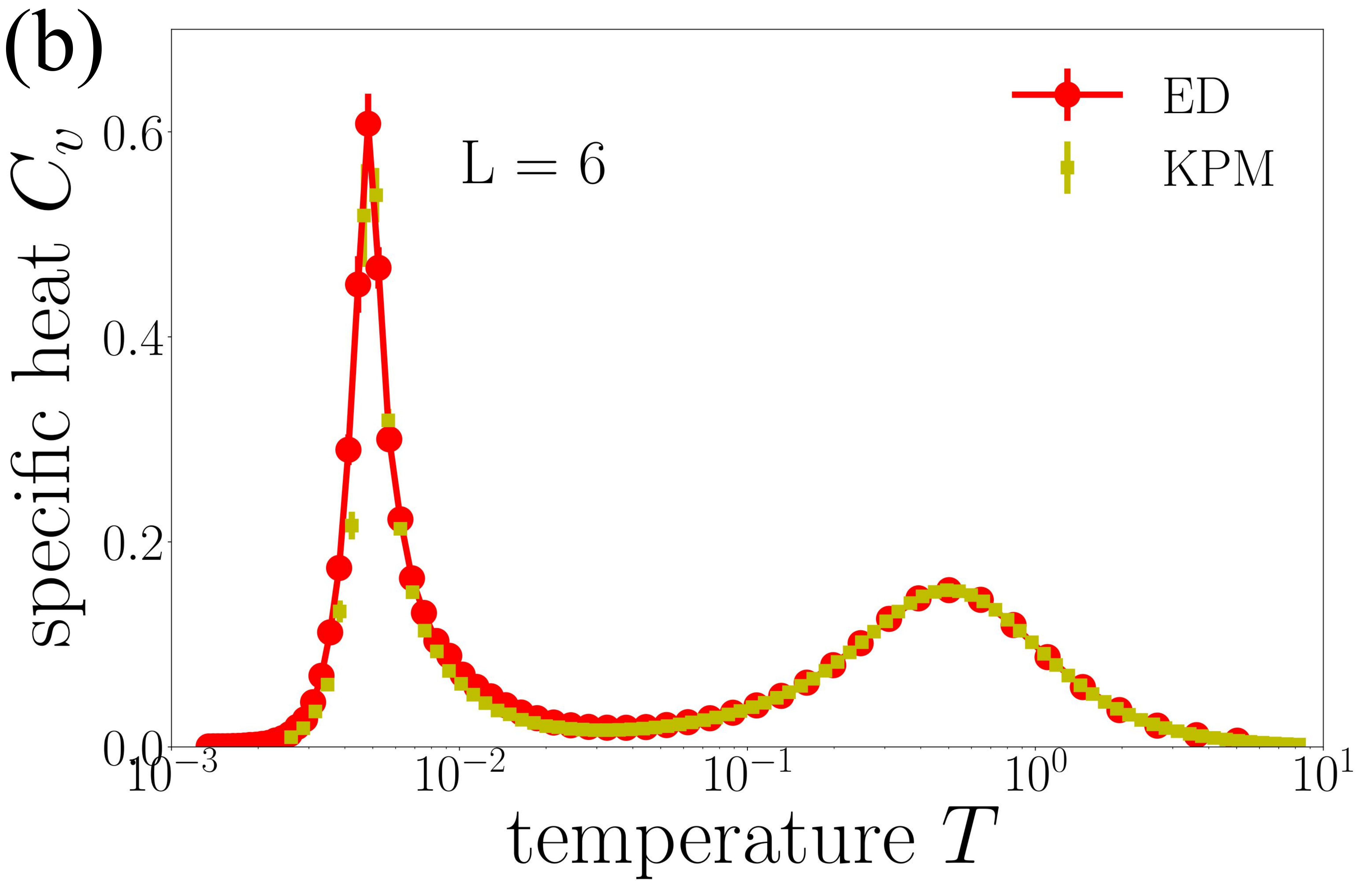}\\
\includegraphics[width=0.49\columnwidth]{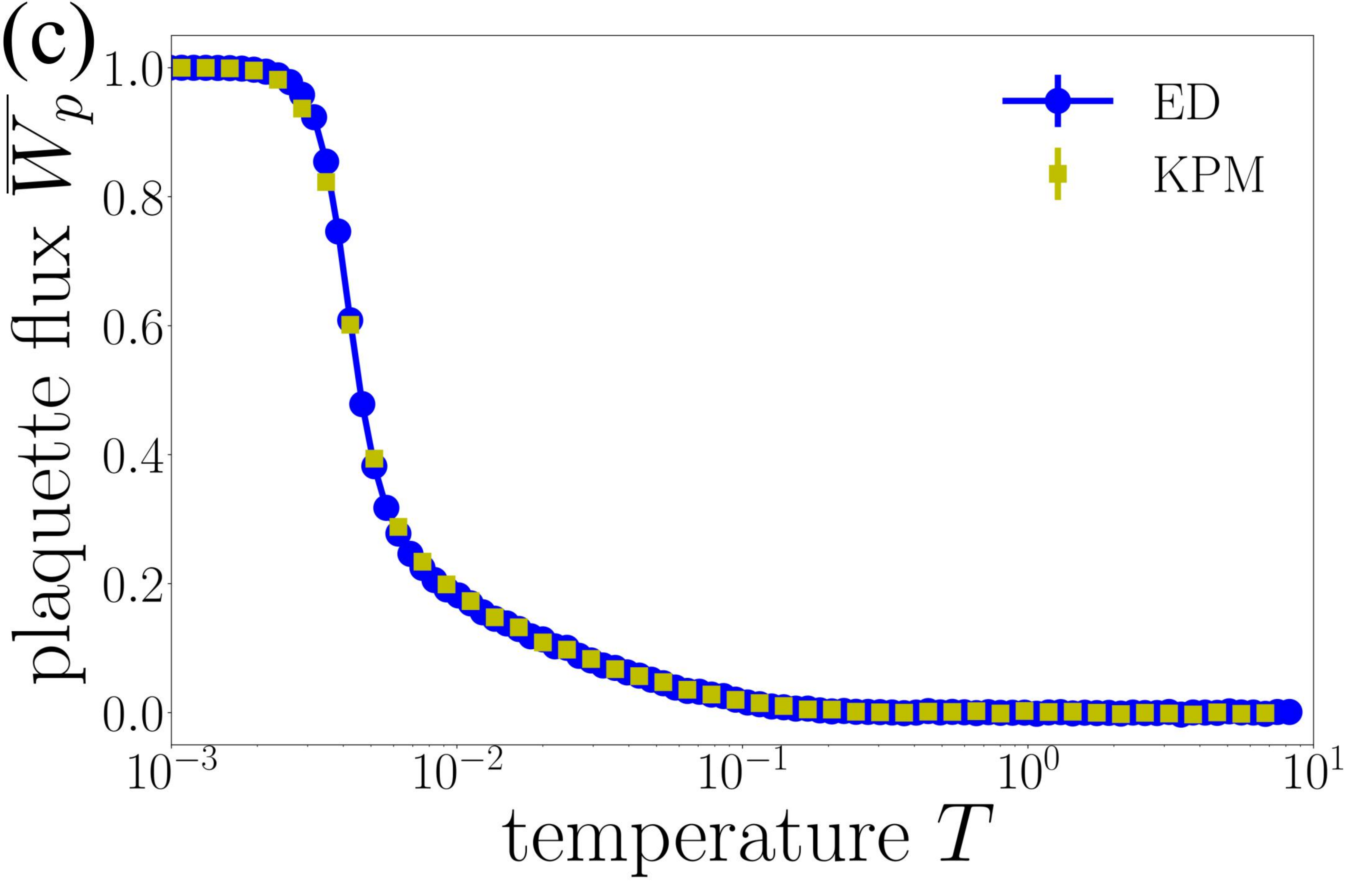} 
\includegraphics[width=0.49\columnwidth]{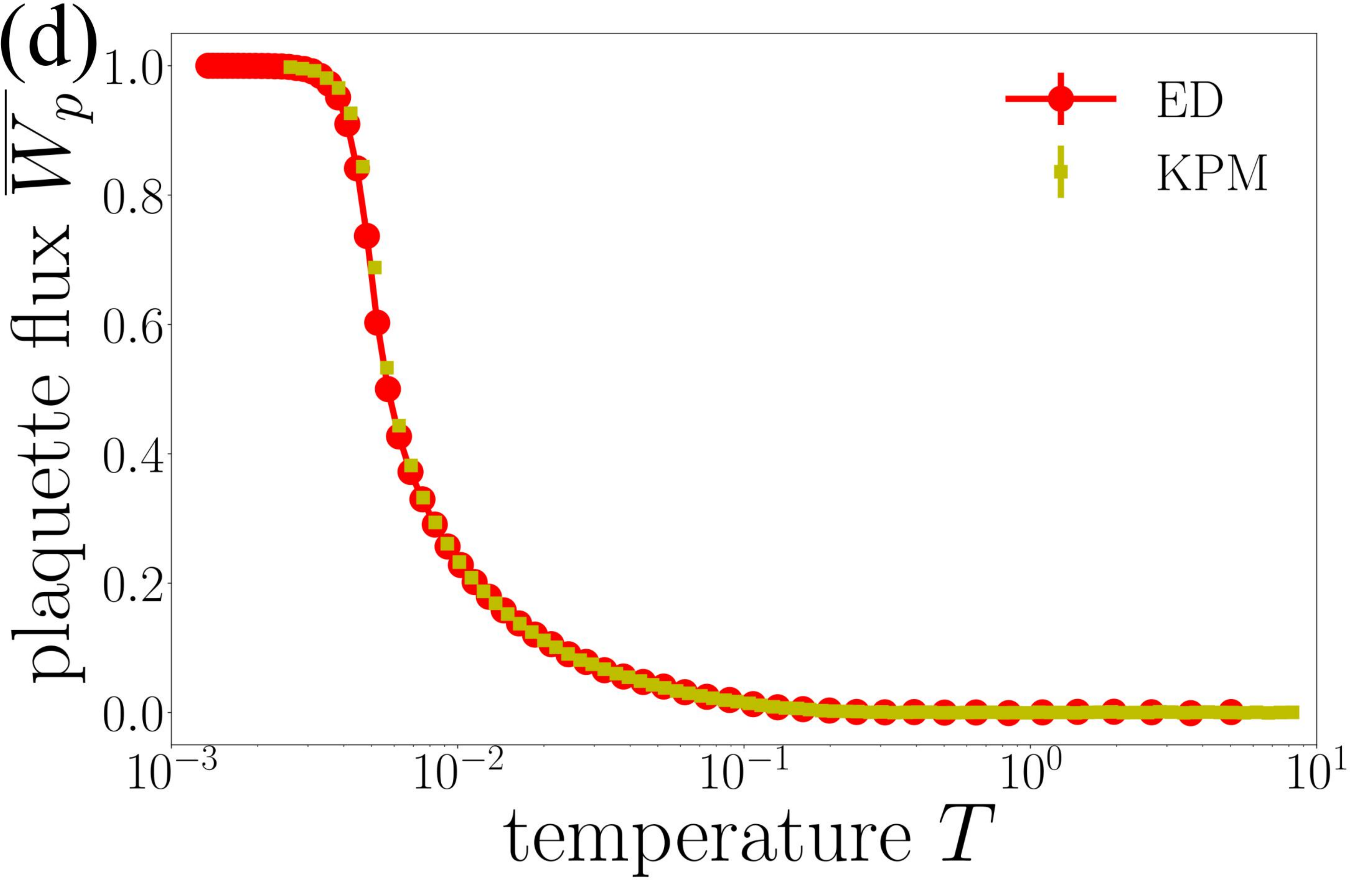}
   \caption{{\bf Comparison of numerical results for the QMC-KPM / QMC-ED method.} Data shown are the specific heat $C_v(T)$ (top) and average plaquette flux $\overline{W}_p$ (bottom) on a (10,3)b Kitaev system (with periodic boundary conditions in the ${\bf a_3}$-direction) with 108 sites, $L = 3$ (a,c) and 864 sites, $L = 6$ (b,d). The number of Chebyshev moments is $M = 512$. Error bars are smaller than the symbol sizes.}
    \label{fig:KPMBenchmark}
\end{figure}

In this scheme, the most time-consuming calculation in each Monte Carlo update is the iteration of the moments $\bra{i} T_m(H/s) \ket{i}$, which is done by subsequent sparse matrix-vector multiplications. This calculation can be most efficiently performed if the matrices and vectors are stored in the compressed-row storage format (CRS). Then, the numerical effort of the matrix-vector multiplication is reduced to $\mathcal{O}(N)$.

A comparison of numerical results for the QMC-KPM and QMC-ED method is presented in Fig. \ref{fig:KPMBenchmark}. It has to be remarked that the GF-KPM method gives sufficiently exact results only for the free energy calculation during the Monte Carlo update and can in practice not be used for the calculation of thermodynamic observables or the Boltzmann weights for replica exchange steps (for the latter, more than four Green functions would be required). Consequently, an exact diagonalization of the Hamiltonian remains necessary after each MC sweep (which, in our case, consists of $N$ MC updates) when observables are evaluated. With the GF-KPM method, lattice system sizes of $N \sim 2000$ become accessible on high-performance computing systems in a reasonable amount of time -- a specific heat plot like the one shown in Fig. \ref{fig:SpHTwoPeaks} takes about 500,000 core hours to calculate. 

The GF-KPM method is not applicable to all lattice systems. Benchmark calculations have shown that it fails with systems whose DOS shows exotic features like delta functions: In our studies, we faced this problem with lattice geometries (10,3)d and (8,3)c, the latter with strong anisotropy in the $J_\gamma$-couplings. 
Therefore, while we performed the QMC simulations for (10,3)d with the QMC-ED method, the simulations for all other systems were done with GF-KPM (in the following also denoted by QMC-KPM). A typical QMC simulation consists of 10,000 sweeps for the thermalization, followed by at least 10,000 sweeps for the measurements. In order to improve the convergence of the simulation at low temperatures, a replica exchange step between nearest neighbor replicas was performed for all temperature points after each MC sweep. 


\section{Thermodynamics} 
\label{Thermodynamics} 

We now turn to a detailed discussion of our results on the thermodynamics of 3D Kitaev systems. We start by revisiting the general two-peak signature in the specific heat \cite{Nasu2014vaporization, Nasu2015thermal, 2019MotomeReviewKitaevMagnets} and then discuss the dependence of these signatures with regard to the underlying lattice geometries. We round off our discussion of the gauge thermodynamics of these 3D Kitaev systems by briefly pointing out two distinct phenomena that occur in two special lattice geometries -- the phenomenon of ``gauge frustration" in the (8,3)c lattice, which was extensively discussed in Ref.~\onlinecite{2019EschmannGaugeFrustration}, and the spontaneous breaking of time-reversal symmetry in the (9,3)a lattice, which has been subject of Ref.~\onlinecite{2020MishchenkoChiralSpinLiquids}.


\subsection{Thermodynamic signatures}

\begin{figure}[t]
   \centering
    \includegraphics[width=\columnwidth]{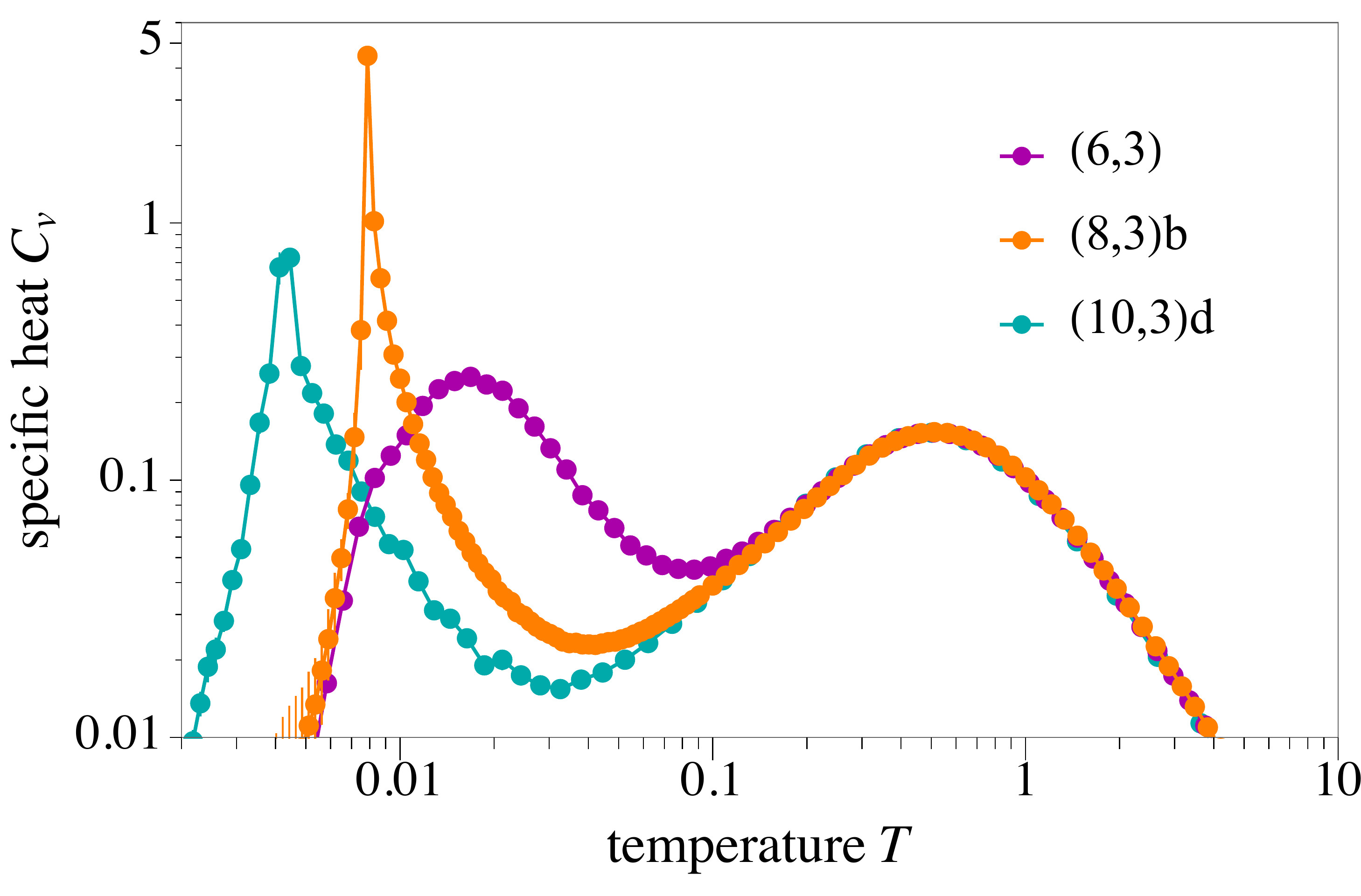}  
   \caption{{\bf Characteristic two-peak signature in the specific heat} $C_{v}$ for 3D Kitaev systems (here: (8,3)b, $L = 7$, and (10,3)d, $L = 5$), and for the 2D Kitaev Honeycomb model, (6,3), $L = 16$. For the latter, the low-temperature peak indicates a thermal crossover\cite{Nasu2015thermal}. Error bars are indicated, but mostly smaller than the symbol sizes.}
    \label{fig:SpHTwoPeaks}
\end{figure}

The principle thermodynamic signature of all Kitaev models, independent of spatial dimension and underlying lattice geometry, is a specific heat with a distinct two-peak structure. Its origin can be rationalized when considering the physics that must play out as one goes from a high-temperature paramagnet to a ground state that is characterized by a Majorana metal (or, more generally, some Majorana band structure) and a statically ordered $\mathbb{Z}_2$ gauge background. This transition from high to low-temperature physics occurs, in fact, in two steps that are closely linked to the two constituents of the parton perspective, where the original spin degrees of freedom fractionalize into emergent (fractional) degrees of freeom -- Majorana fermions and $\mathbb{Z}_2$ gauge fields. At some temperature $T' \sim J$ (with $J$ being the coupling constant of the Kitaev interactions) this fractionalization actually happens, while at some lower temperature $T_c$ -- which is numerically found to be of the order of $T_c \sim J/100$ -- the $\mathbb{Z}_2$ gauge field orders. An example of this general two-peak structure is shown (on a doubly-logarithmic scale) in Fig.~\ref{fig:SpHTwoPeaks} for two representative 3D lattices and, for comparison, the 2D honeycomb lattice. Here and in the following, the energy scale is set by the coupling strength $J_x = J_y = J_z = 1/3$. 

\begin{figure}[t]
   \centering
    \includegraphics[width=\columnwidth]{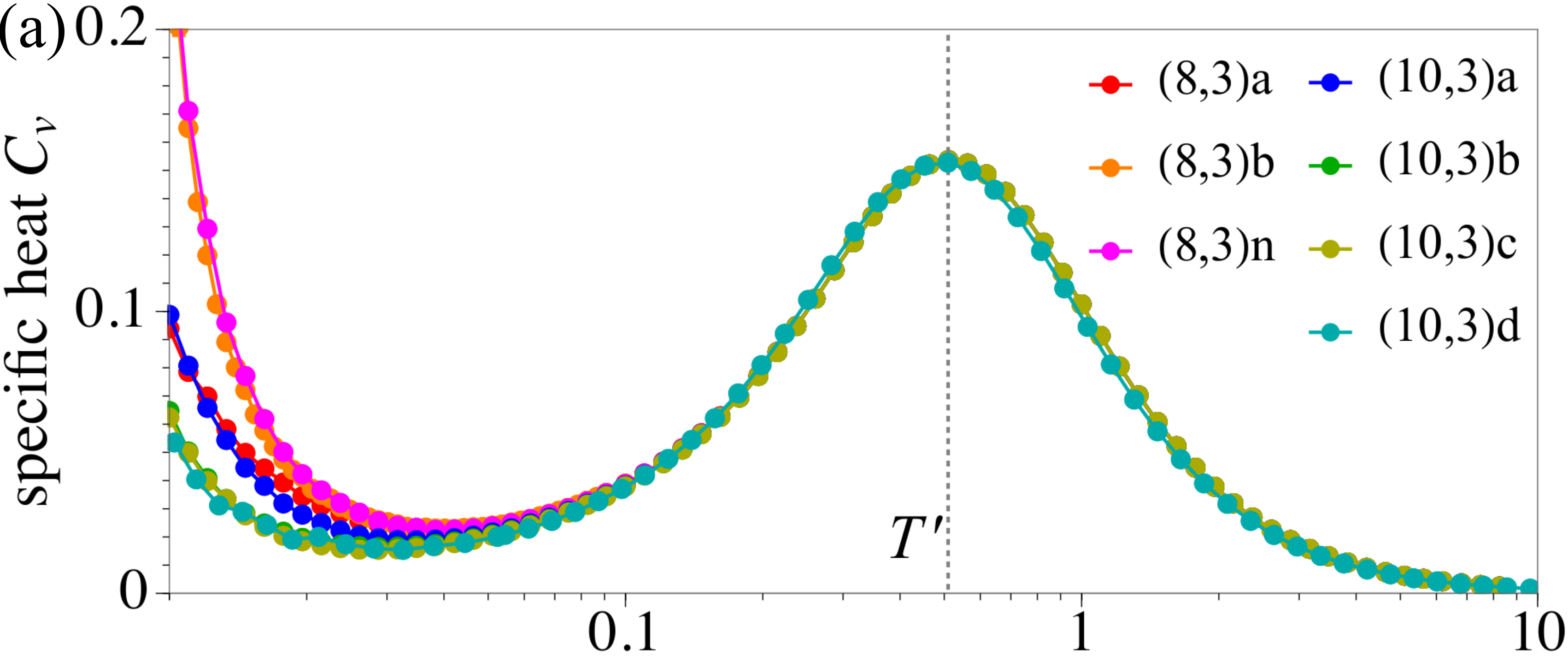}\\ 
    \includegraphics[width=\columnwidth]{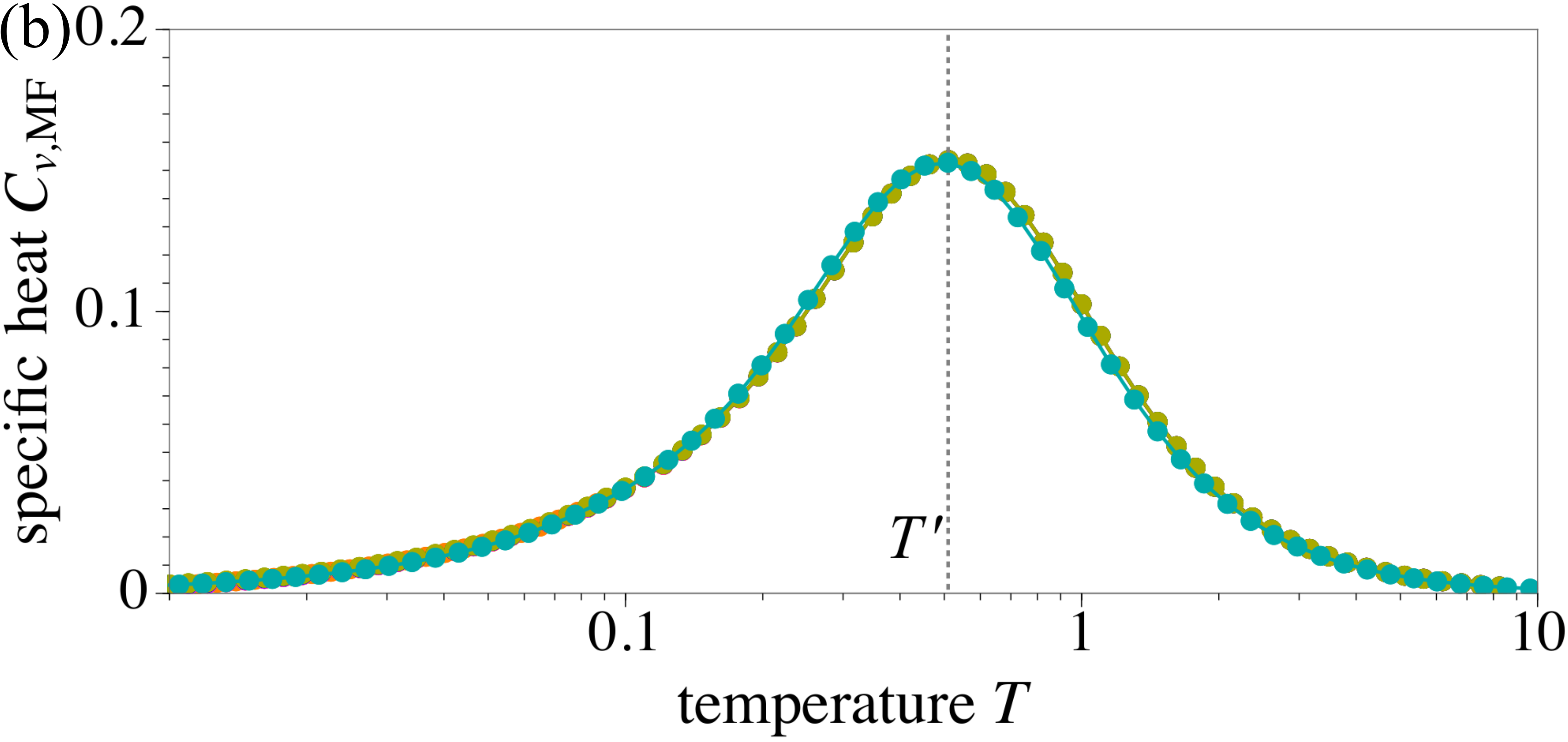}  
      \caption{{\bf High-temperature crossover} in the specific heat $C_v$. Panel (a) shows the full specific heat, while panel (b) only the contribution arising in the Majorana fermion sector $C_{v,{\rm MF}}$. This high-temperature peak, which exhibits no finite-size scaling, indicates a thermal crossover, caused by the (local) fractionalization of spins into (itinerant) Majorana fermions and a (static) $\mathbb{Z}_2$ gauge field. The position and shapes of the peaks are (nearly) equal for all lattice systems considered in this study, which again underlines the strictly local character of the thermal crossover. Data shown is for the linear system size $L=7$, except for (8,3)n and (10,3)d (here, $L = 5$). The dotted line indicates the crossover temperature $T' = 0.51(5)$. Error bars are smaller than the symbol sizes.}
      \label{fig:SpH_all_highT}
\end{figure}

\begin{figure}[b]
   \centering
    \includegraphics[width=\columnwidth]{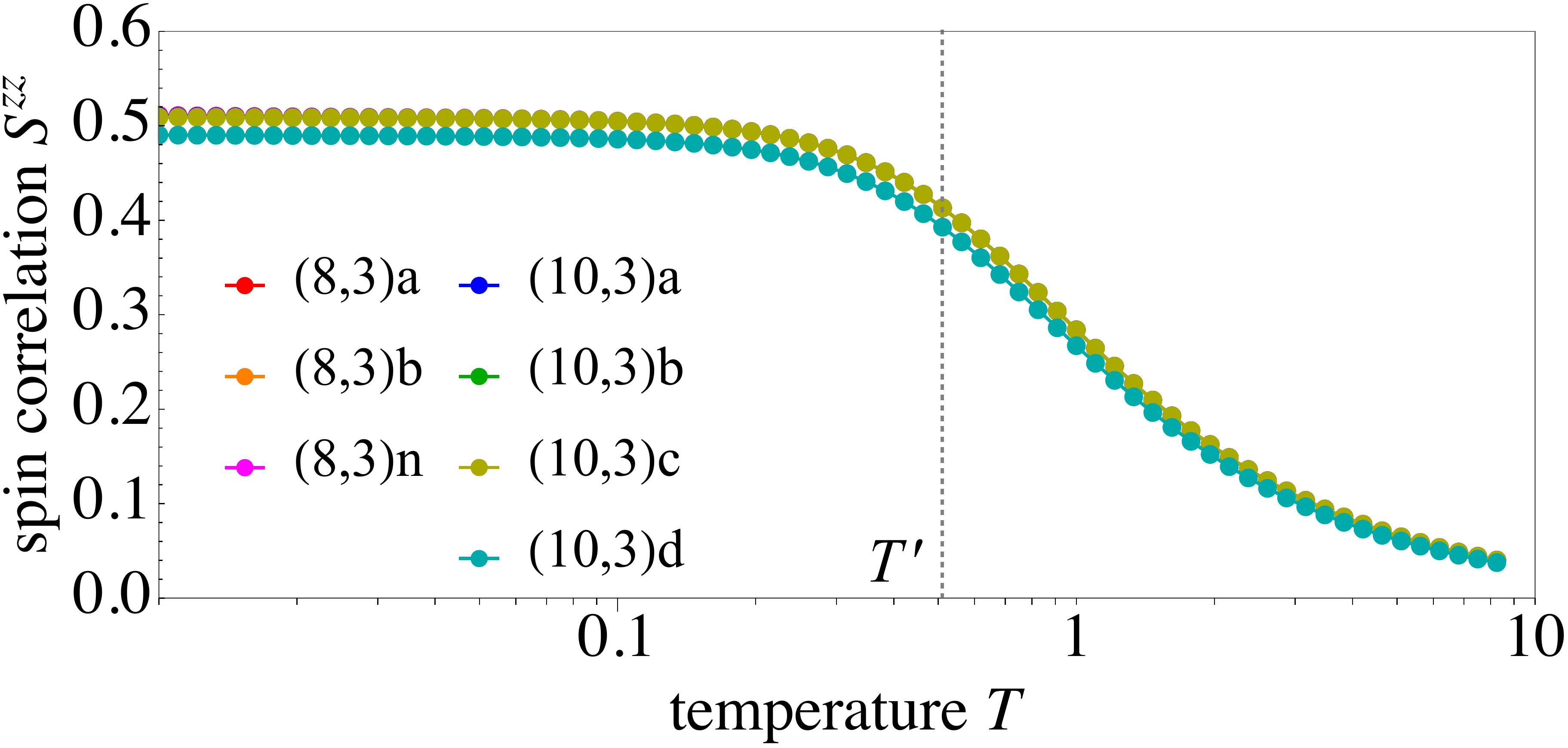}  
   \caption{{\bf Spin correlation $S^{zz}$}. 
   Below the thermal crossover at $T' = 0.51(5)$ (indicated by the dotted line), the correlator assumes a finite plateau value for all lattice systems. 
   Data shown is for the linear system size $L=7$, except for (8,3)n and (10,3)d (here, $L = 5$). Error bars are smaller than the symbol sizes.}
    \label{fig:SpinCorrelator}
\end{figure}


\subsection{Thermal crossover and local spin fractionalization}

Let us first turn to the higher temperature feature -- the fractionalization of the original spin degrees of freedom. This is a purely local phenomenon and therefore results in a {\sl thermal crossover}, i.e., a peak-like feature in the specific heat that is largely insensitive to the specifics of the  lattice geometry (and the system size), as documented in Fig.~\ref{fig:SpH_all_highT} where specific heat traces from different lattice geometries almost perfectly collapse onto one another. They only start to slowly differ at temperatures below $T \sim 0.1$, an order of magnitude below the scale of the actual crossover phenomenon. 

\begin{figure}[t]
   \centering
    \includegraphics[width=\columnwidth]{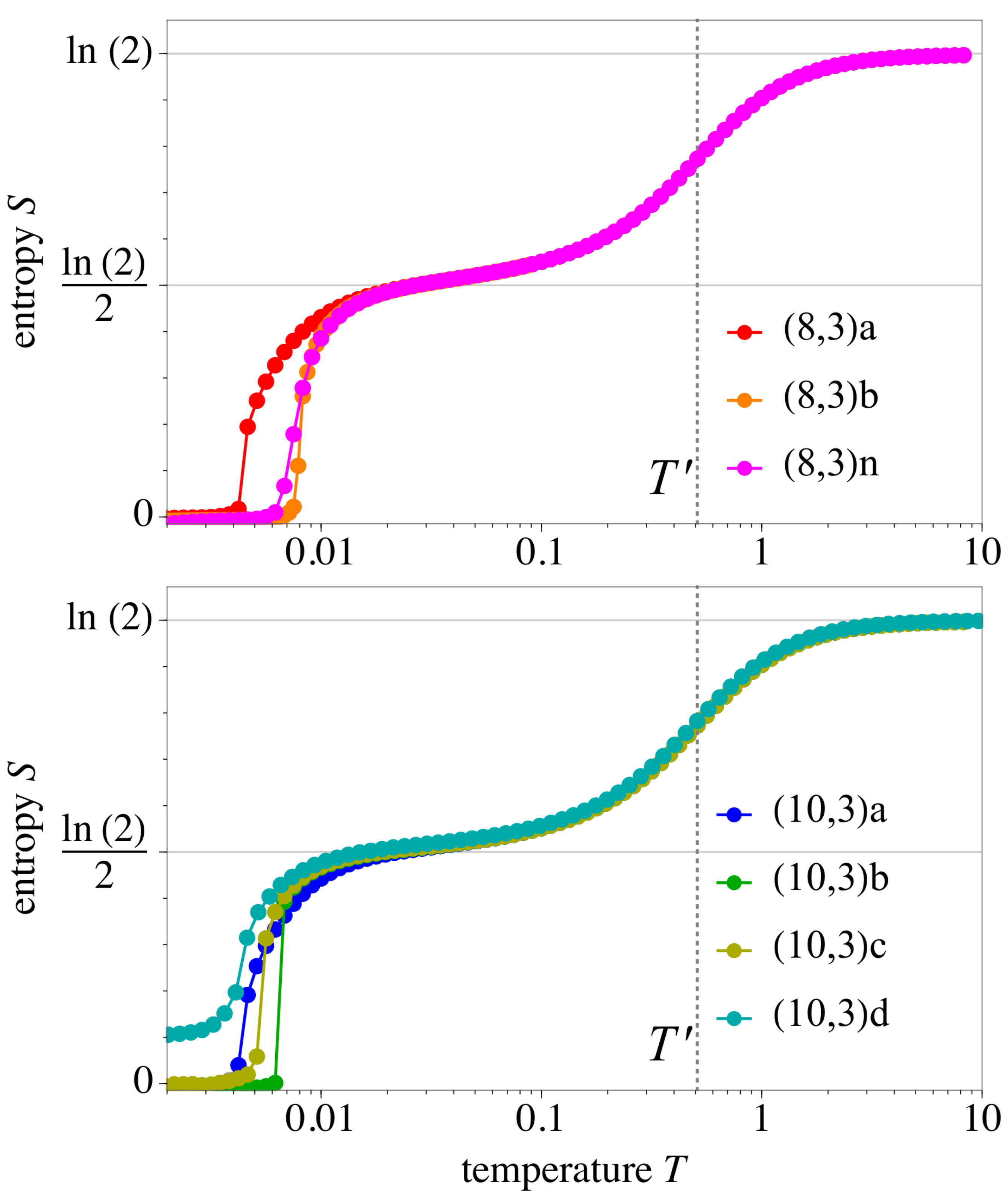}  
   \caption{{\bf Entropy $S$ per site.} The entropy in the paramagnetic (high-temperature) region is $\ln 2$, indicating two possible states per spin. At the thermal crossover, the system releases half of its entropy due to the fractionalization of the spins into (itinerant) Majorana fermions and a (static) $\mathbb{Z}_2$ gauge field. The latter remains disordered in this intermediate region. It is at the thermal phase transition at $T \sim \mathcal{O}(J/100)$ that the system releases its rest entropy, and the $\mathbb{Z}_2$ gauge field freezes into an ordered configuration, indicated by a sharp drop. For (10,3)d, open boundary conditions in two spatial directions cause a residual entropy for $T \rightarrow 0$, since $\mathbb{Z}_2$ gauge degrees of freedom on the edge bonds may fluctuate without a cost in energy.  The residual entropy approaches 0 when the system size is increased. Data shown is for the linear system size $L=7$, except for (8,3)n and (10,3)d (here: $L = 5$). Error bars are smaller than the symbol sizes.}
    \label{fig:Entropy}
\end{figure}

What sets the temperature scale for this thermal crossover? One hint comes from the uniform behavior of the spin-spin-correlation function $S^{\gamma \gamma} = \frac{2}{N} \sum_{\langle j,k \rangle_\gamma} \langle \sigma_j^\gamma \sigma_k^\gamma \rangle$ for all lattice geometries: It strictly vanishes above the crossover scale, but quickly saturates to its finite, low-temperature value right at this crossover temperature $T'$ (Fig. \ref{fig:SpinCorrelator}). Note that this spin-spin-correlation function is precisely equivalent to the kinetic energy $-i\langle c_i c_j \rangle_\gamma$ of the emergent Majorana fermions \cite{Nasu2015thermal,2019MotomeReviewKitaevMagnets}.  
Indeed, all observable features of the system in this crossover regime  are almost entirely governed by the physics of these Majorana fermions.
For instance, if one measures the specific heat contribution only of the Majorana fermions via the derivative of the internal Majorana energy $E_f$ with respect to the inverse temperature (see Appendix \ref{sec:Observables} for a detailed derivation)
\begin{equation}
C_{v,\text{MF}}(T) = -\frac{1}{T^2} \left \langle \frac{\partial E_f(\{u_{jk}\})}{\partial \beta} \right \rangle_{\rm MC},
\end{equation}
where $\langle \cdot \rangle_{\rm MC}$ denotes a numerical Monte Carlo average (which, technically, corresponds to a sample average of $\mathbb{Z}_2$ gauge field configurations $\{u_{jk}\}$), one finds that one can capture the complete crossover signature of the {\sl entire} spin system. This is illustrated in the lower panel of Fig.~\ref{fig:SpH_all_highT}, which shows only this Majorana contribution to the specific heat. The associated entropy release is plotted in Fig.~\ref{fig:Entropy}, which shows that the system releases precisely {\sl half} of its entropy at the thermal crossover, dropping from $\ln 2$ in the high-temperature regime to a plateau of $\frac{1}{2}\ln 2$ below the crossover temperature. This implies that all entropy associated with the Majorana fermion degrees of freedom is released -- the Majorana fermions enter their low-temperature state, i.e., they form a Majorana band structure whose details, however, might still depend on the (still disordered) $\mathbb{Z}_2$ gauge background. 
This also explains that the shape of the crossover peak is somewhat sensitive to a variation of the underlying coupling parameters in the Kitaev model, which in turn alter the characteristic energy scale of the Majorana fermions. For instance, if one moves away from the isotropic coupling point to strongly anisotropic couplings, e.g., $J_z \gg J_x, J_y$, this will shift the specific heat peak --  consistent with a change of the center of mass in the DOS of the Majorana fermion band \cite{Nasu2015thermal, 2019MotomeReviewKitaevMagnets}.

The complete insensitivity of the thermal crossover feature on the underlying lattice geometry and system size (which we also checked) illustrates that the associated fractionalization phenomenon is a generic feature of all Kitaev systems. It goes beyond the 3D systems at the heart of the current study and is found also in 2D geometries~\cite{Nasu2015thermal}, non-bipartite lattices~\cite{Nasu2015, 2020MishchenkoChiralSpinLiquids}, and even for generalized (higher-spin) Kitaev systems on lattice geometries with higher coordination numbers~\cite{DwivediCornerModes2018}.


\subsection{Thermal phase transition and $\mathbb{Z}_2$ gauge field ordering}

\begin{figure}[t]
   \centering
    \includegraphics[width=\columnwidth]{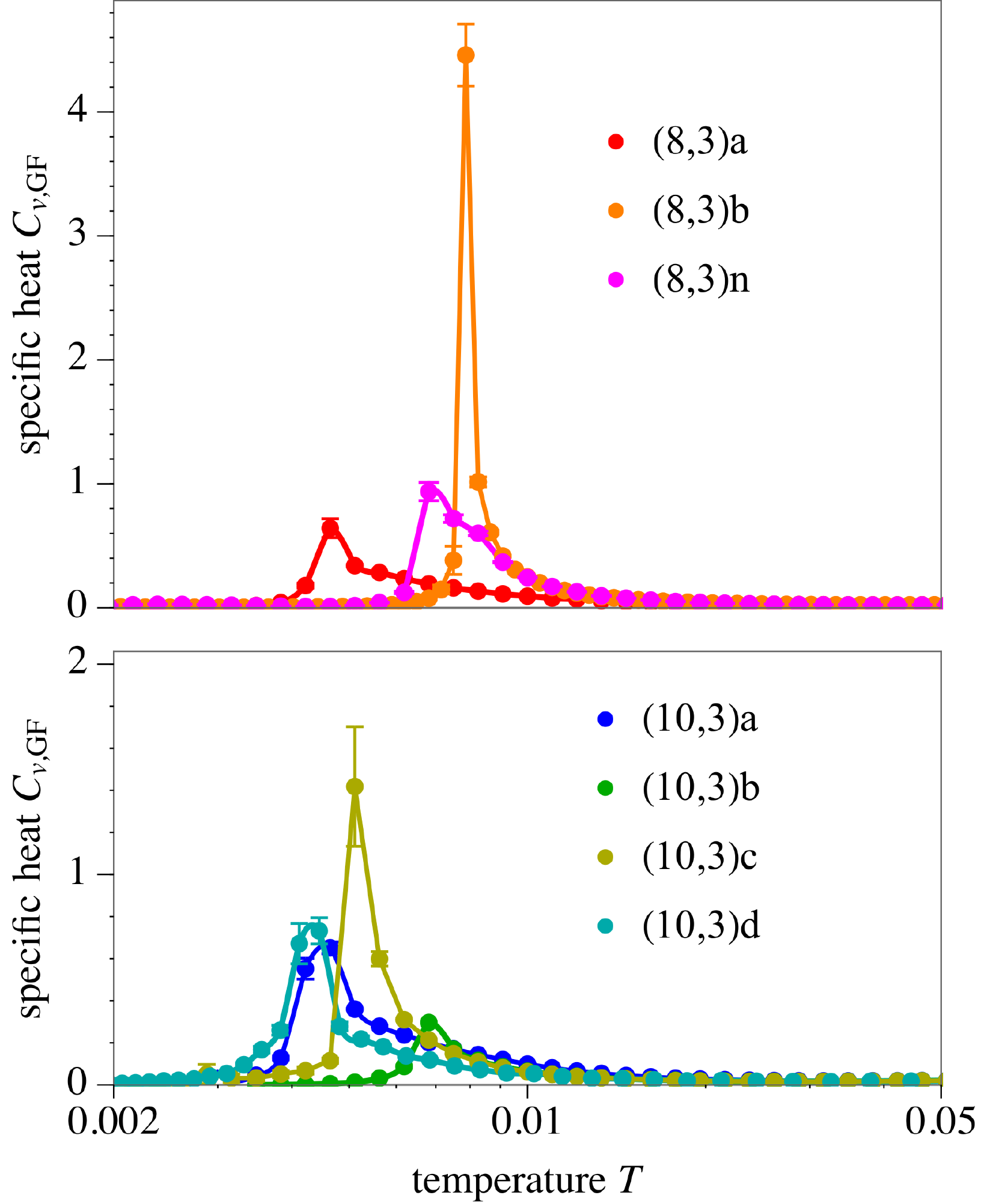}  
   \caption{{\bf Signature of thermal ``gauge ordering" transition} in the specific heat $C_{v,\text{GF}}$ (gauge field contribution) in the low-temperature region. The low-temperature peak indicates a thermal phase transition, caused by the ordering of the $\mathbb{Z}_2$ gauge field. Its position in temperature space is lattice-specific and correlated with the size of vison gap $\Delta$ (see Fig. \ref{fig:CorrelationPlot}). Data shown is for the linear system size $L=7$, except for (8,3)n and (10,3)d (here: $L = 5$).}
    \label{fig:SpH_all_lowT}
\end{figure}

\begin{figure}[t]
   \centering
    \includegraphics[width=0.49\columnwidth]{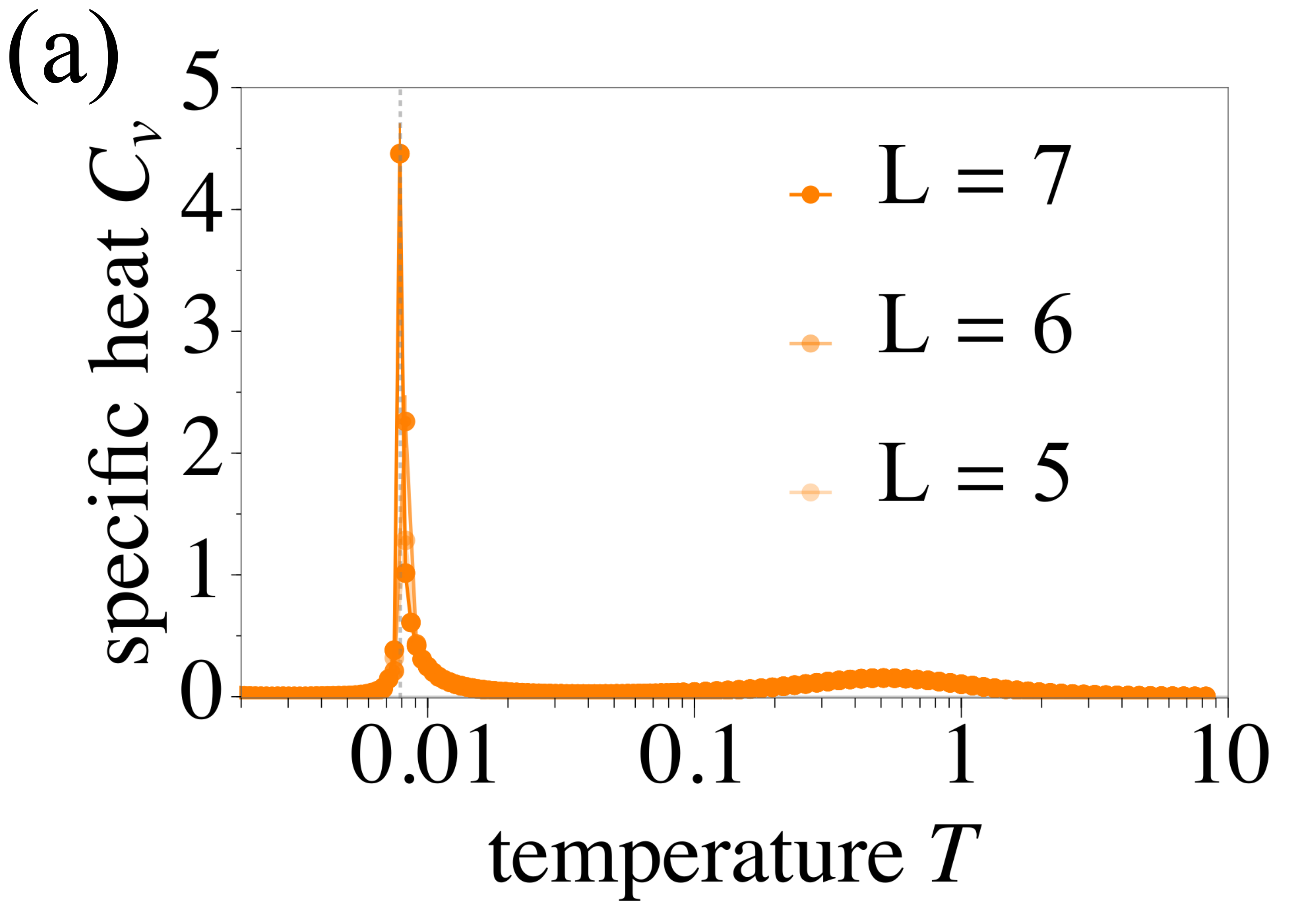}
    \includegraphics[width=0.49\columnwidth]{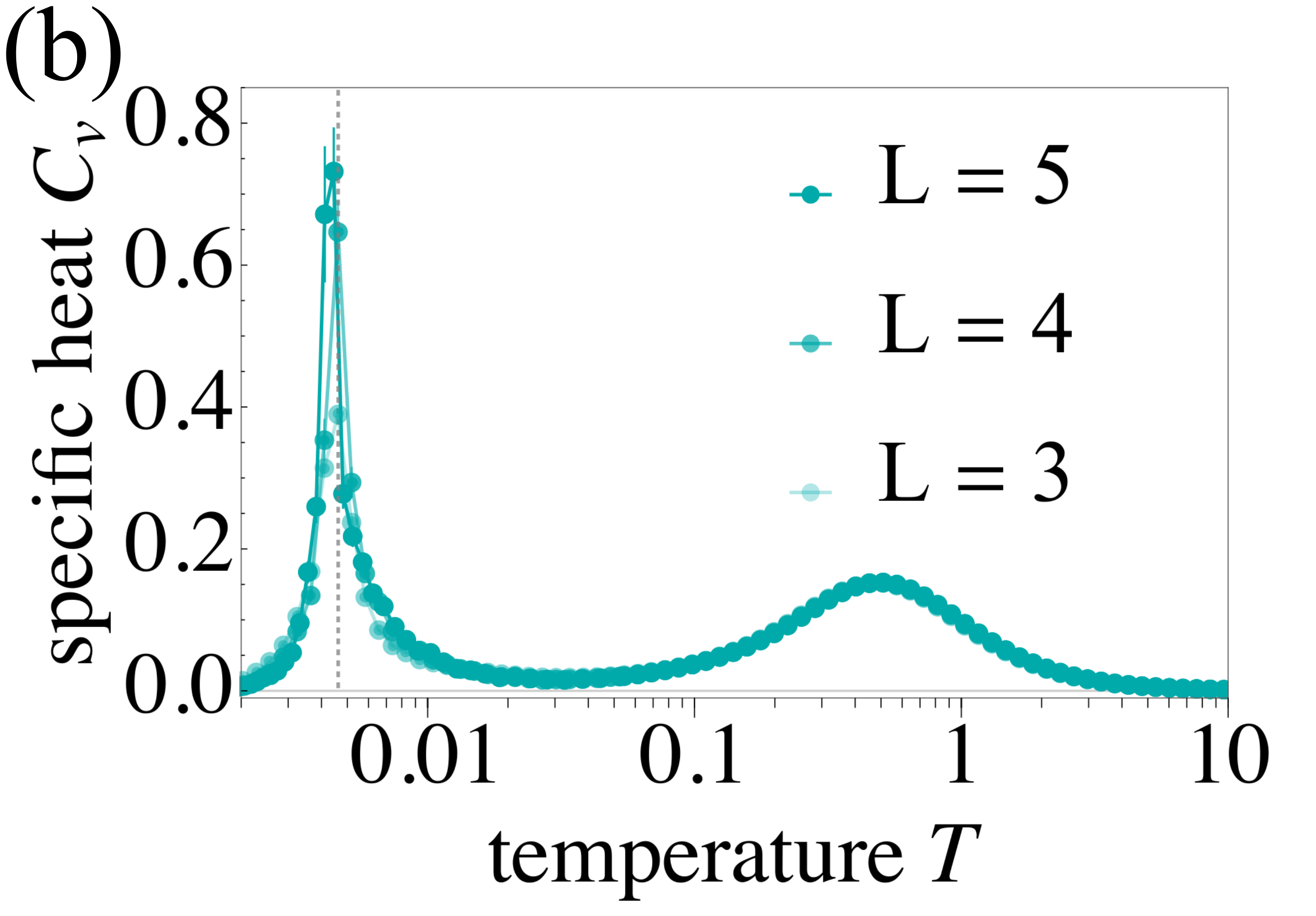}\\
    \includegraphics[width=0.49\columnwidth]{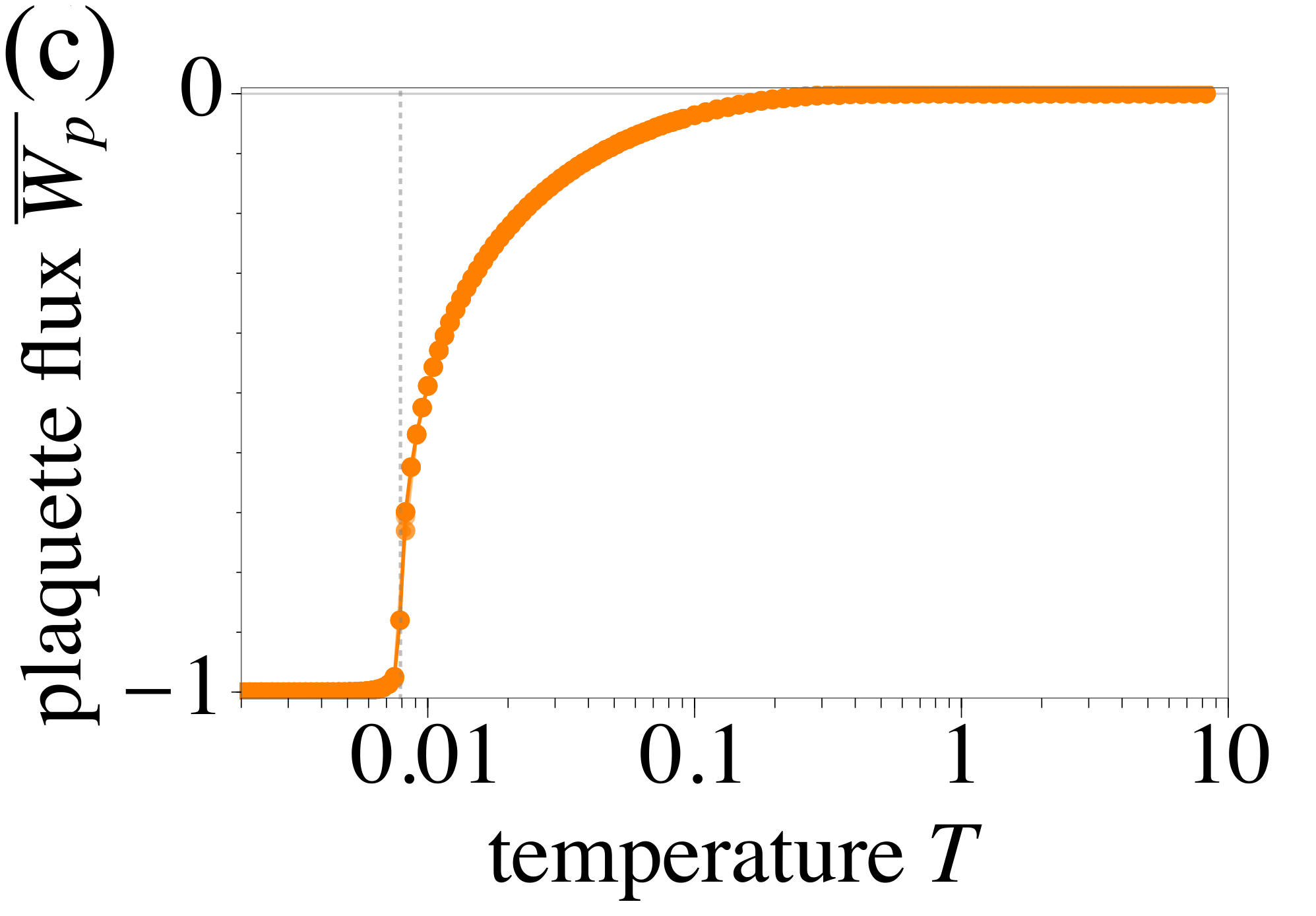}  
    \includegraphics[width=0.49\columnwidth]{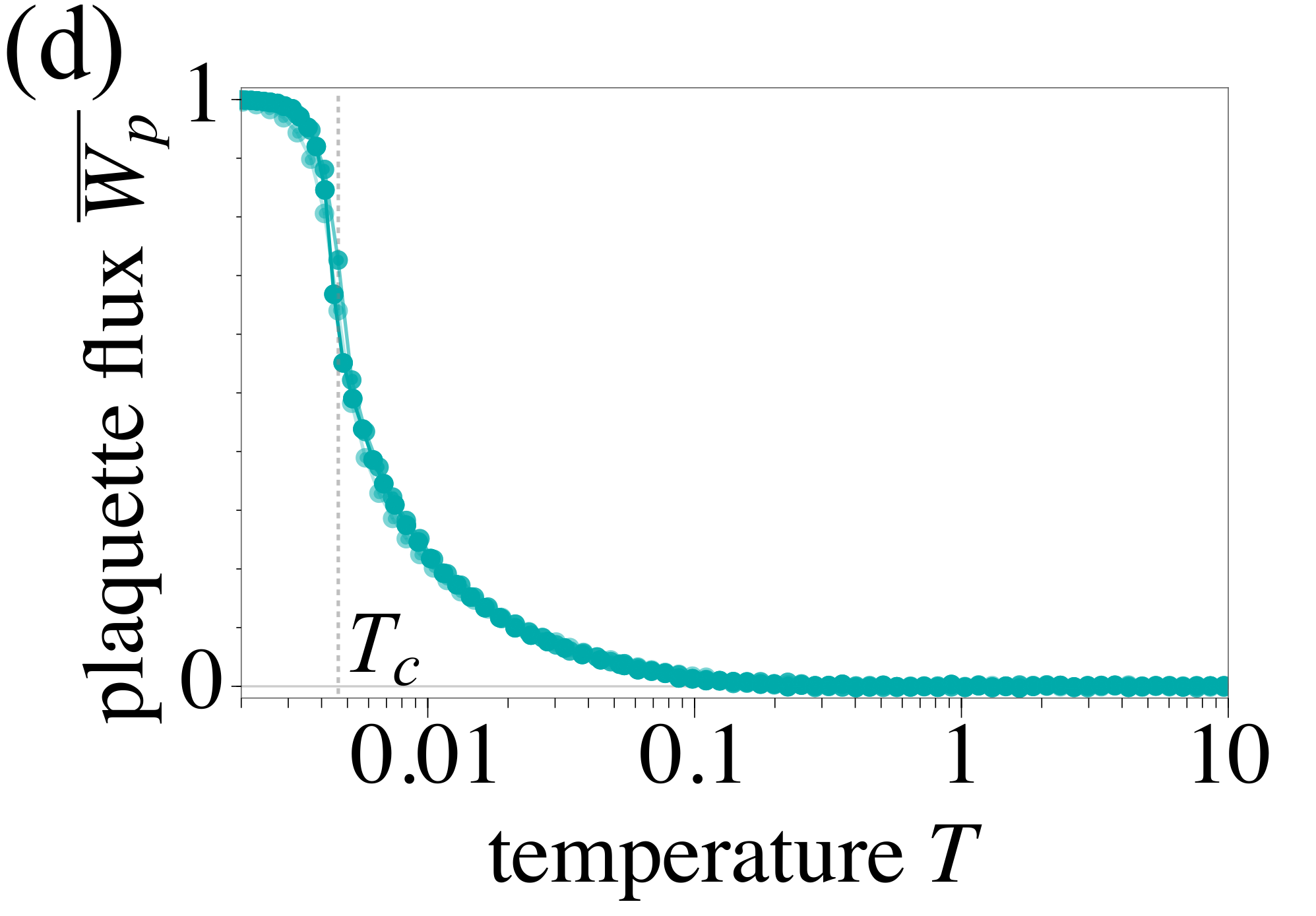}     
   \caption{{\bf Finite-size dependence of the specific heat $C_v$ and average plaquette flux $\overline{W_p}$.} While the high-temperature peak of the specfic heat (a,b) is unchanged for increased system sizes, the low-temperature peak diverges in the thermodynamic limit. This is a clear signature of a thermal phase transition, which is associated with the ordering of $\mathbb{Z}_2$ fluxes (c,d). Data shown is for the lattices (8,3)b (a,c) and (10,3)d (b,d).
	}
    \label{fig:FiniteSizeScaling}
\end{figure}

In contrast to the thermal crossover regime the specific heat curves start to significantly differ at lower temperatures when considering the various lattice geometries. Here, the physical behavior of the system is entirely governed by the $\mathbb{Z}_2$ gauge field $\{u_{jk}\}$ and its fluctuations. This can be seen by explicitly calculating the contribution of the $\mathbb{Z}_2$ gauge field to the specific heat, which is captured by the {\sl variance} of the internal energy $E_f$ (see Appendix \ref{sec:Observables} for a detailed derivation)
\begin{equation}
C_{v,\text{GF}}(T) = \frac{1}{T^2} \left ( \langle E_f^2(\{u_{jk}\}) \rangle_\text{MC} - \langle E_f(\{u_{jk}\}) \rangle_\text{MC}^2 \right ) \,.
\end{equation}
As plotted in Fig.~\ref{fig:SpH_all_lowT}, one can see that all (bipartite) 3D Kitaev systems show a single, relatively sharp low-temperature peak, which is diverging for increasing system sizes (see Fig.~\ref{fig:FiniteSizeScaling}). These peaks reflect the release of entropy associated with the  ordering of the $\mathbb{Z}_2$ gauge field -- a true thermal phase transition. The peak heights, shapes and locations of these peaks in temperature space strongly differ for different lattice geometries. The sharpest low-temperature peak is found for lattice geometry (8,3)b (which also has the highest transition temperature $T_c = 0.0071(3)$), the broadest peak for lattice geometry (10,3)a (which also has the lowest $T_c = 0.00405(9)$). There is no apparent correlation between the elementary loop length and the critical temperature, nor the peak size: Lattice geometry (8,3)a, which exhibits vison loops of length 2, has a transition temperature that is only slightly higher than the one for lattice geometry (10,3)a, which exhibits vison loops of length 10. 
\footnote{
Interestingly, these two lattices are the ones that host a Majorana metal ground state, which harbors topological Fermi surfaces (i.e., Fermi surfaces that encapsulate Weyl nodes at finite energy). It was laid out within the classification paper of the Kitaev ground states \cite{Obrien2016classification} that the occurrence of these Fermi surfaces is geometrically determined by a nontrivial sublattice symmetry of the underlying lattices, in combination with the absence of inversion-symmetry.}

\begin{figure}[h!]
    \centering
    \includegraphics[width=\columnwidth]{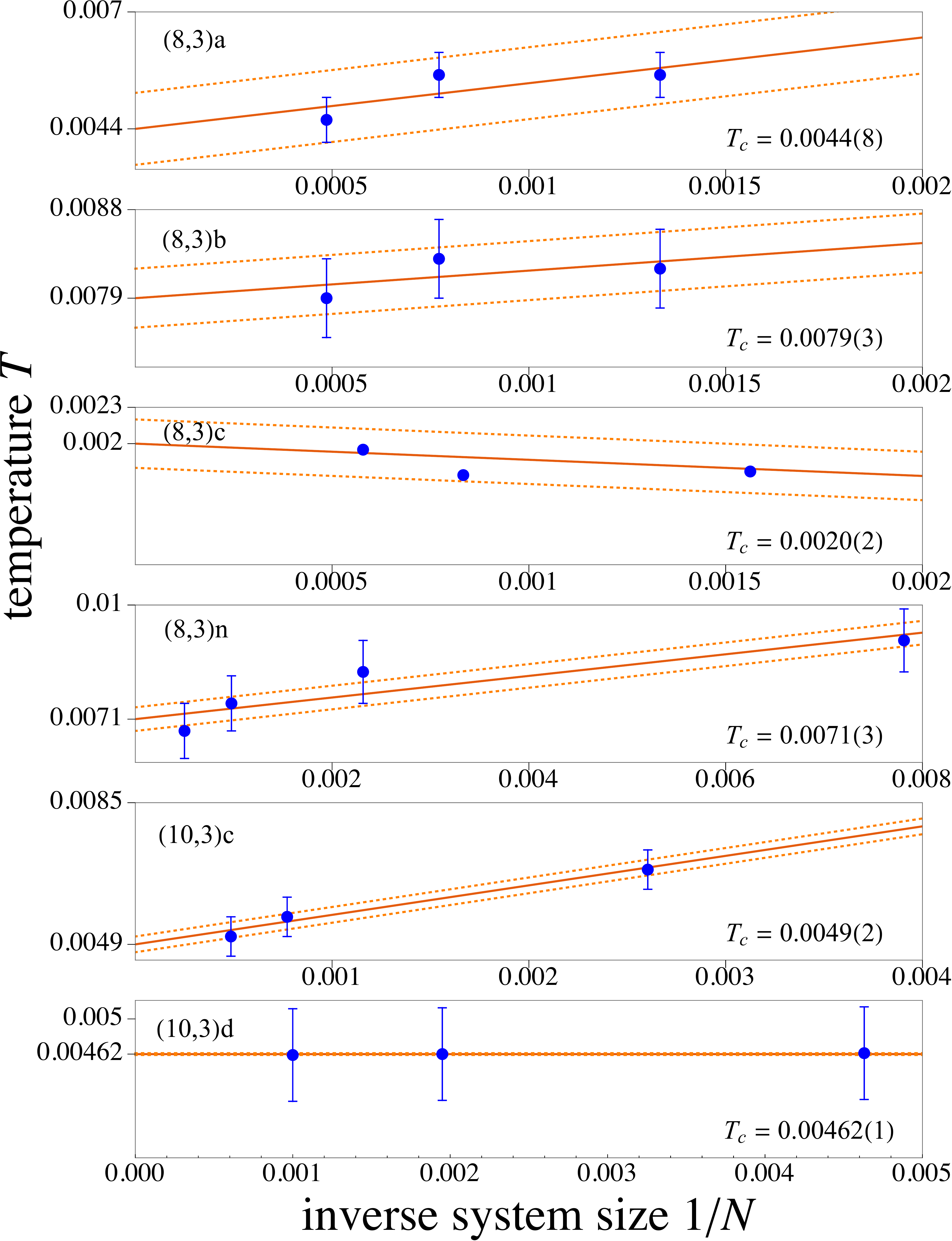}  
      \caption{{\bf Finite-size extrapolation of the transition temperature} as function of the inverse system size $1/N$.
      		Plotted are the gauge-ordering transition temperatures $T_c$ for different system size (blue dots) and their extrapolation (red line) 
		to the limit $1/N \rightarrow 0$. The quality of the extrapolation is marked by the red dashed line, which indicates the standard deviation
		of the fit.}
    \label{fig:CriticalTemperaturesPlot}
\end{figure}

Instead the most pronounced correlation that the transition temperature exhibits is the one with the size of the vison gap, illustrated in Fig.~\ref{fig:CorrelationPlot}. This correlation stems from the (somewhat unconventional) mechanism that causes the phase transition -- a proliferation of the vison loops in a ``topological" phase transition. While the creation of a vison at low temperatures results in the formation of a small loop, these loops can gain in size as one allows for an increase in thermal fluctuations at more elevated temperatures -- the well-known trade off between energy (loss) and entropy (gain) at finite temperatures. Eventually, when the loops gain a spatial extent comparable to the (finite) system size, it becomes favorable to wrap around system -- the closed loops break open and reconnect across the (periodic, torus-shaped) lattice geometry, a topological phase transition \cite{Nasu2014vaporization}. This perspective on the phase transition of the $\mathbb{Z}_2$ gauge field corresponds to a description of the phase transition of the related, well-known Ising model \cite{PhysRevB.89.115125}. As such, we can readily see that the thermal phase transition of the $\mathbb{Z}_2$ gauge field can be cast as an inverse Ising transition (in which the roles of high and low temperatures are switched compared to the conventional Ising transition). For those model systems, for which we observe a continuous phase transition in our numerics, 
we therefore expect a transition in the inverted 3D Ising universality class.
However, for most lattice geometries the system sizes accessible in our numerics  do not allows us to perform a full finite-size scaling analysis  and to extract the expected critical exponents (though we can firmly extrapolate the critical temperature, see Fig.~\ref{fig:CriticalTemperaturesPlot}).
The only scenario for which critical exponents could be effectively determined~\cite{PhysRevB.89.115125} has been for the anisotropic limit (``toric code" limit) of the hyperhoneycomb Kitaev model \footnote{It was later realized that such a ``toric code'' model also applies to the anisotropic limit of the hyperoctagon Kitaev model~\cite{mishchenko_prb_96_2017}.}.

Between the two thermal transitions, there is an intermediate temperature regime that spans about two orders of magnitude, $J/100 \sim T_c < T < T' \sim J$ and might, in fact, be the most relevant temperature regime in experimental probes of Kitaev materials. In this regime, one expects to observe the first signatures of fractionalization -- with the original spins already broken apart into Majorana fermions and a $\mathbb{Z}_2$ gauge field. The latter, however, is still highly disordered in the intermediate regime which prevents the formation of a ``clean" Majorana band structure (as it is the case at strictly zero temperature or, more precisely, the low-temperature transition). Instead one expects to see a ``disordered Majorana metal"\cite{Nasu2015thermal, 2019MotomeReviewKitaevMagnets} or ``thermal metal" \cite{Chalker2001,Laumann2012}, which for certain 2D settings has indeed been seen in numerical simulations \cite{Self2019}.


\subsection{Gauge frustration}
\label{(8,3)c}

The two-step thermodynamic scenario of (high-temperature) spin fractionalization and (low-temperature) gauge ordering transition
laid out above is rather generic and applies to almost all 3D Kitaev models. One notable exception is found in lattice
geometry (8,3)c, which exhibits a phenomenon that we have dubbed ``gauge frustration" in Ref.~\onlinecite{2019EschmannGaugeFrustration}, which discusses the
peculiarities of the associated Kitaev model in full detail. Here we provide a brief summary to make this study comprehensive in its
own right. The special situation encountered in this lattice geometry is that the $\mathbb{Z}_2$ gauge field itself is subject to 
{\sl geometric frustration}, which in turn substantially suppresses the low-temperature ordering transition. In a certain sense, the 
Kitaev model on the (8,3)c lattice is therefore ``doubly frustrated" -- on the level of the original spin degrees of freedom and on the
level of the emergent gauge field.

\begin{figure}[b]
   \centering
    \includegraphics[width=\columnwidth]{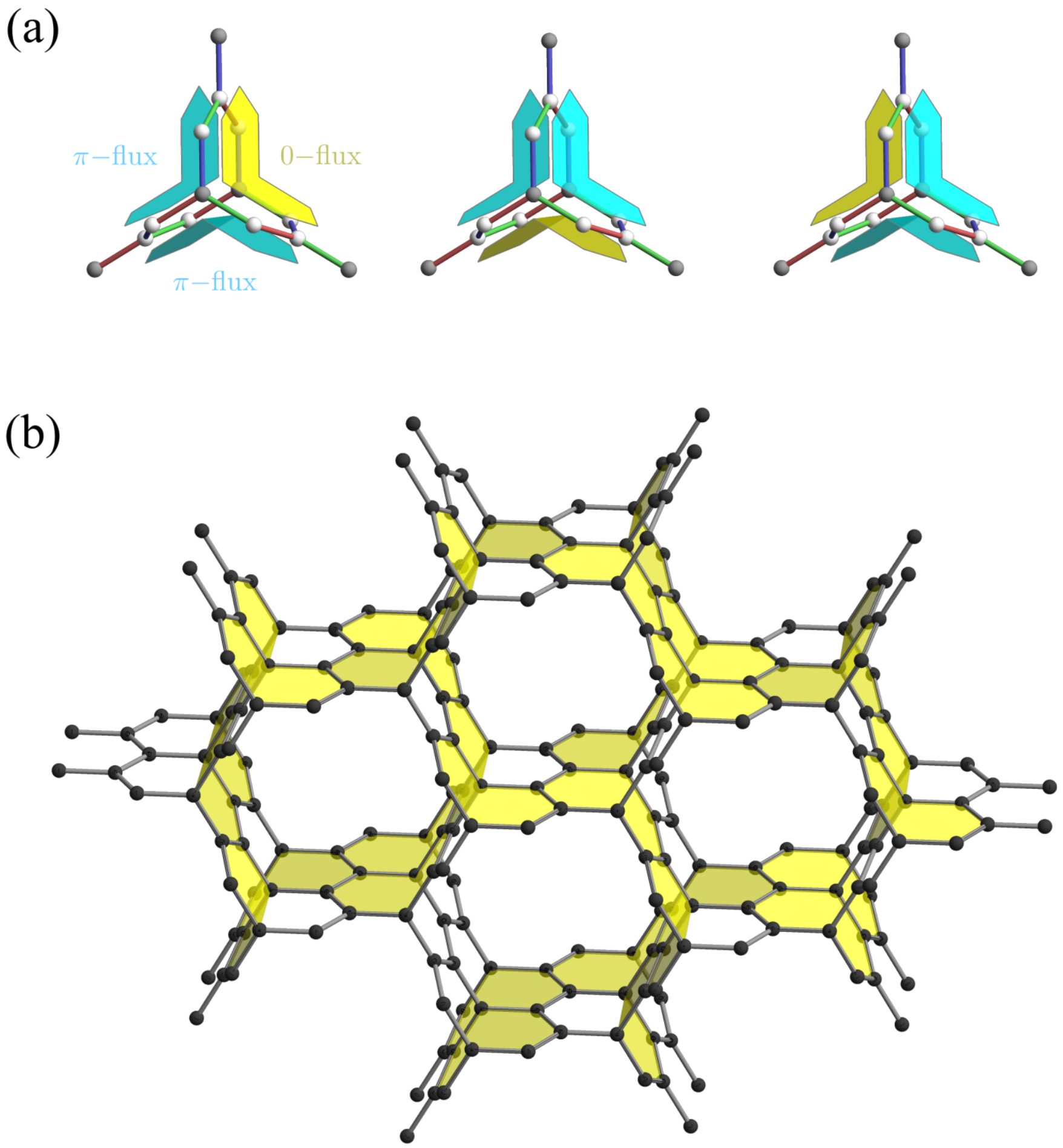}
  \caption{{\bf Gauge frustration in the (8,3)c lattice.} Because of the elementary plaquette length $p = 8$, a $\pi$-flux per plaquette is energetically preferred. However, the lattice consists of plaquette triplets which form closed boundaries. Therefore, the product of the loop operator eigenvalues $W_p$ is constrained to $\prod_p W_p = 1$ for each triplet. In consequence, one plaquette in each triplet must carry a $0$-flux, which leads to three degenerate configurations per triplet (a), and a macroscopically frustrated manifold of $\mathbb{Z}_2$ flux configurations for the entire lattice -- a phenomenon which we have dubbed {\it gauge frustration}. Panel (b) shows one flux configuration from the gauge-frustrated manifold. Here, the yellow plaquettes are flux-free.   }
    \label{fig:8c}
\end{figure}

How this unusual phenomenon comes about can readily be understood. As a lattice with an elementary plaquette of length 8, 
one expects -- via the intuition gained from the broader application of Lieb's theorem and its subsequent numerical verification --
that each plaquette carries a $\pi$-flux. The crucial ingredient then arises from the lattice geometry where {\sl three} such plaquettes
are constraining one another around tricoordinated junctions, see Fig.~\ref{fig:8c}, to the following effect: With flux conservation strictly
required (similar to a divergence-free condition familiar from Maxwell theory), only two out of these three plaquettes can actually carry
a $\pi$-flux with the third plaquette ending up flux-less. But which one of the three plaquettes does not reach the lowest energy $\pi$-flux state
remains open and the origin for a residual entropy (i.e., an extensive manifold of states) in the gauge sector -- the hallmark of (geometric) frustration.

\begin{figure}[t]
   \centering
    \includegraphics[width=\columnwidth]{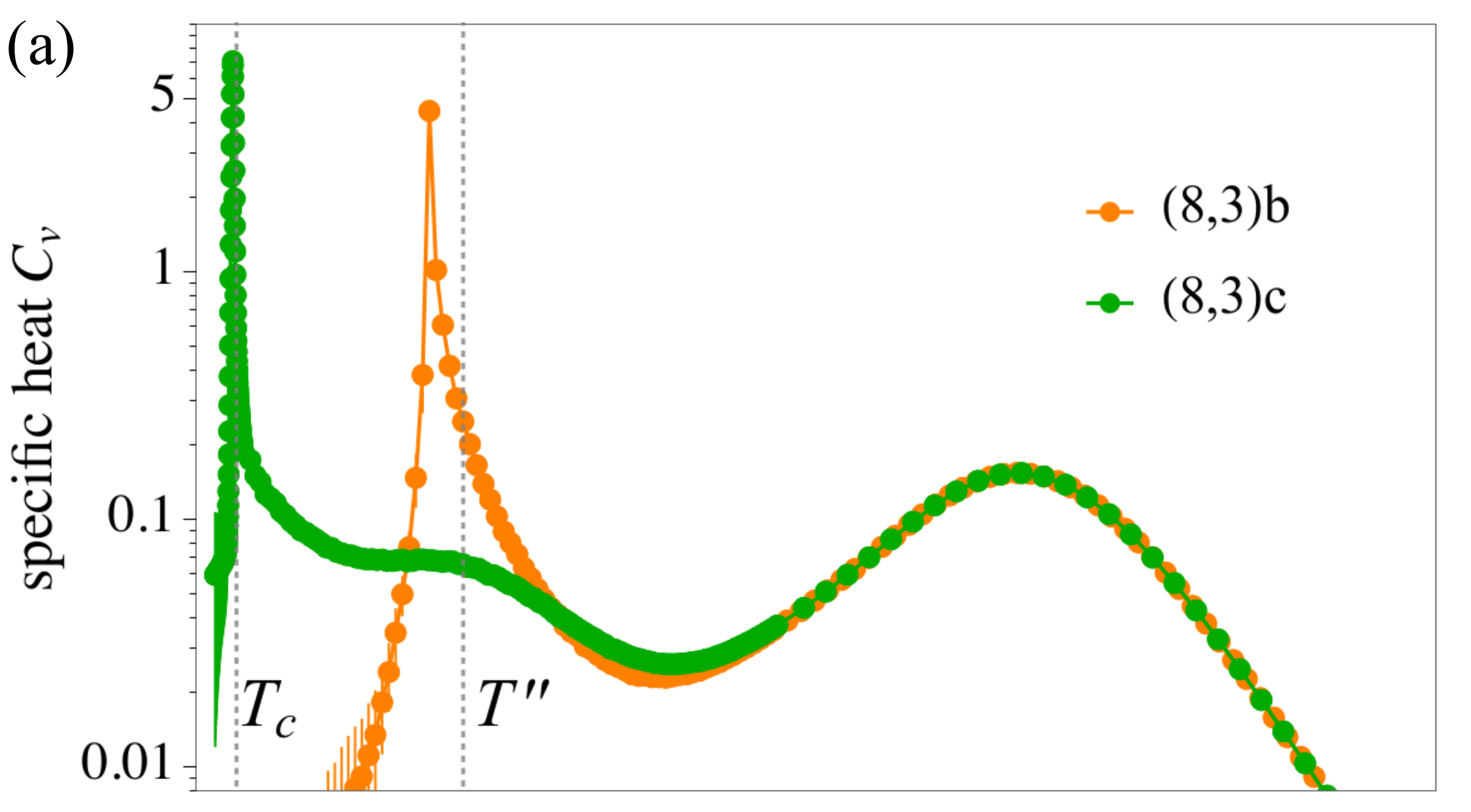}
     \includegraphics[width=\columnwidth]{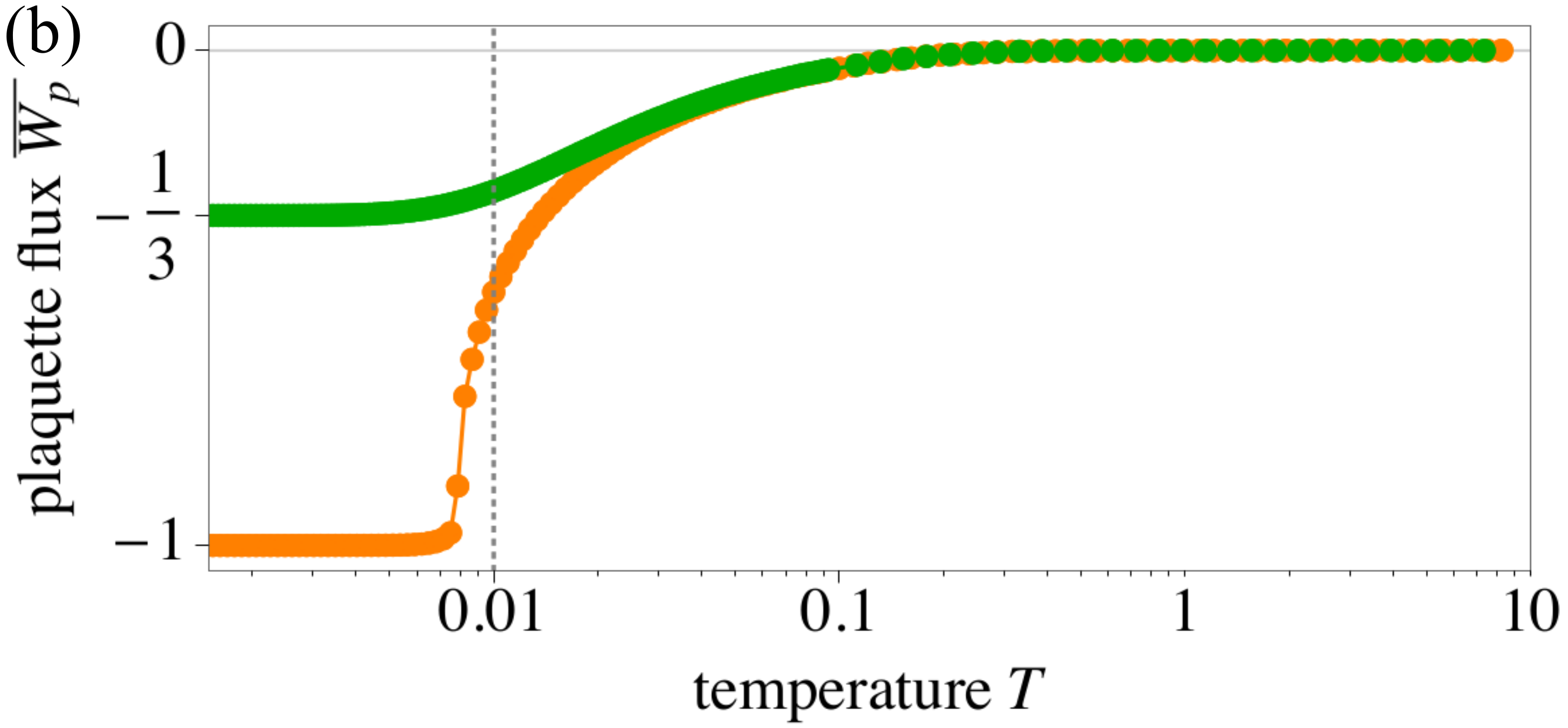}
   \caption{{\bf Thermodynamics of (8,3)c.} (a) Due to the gauge frustration, the thermal phase transition of the (8,3)c Kitaev model is suppressed by almost one order of magnitude. At the usual transition temperature scale $T'' \sim 0.01$, we encounter a broad shoulder in the specific heat $C_v$. This signature indicates a thermal crossover, which is associated with the system entering the frustrated gauge manifold. At $T_c = 0.0020(2)$, the specific heat of the (8,3)c model shows the characteristic phase transition peak. (b) Uniform ground state flux configurations are accompanied by an average flux operator eigenvalue $\overline{W_p} = \pm 1$. Here instead, we find that $\overline{W_p} = -1/3$ at low temperatures. This indicates that two plaquettes out of three acquire a $\pi$-flux ($W_p = -1$), and one plaquette remains flux-free ($W_p = 1$), which is a consequence of the interplay between the elementary plaquette length $p = 8$ and a particular volume constraint in the (8,3)c lattice. Data shown is for the linear system sizes $L = 7$, (8,3)b, and $L = 6$, (8,3)c. 
   }
    \label{fig:8cTD}
\end{figure}

The existence of such a gauge-frustrated low-energy manifold of states and the associated suppression of thermal gauge ordering is indeed readily visible in our numerical simulations. As shown in Fig.~\ref{fig:8cTD} the low-temperature specific heat peak is shifted towards considerably lower temperature in comparison to other 8-loop lattice geometries. Upon closer inspection, one does indeed find that the average plaquette flux $\overline {W_p}$ does not converge to a value of $+1$ or $-1$ at low temperatures, which would indicate a uniform flux configuration, but to $\overline{W_p} = -1/3$ (see Fig. \ref{fig:8cTD}): Two plaquettes out of three acquire a $\pi$-flux ($W_p = -1$), and one plaquette remains flux-free ($W_p = 1$), leading to an average of $(-2 + 1)/3 = -1/3$. This also means that the system prefers to adopt a plaquette flux configuration that does {\it not} preserve all the lattice symmetries (the only allowed configuration preserving all symmetries would be a uniform configuration of 0-fluxes). This regime is also visible in the specific heat plot of Fig.~\ref{fig:8cTD} as a shoulder below the crossover peak.

Upon further lowering the temperature also the (8,3)c Kitaev model exhibits an actual ordering transition. 
As detailed in Ref.~\onlinecite{2019EschmannGaugeFrustration}, it is a subtle interplay with the formation of a nodal-line semi-metal of the itinerant Majorana fermions
that leads to a columnar zig-zag order of the 0-fluxes.

\subsection{Spontaneous time-reversal symmetry breaking}
\label{(9,3)a}

The second lattice geometry in this classification that is destined to give rise to physics different from the above two-step thermodynamic scenario
is the ``hypernonagon" lattice (9,3)a -- the only lattice geometry with an {\sl odd} length elementary plaquette. This has dramatic consequences, as Majorana fermions hopping around such a plaquette will pick up a phase of $\pm \pi/2$, corresponding to the flux through the plaquette
\footnote{These two flux assignments correspond to imaginary loop operator eigenvalues $\pm i$. According to the definition of the loop operator in the $\mathbb{Z}_2$ gauge field representation, $W_p = \prod (-i u_{ij})$, its sign depends on the {\sl direction} of measurement: If a measurement of $W_p$ on a given plaquette gives a result $W_p = + i$ if the plaquette is defined in a clockwise direction, a measurement in counter-clockwise direction would give the opposite result $W_p = -i$. This direction dependence is not the case for even-length plaquettes, where the loop operator eigenvalues are not imaginary. In the consequence, on plaquettes with odd loop length, there is no conjecture on the flux ground state as Lieb's theorem.}. 
But only one of the possibilities will prevail, implying a spontaneous breaking of time-reversal symmetry (which connects the two choices). This in turn means that there must be a thermal phase transition associated with this breaking of time-reversal symmetry (independent of spatial dimensionality) and that the low-temperature free (Majorana) fermion system falls into symmetry class D~\cite{Altland1997classification,Zirnbauer1996}.  

\begin{figure}[t]
   \centering
   \includegraphics[width=0.24\columnwidth]{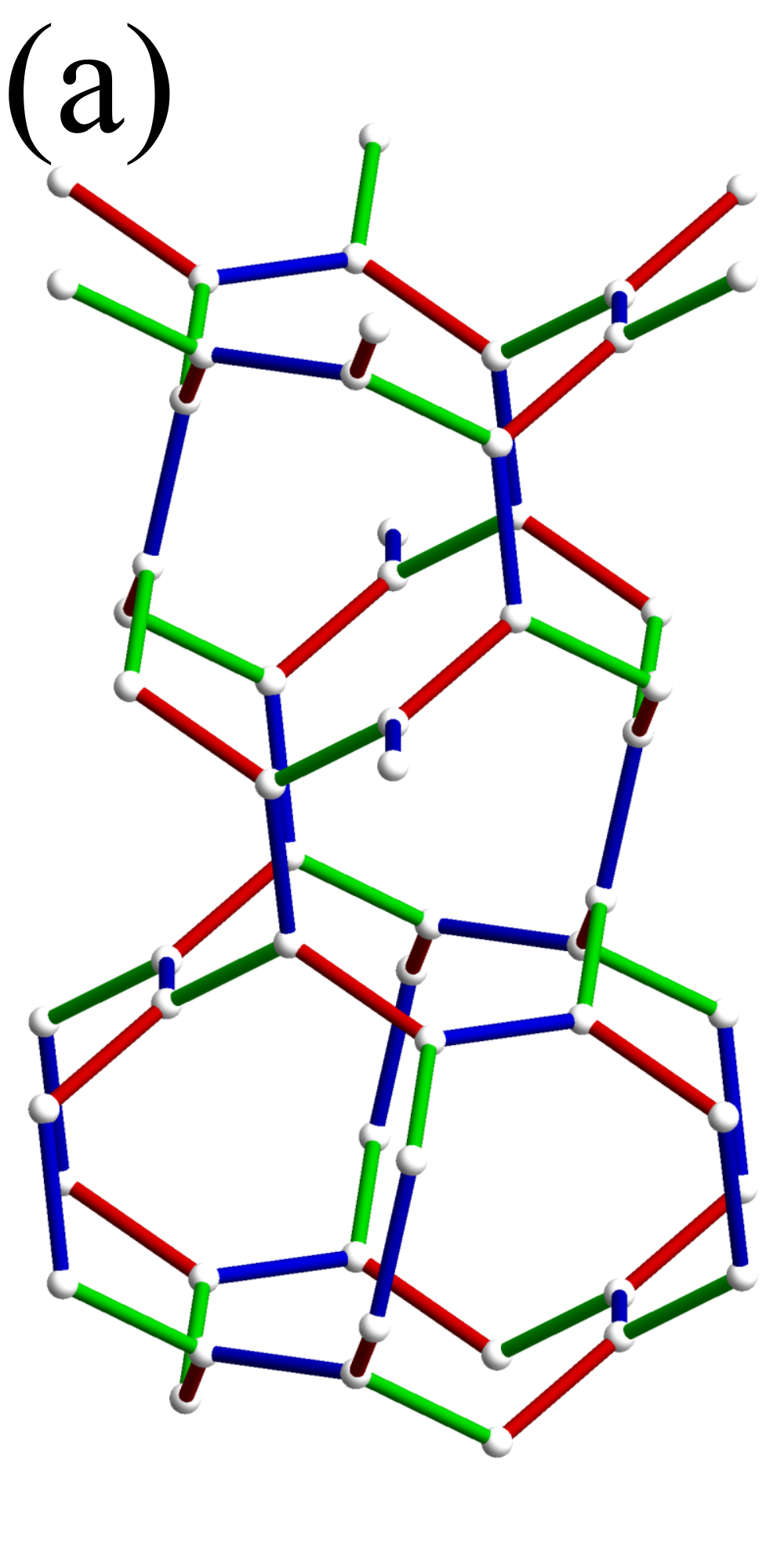}  
   \includegraphics[width=0.745\columnwidth]{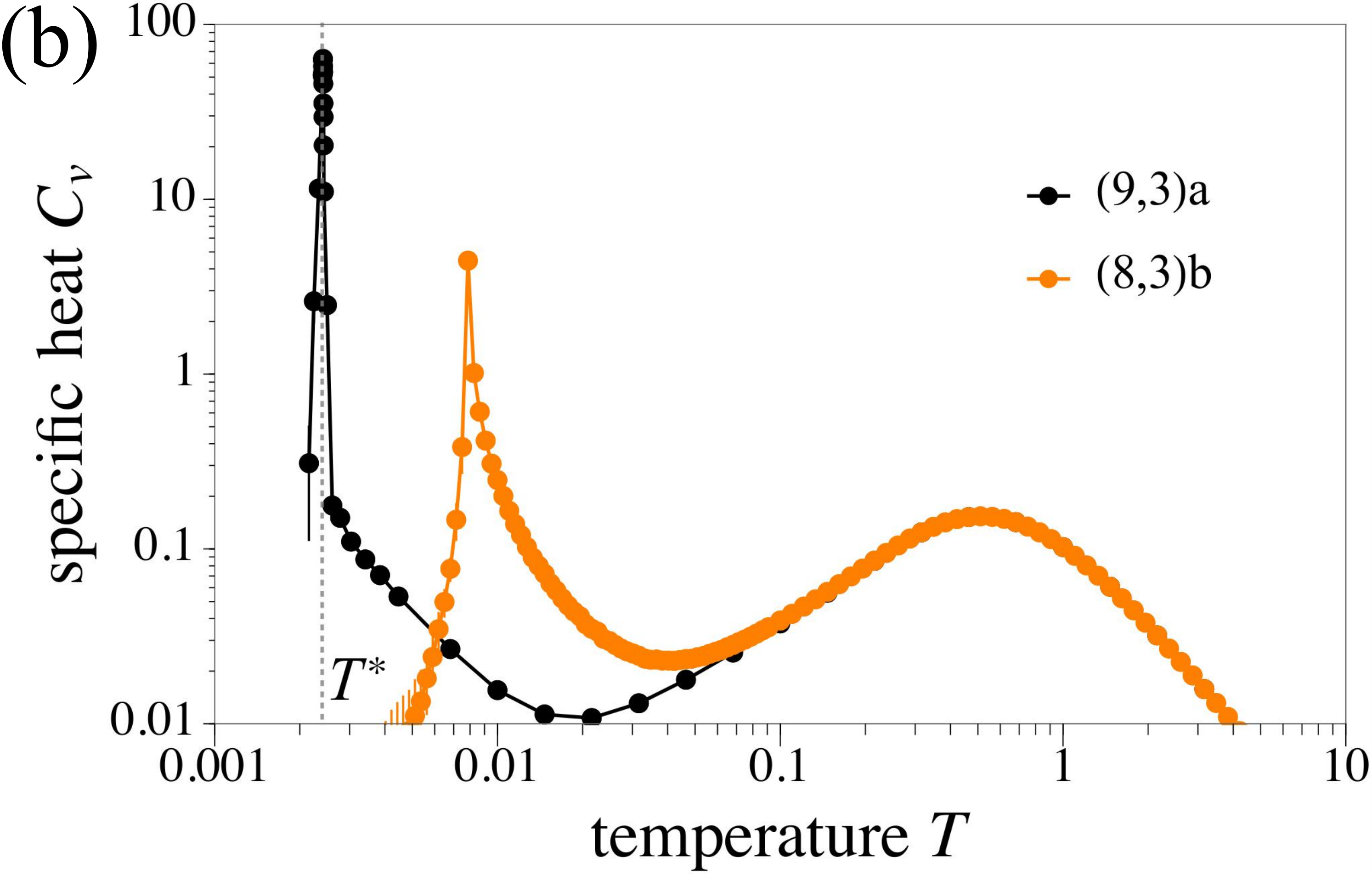}  
        \caption{{\bf Lattice geometry and thermodynamics of (9,3)a.} At $T^* \sim 0.0024$, the system undergoes a first-order phase transition which is associated with vison loop proliferation, time-reversal symmetry breaking and the breaking of point-group symmetries. The signature of this phase transition is a sharp low-temperature peak in the specific heat, with a height that is one order of magnitude larger than for other 3D Kitaev systems -- here: (8,3)b. Data shown is for the linear system sizes $L = 6$, (9,3)a, and $L = 7$, (8,3)b \cite{2020MishchenkoChiralSpinLiquids}.}
    \label{fig:9aThermodynamics}
\end{figure}

In 2D Kitaev systems with odd loop length, such as, e.g., the decorated honeycomb / Yao-Kivelson lattice, the symmetry-based classification of topological states \cite{Schnyder2008classification,Kitaev2009periodic} allows for a topologically non-trivial state in symmetry class D. In the language of spin systems this topological state is referred to as a gapped {\sl chiral spin liquid}, which has been the subject of  analytical~\cite{yao-kivelson} and numerical\cite{Nasu2015} studies. In three spatial dimensions, on the other hand, symmetry class D does {\sl not} allow for the formation of a topological chiral spin liquid ground state \cite{Schnyder2008classification,Kitaev2009periodic}. However, here,  it is a combination of time-reversal symmetry breaking, the vison loop proliferation mechanism presented for other 3D Kitaev models {\it and} the breaking of additional point-group symmetries, resulting in a non-uniform ground state flux pattern, that determine the thermodynamics as detailed in Ref.~\onlinecite{kato_prb_96_2017, 2020MishchenkoChiralSpinLiquids}. Here we provide a brief summary of these results from large-scale QMC simulations and variational calculations. In short, these studies have shown that all three mechanisms in the system result not in a series of multiple thermal phase transitions, but in a {\sl single first-order phase transition}, happening at a transition temperature $T^* \sim 0.0024$ (for isotropic coupling parameters, Fig. \ref{fig:9aThermodynamics}). The first-order nature of the transition was verified via a careful analysis of the histograms of the internal energy in the temperature region close to the transition. It was also shown that the ratio of critical temperature and vison gap $T_c/\Delta$ is particularly low for this system (with a value of $T_c / \Delta \sim 0.07$, compared to a typical range of $0.11 - 0.22$ for other 3D Kitaev systems, see Fig.~\ref{fig:CorrelationPlot} and Table \ref{TableVisonGaps}), which might be attributed to the first-order nature of the transition. Note that the high-temperature crossover transition, however, does not show any substantial deviation from the other 3D Kitaev systems.

What further sets the low-temperature phase of the hypernonagon Kitaev model apart from its other 3D cousins is that is exhibits a {\sl crystalline $\mathbb{Z}_2$ gauge order}. For isotropic coupling parameters $J_x = J_y = J_z = 1/3$, this crystalline order has a generalized antiferromagnetic structure in terms of columnar arrangements of the $\pm \pi/2$ plaquette fluxes. In Ref.~\onlinecite{2020MishchenkoChiralSpinLiquids} this crystalline gauge order is denoted as ``AFII" and shown to come along with the formation a nodal-line semi-metal in the Majorana sector. 

The AFII is in fact only one example of a larger variety of different nonuniform flux configurations, which govern the whole ground state phase diagram of the system when considering a variation of (anisotropic) coupling parameters. Variational calculations have shown that there are, in total, five distinct flux patterns \cite{kato_prb_96_2017, 2020MishchenkoChiralSpinLiquids}. What is common to all but one of these ground state flux configurations is that at least one point-group symmetry is broken. 
The hypernonagon Kitaev model is as such a prototypical example for a frustrated spin system that harbors chiral spin liquid ground states with crystalline gauge order
\footnote{These phases should not be confused with the ``crystalline Kitaev spin liquid'' put forward \cite{Yamada2017} in the context of lattice geometry (10,3)d where a topological semimetal forms in the Majorana sector that is protected by certain crystalline symmetries, akin to the formation of a topological crystalline insulator \cite{Fu2011tci}.}.


\section{Majorana density of states}

We round off our discussion of the 3D Kitaev models in this manuscript by taking a brief look at the emergent Majorana fermion sector
and calculate the distinct form of the DOS for all lattices -- both numerically and analytically. In doing so, our focus has  been on the low-temperature behavior -- where the DOS exhibits distinct signatures for the various lattice geometries -- and less on higher temperature thermodynamic signatures. 

In our QMC simulations, the DOS corresponds to the distribution of eigenvalues (i.e., single-particle energies) of the Majorana Hamiltonian, which are calculated after each Monte Carlo sweep.  In order to obtain an approximate version of the ground-state DOS, we choose a temperature well below the thermal phase transition, but still high enough for the simulation not to be frozen in a single gauge field configuration. Typically, we report results for the numerical DOS at a temperature $T \sim 0.0016$. 
In addition, we have performed analytical calculations of the Majorana DOS at zero temperature. To this end, we have transformed the Majorana Hamiltonian of each lattice into reciprocal space and calculated the analytic DOS of the exact ground state (generated by the appropriate $\mathbb{Z}_2$ gauge field configuration $\{u_{jk}\}$) in the discretized Brillouin zone, here consisting of $L_k^3$ momentum points (where we typically choose $L_k = 400$). 

\begin{figure}[b]
   \centering
    \includegraphics[width=0.49\columnwidth]{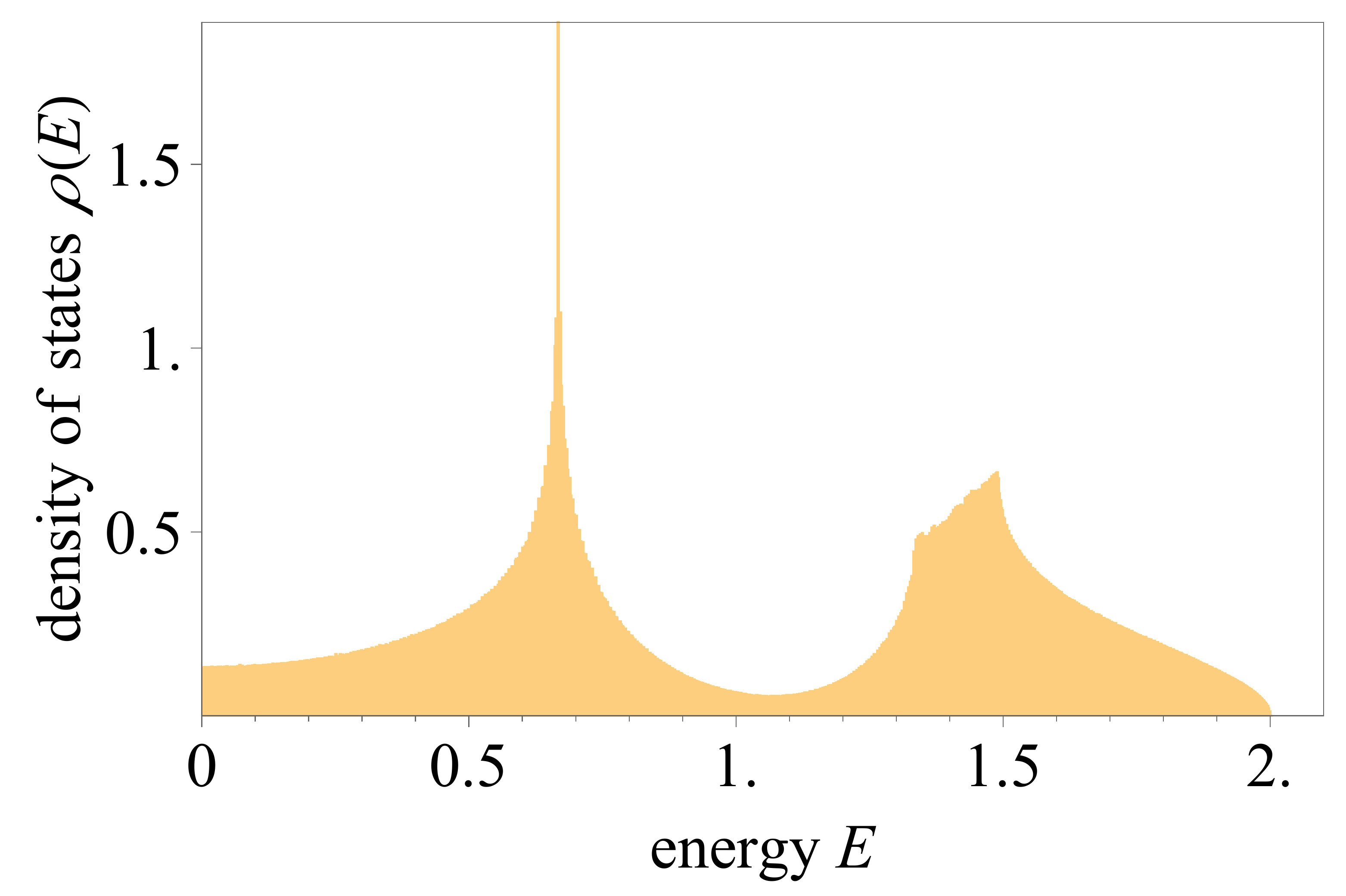}
    \includegraphics[width=0.49\columnwidth]{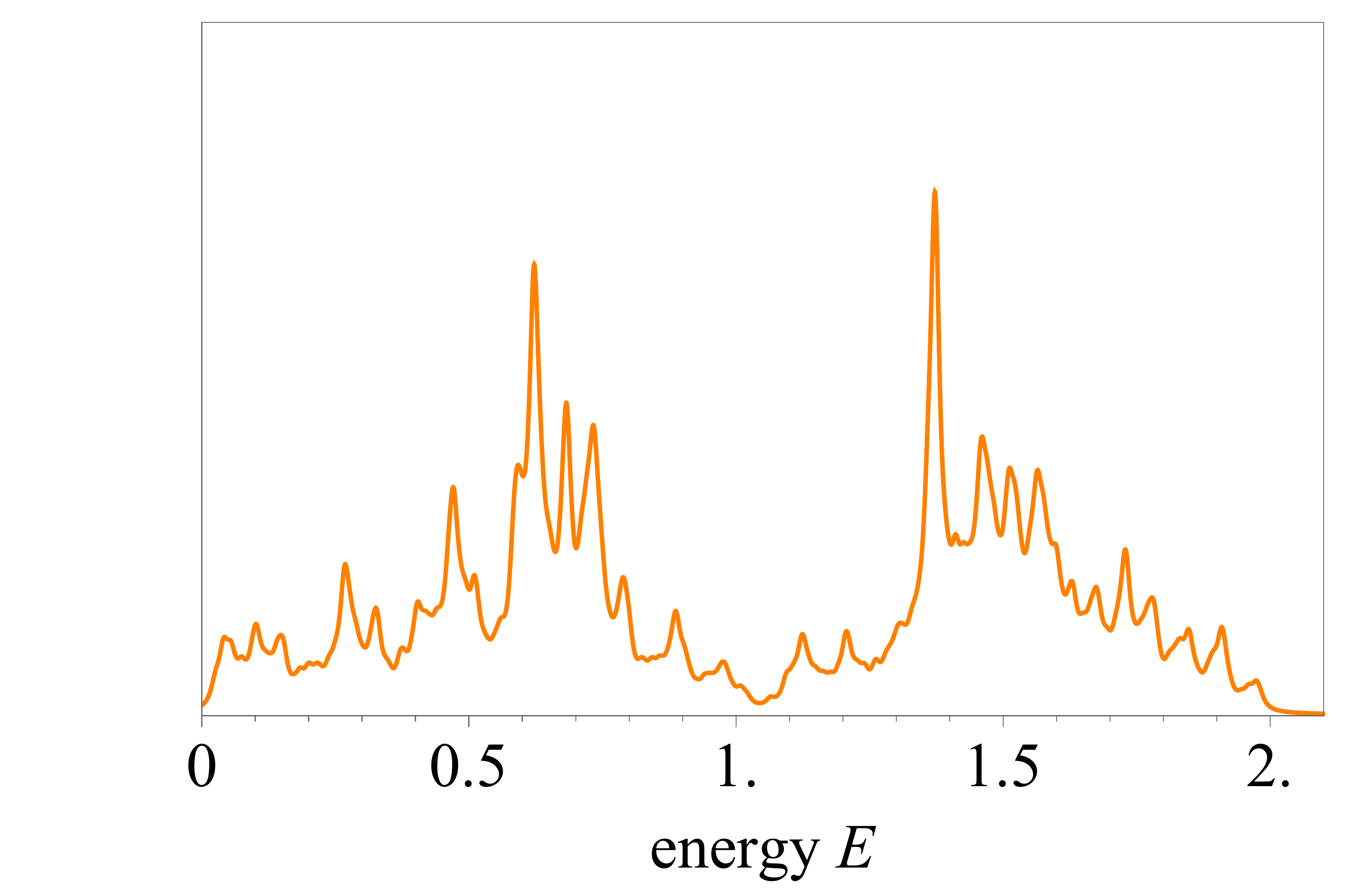}\\
    \includegraphics[width=0.49\columnwidth]{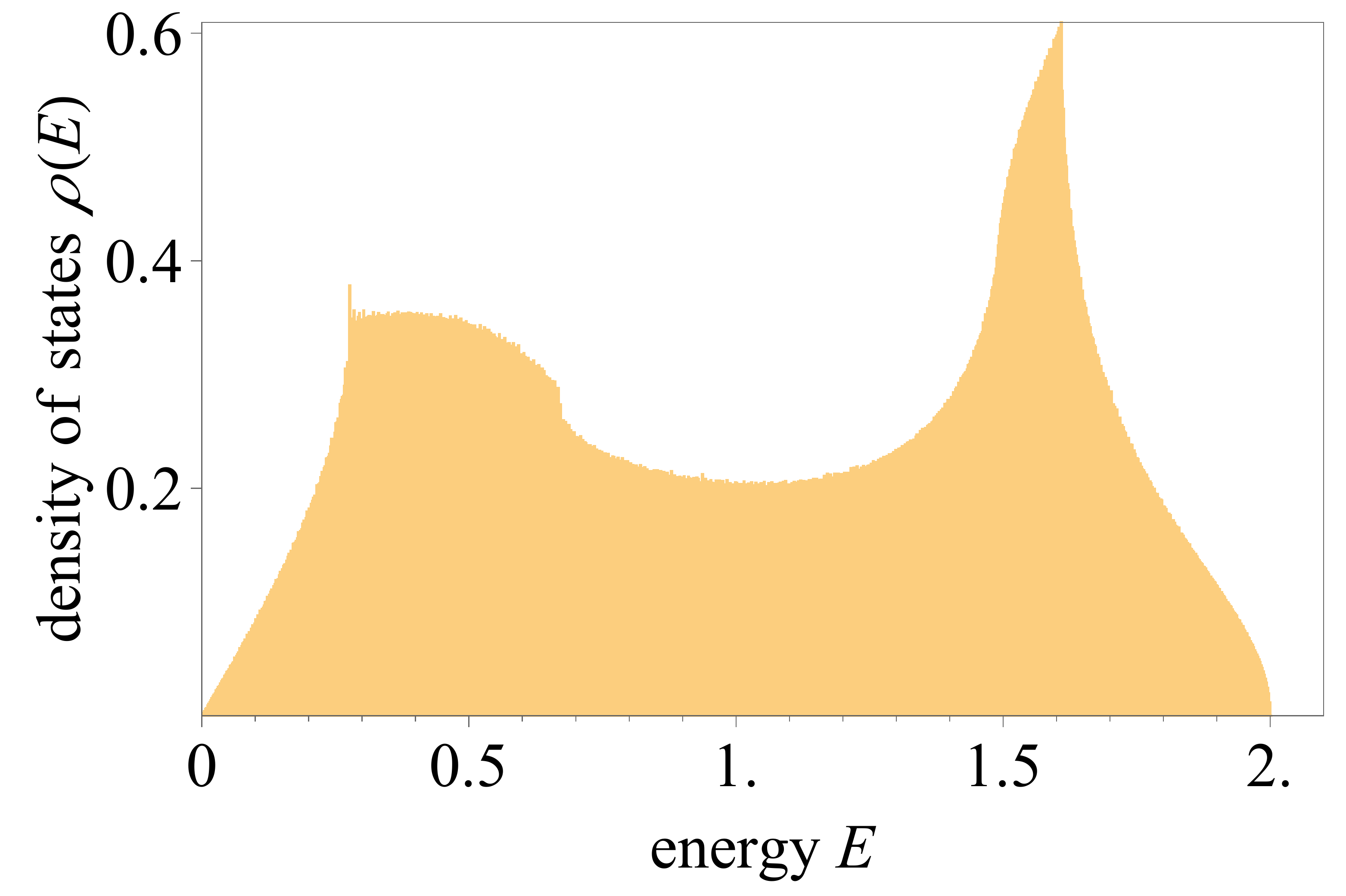}
    \includegraphics[width=0.49\columnwidth]{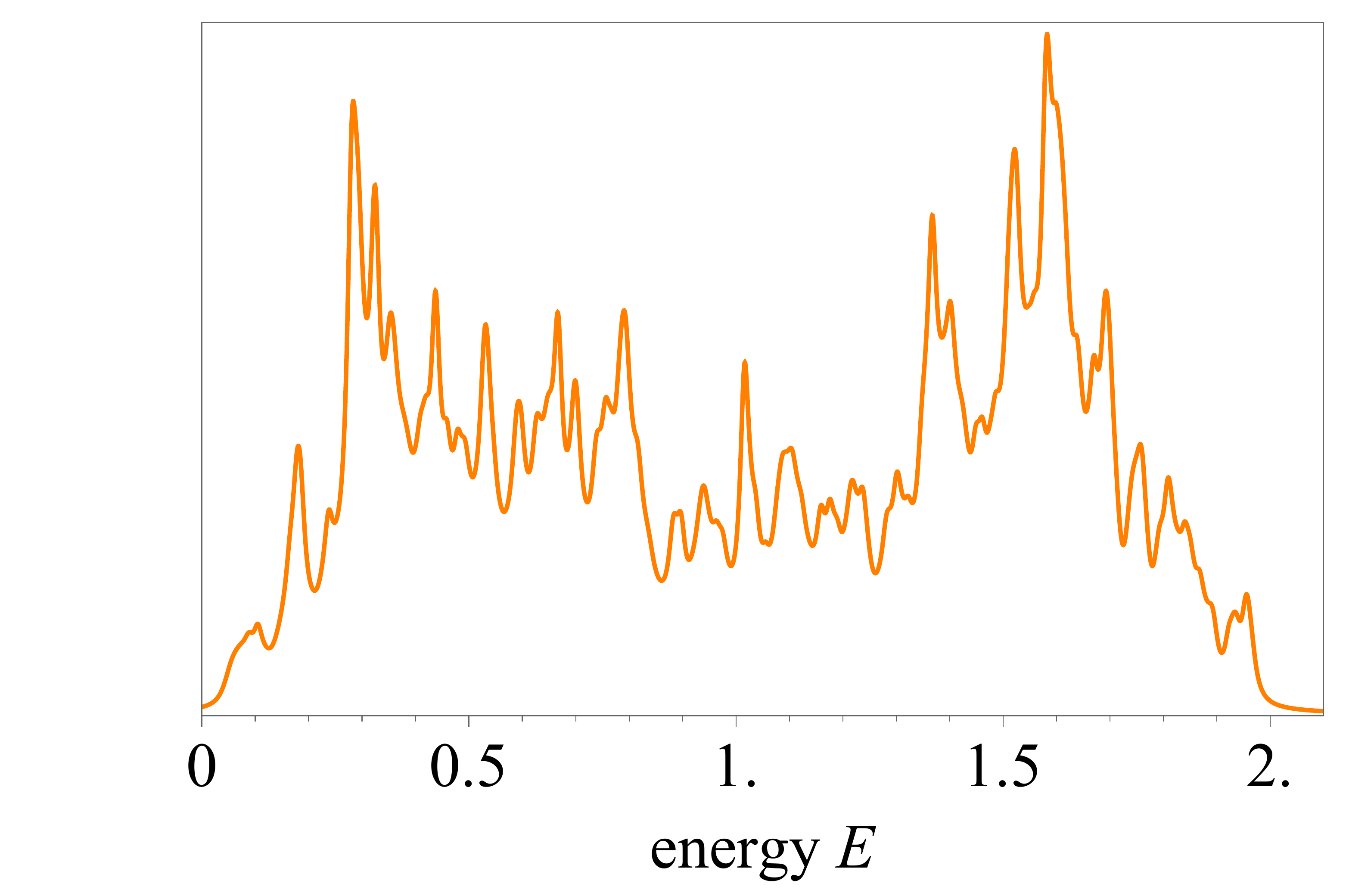}\\
    \includegraphics[width=0.49\columnwidth]{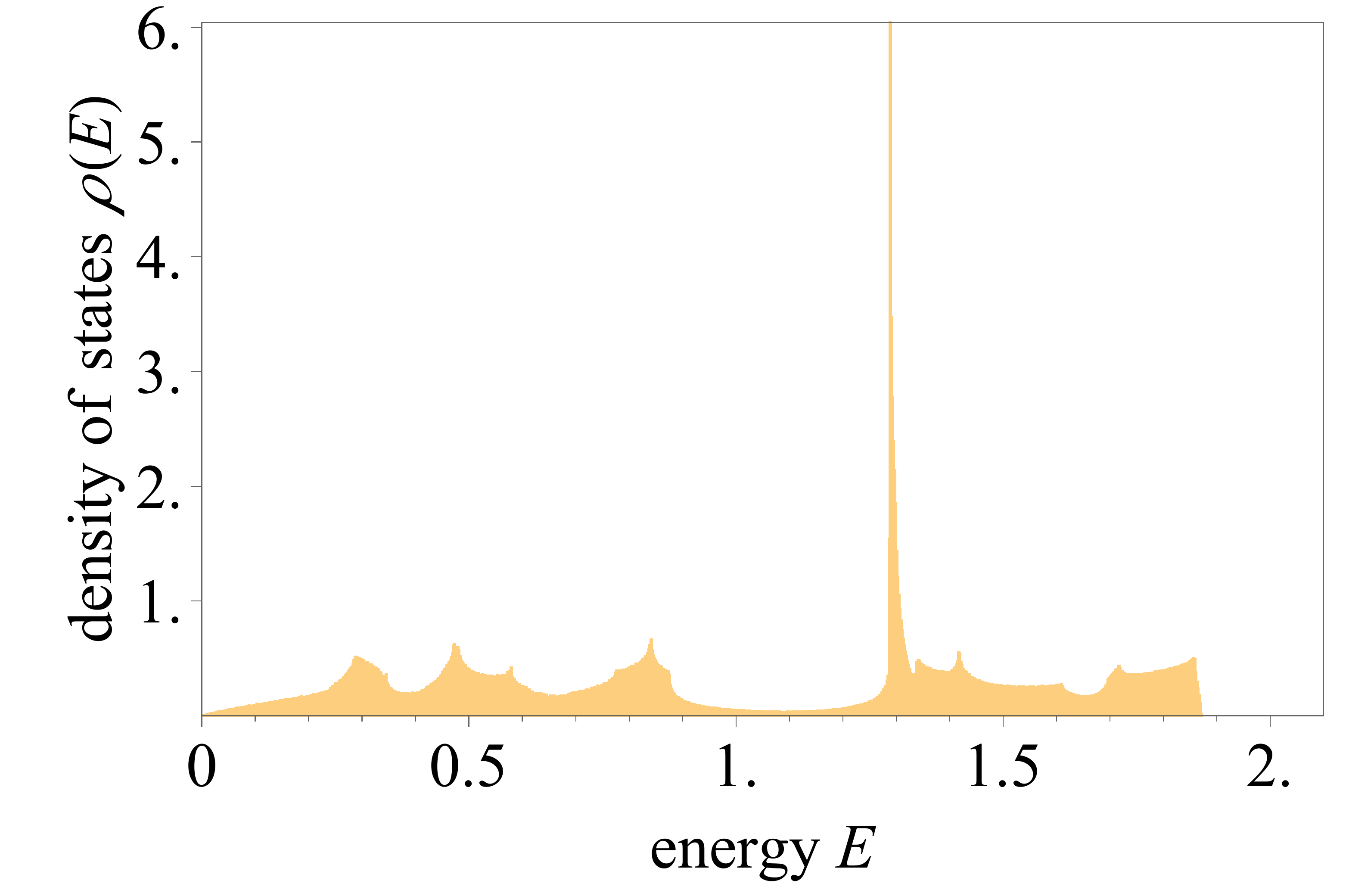}
    \includegraphics[width=0.49\columnwidth]{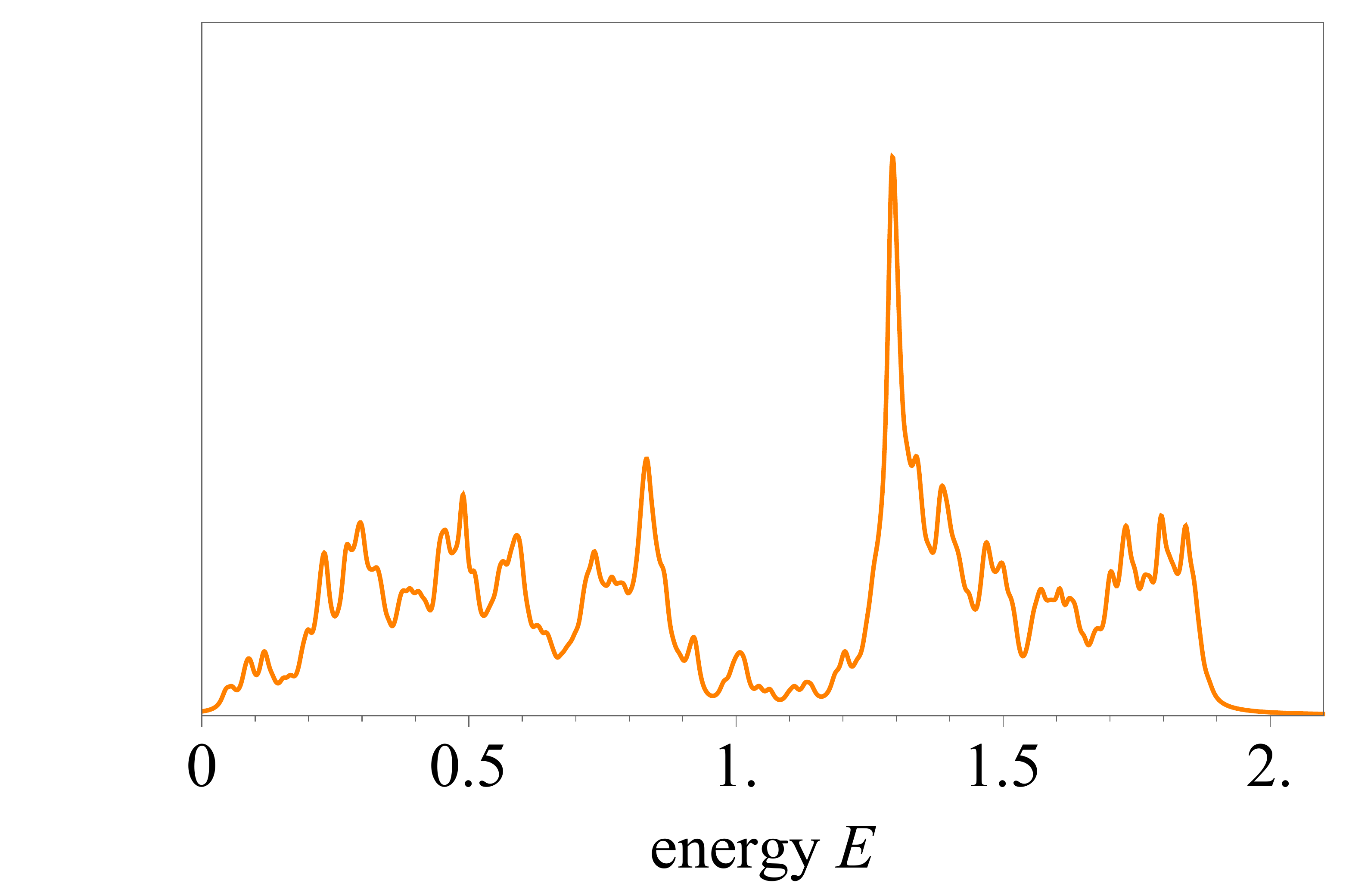}\\
    \includegraphics[width=0.49\columnwidth]{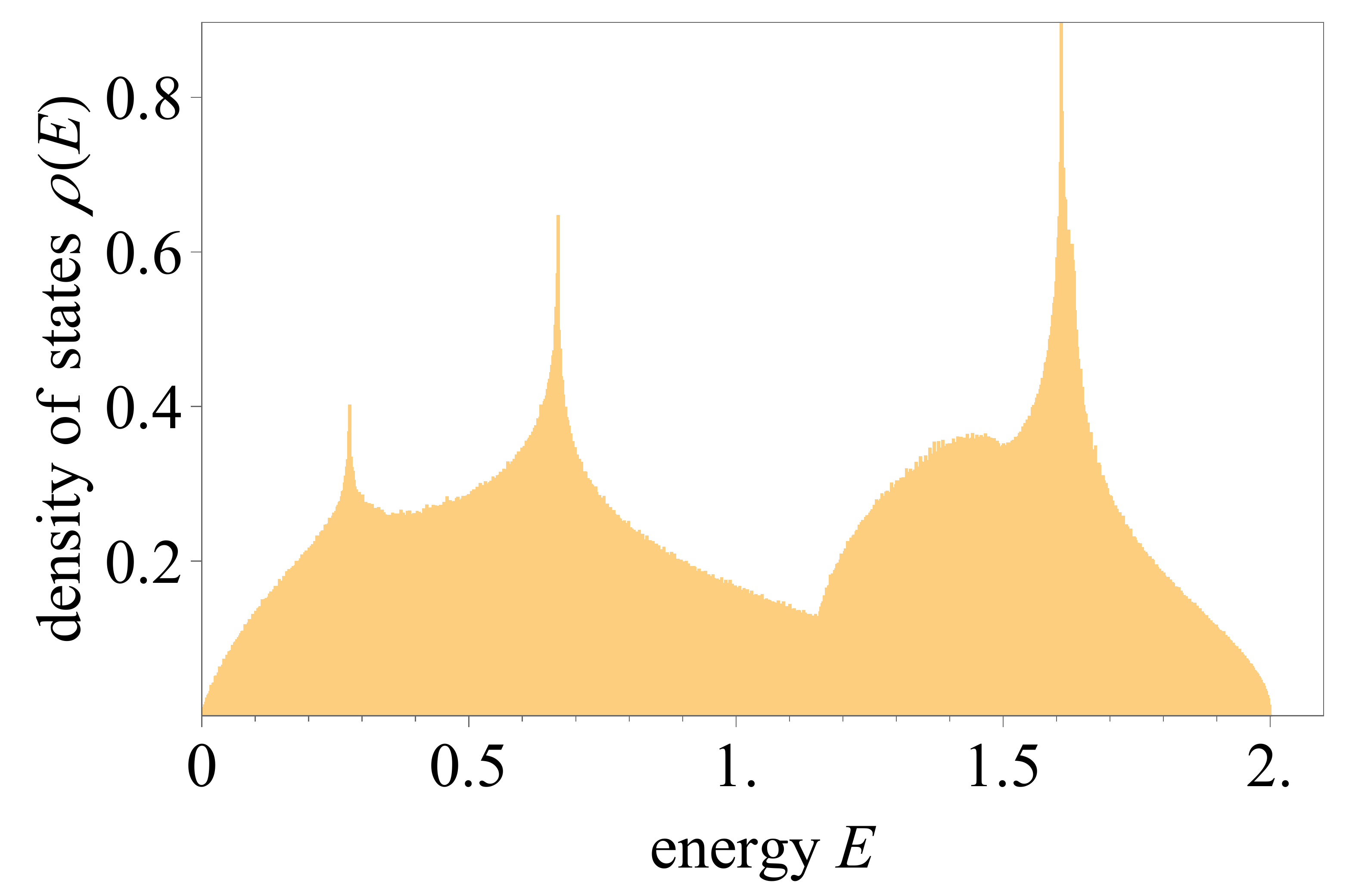}
    \includegraphics[width=0.49\columnwidth]{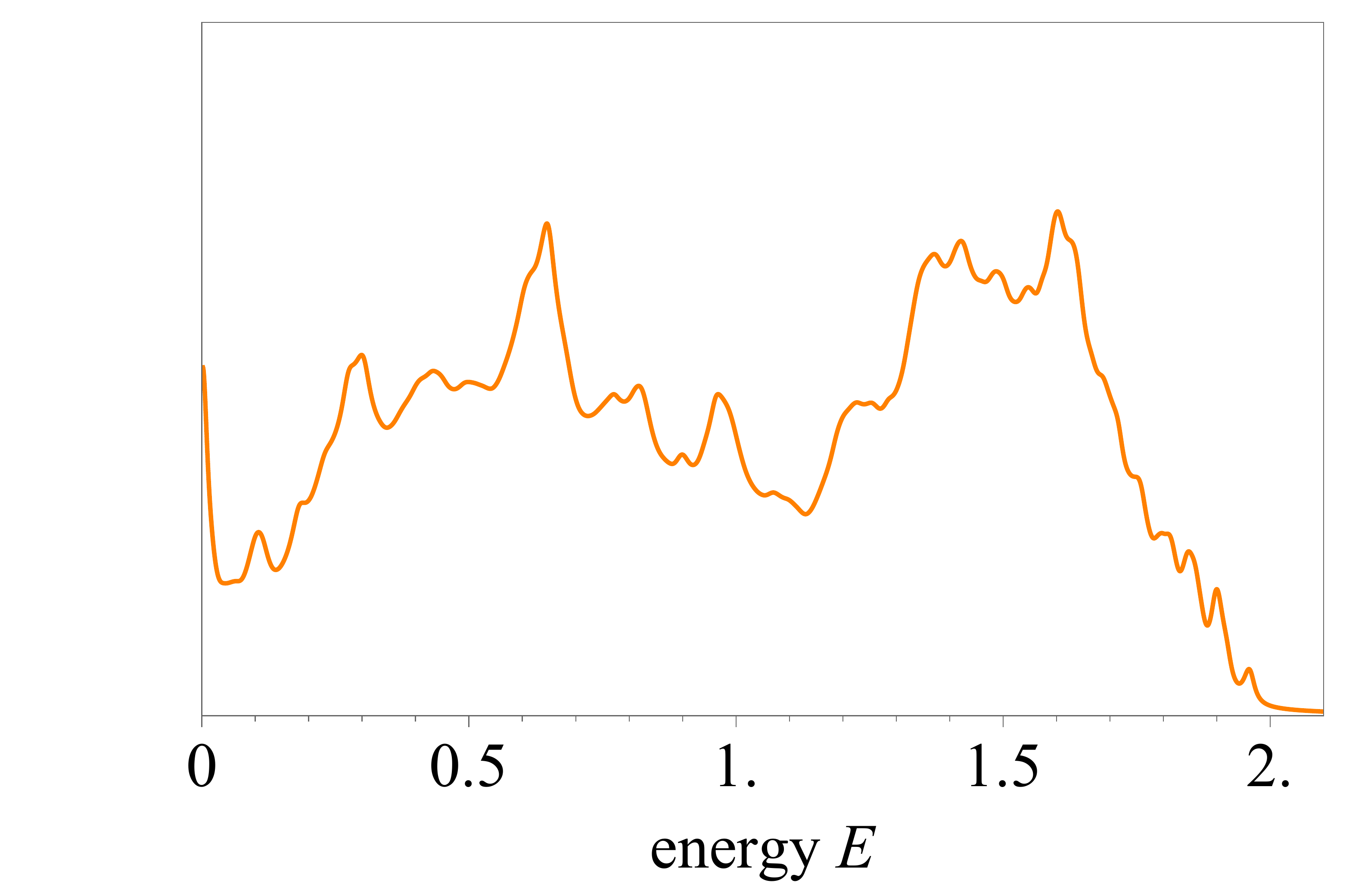}\\
         \caption{{\bf Low-temperature density of states for the (10,3)$x$ lattice systems.} The analytic DOS on the left hand side was calculated via exact diagonalization of the Majorana Hamiltonians in reciprocal space. The numerical DOS on the right hand side was obtained from QMC simulations of finite systems in real space. Data shown is for the linear system size $L=7$, except for (10,3)d (here: $L = 5$) The peak at $E = 0$ for (10,3)d is an artifact from the open boundary condition that was used in the simulation of this lattice.
        }
    \label{fig:DOS10}
\end{figure}

\begin{figure}[t]
   \centering
    \includegraphics[width=0.49\columnwidth]{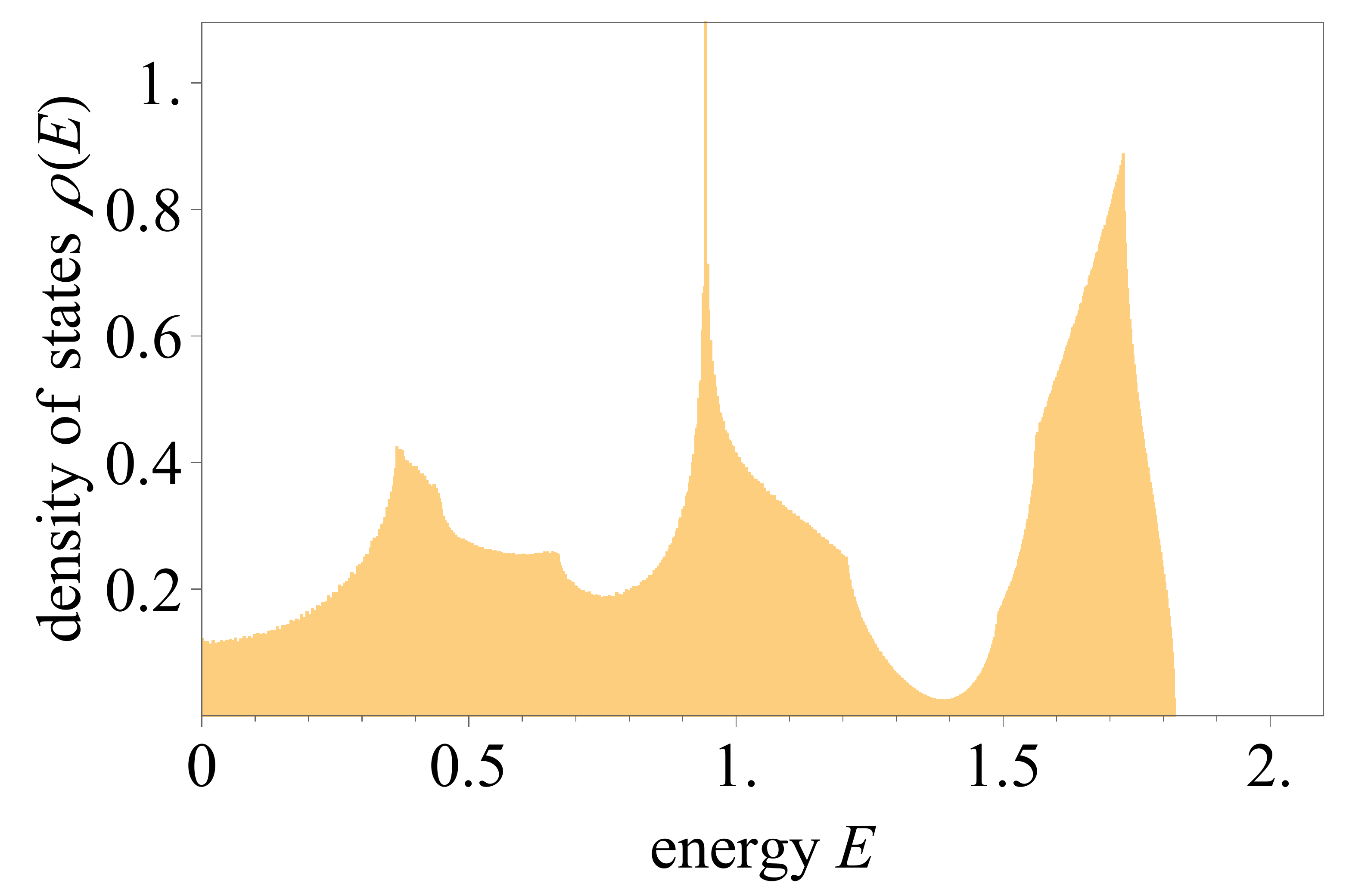}
    \includegraphics[width=0.49\columnwidth]{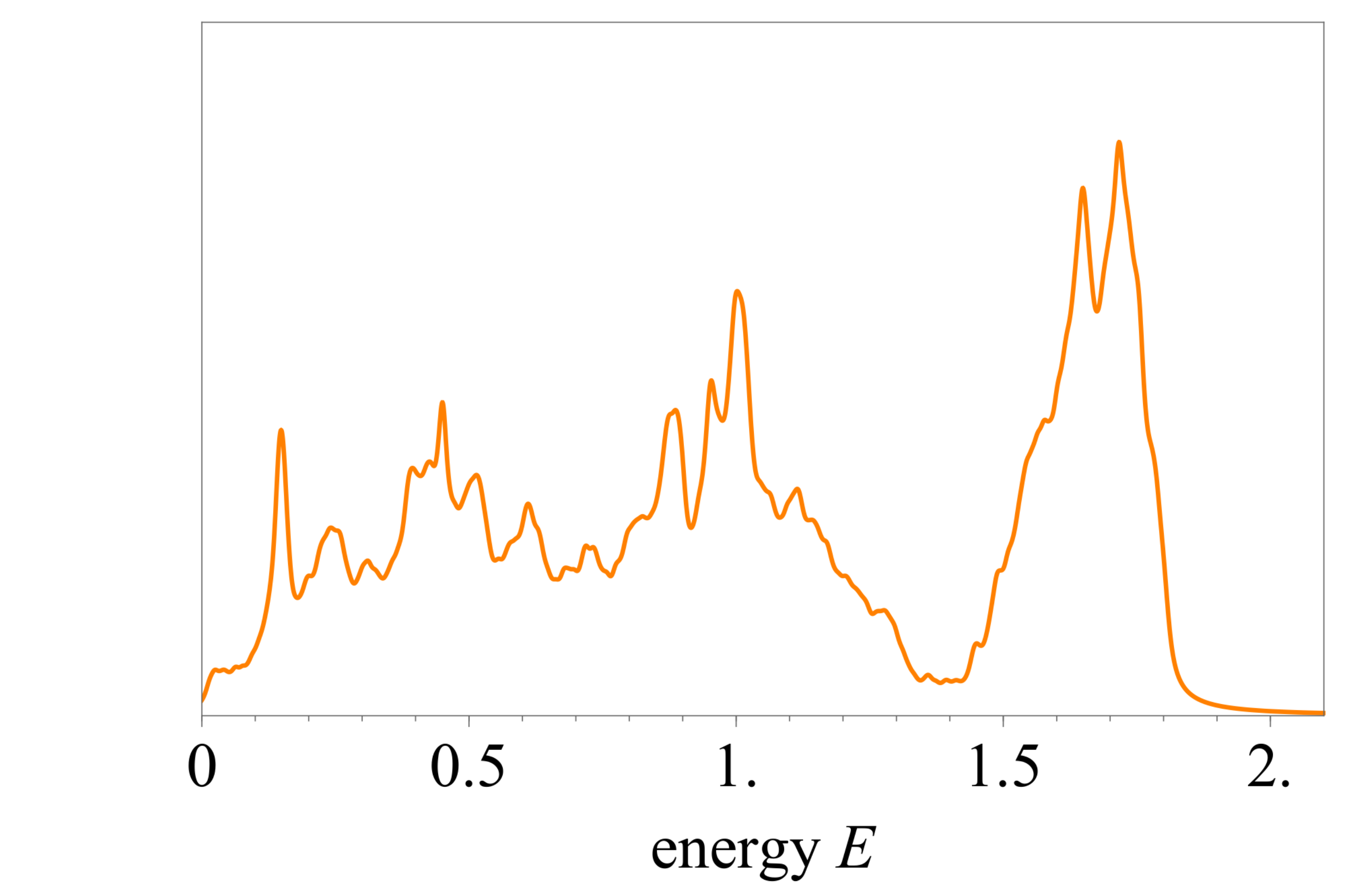} \\  
    \includegraphics[width=0.49\columnwidth]{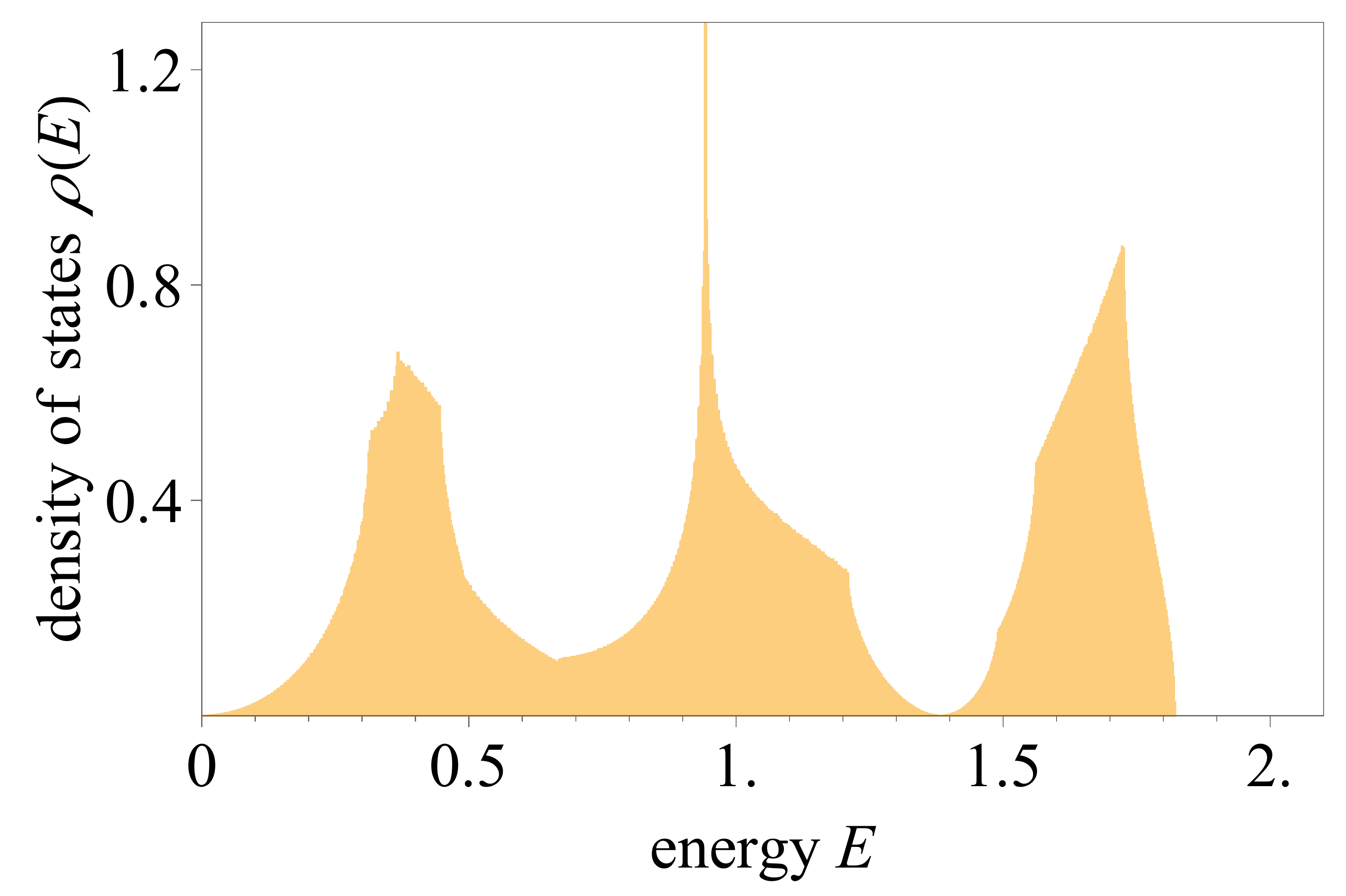}
    \includegraphics[width=0.49\columnwidth]{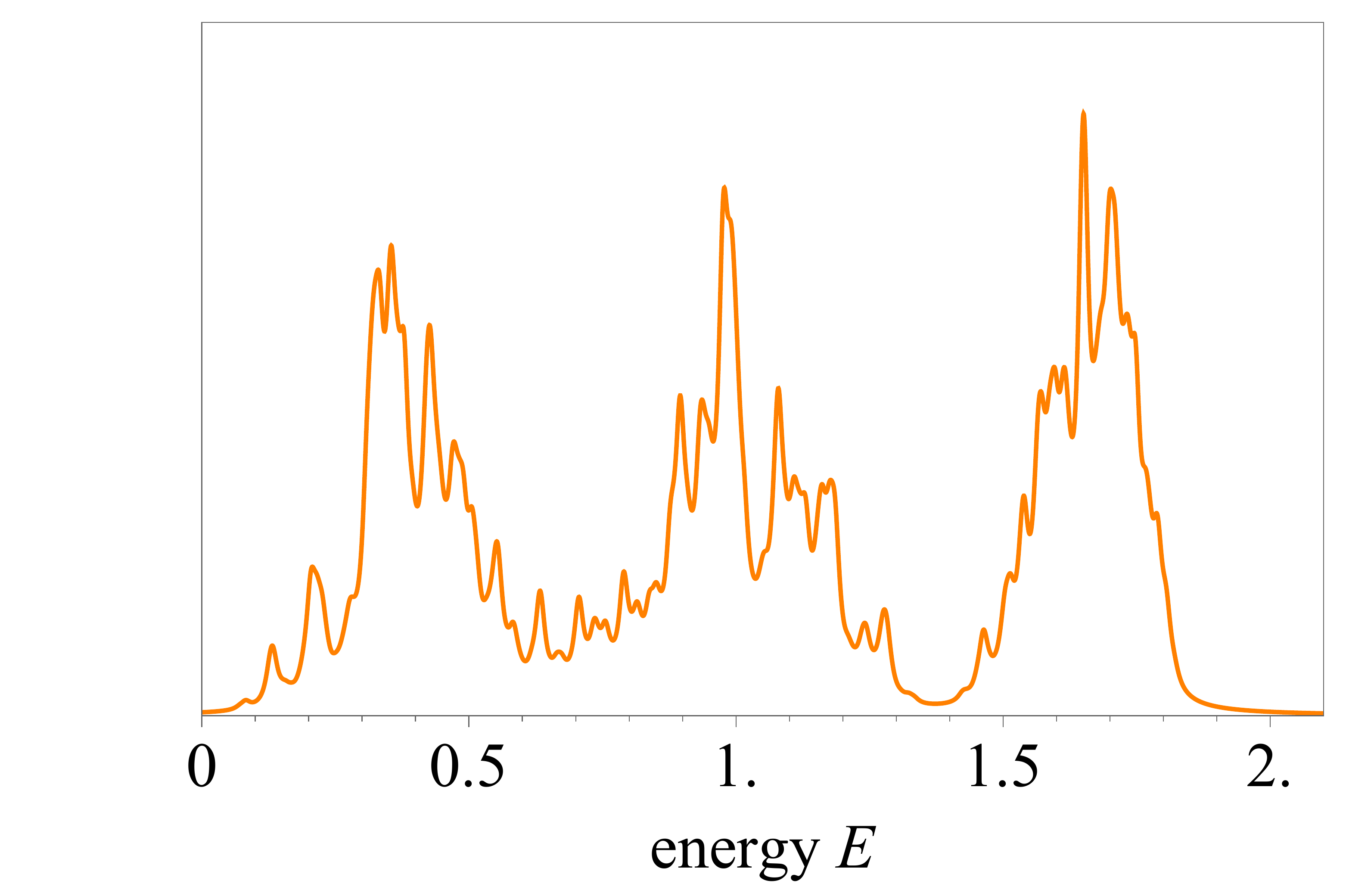} \\ 
    \includegraphics[width=0.49\columnwidth]{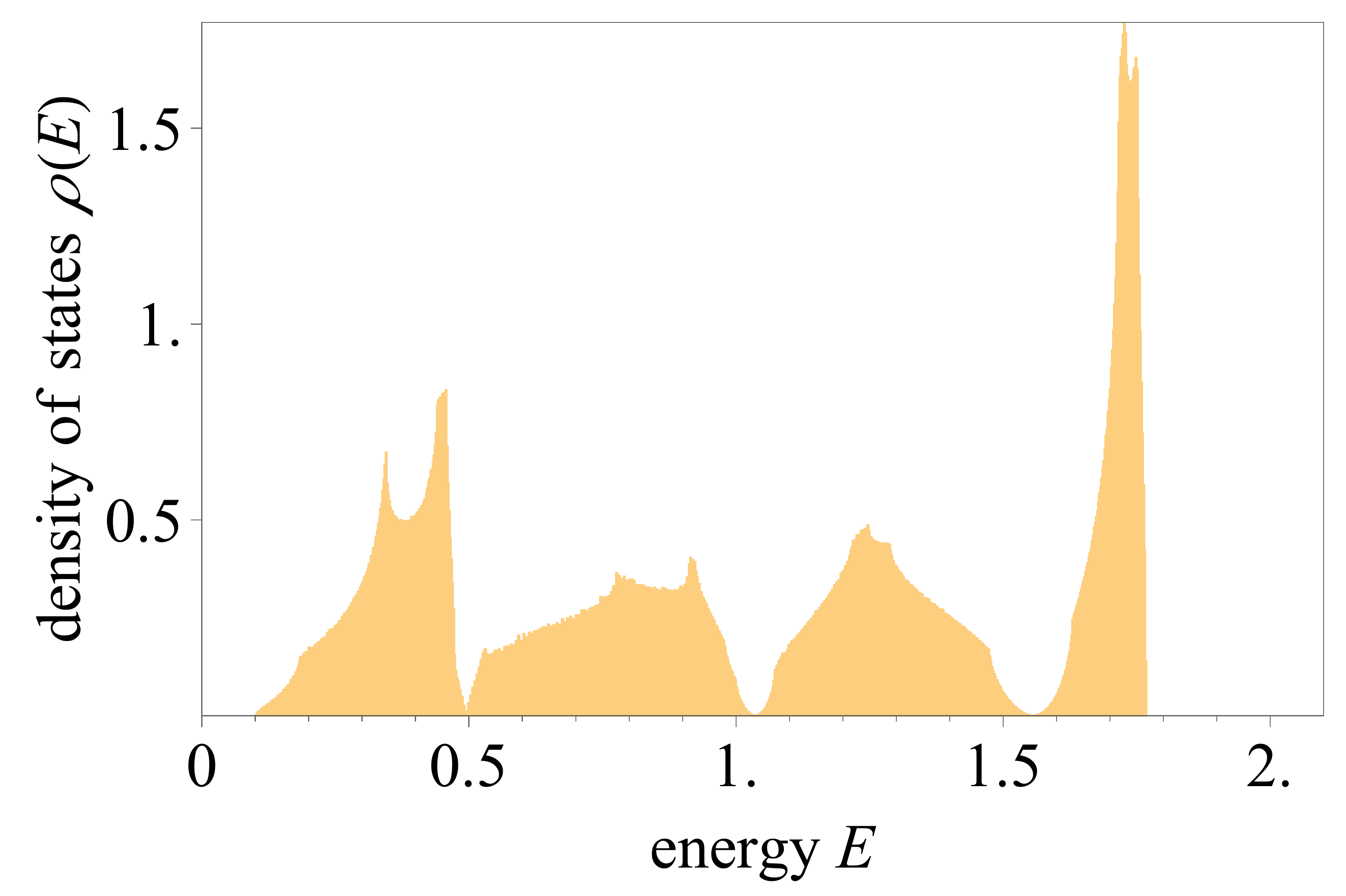}
    \includegraphics[width=0.49\columnwidth]{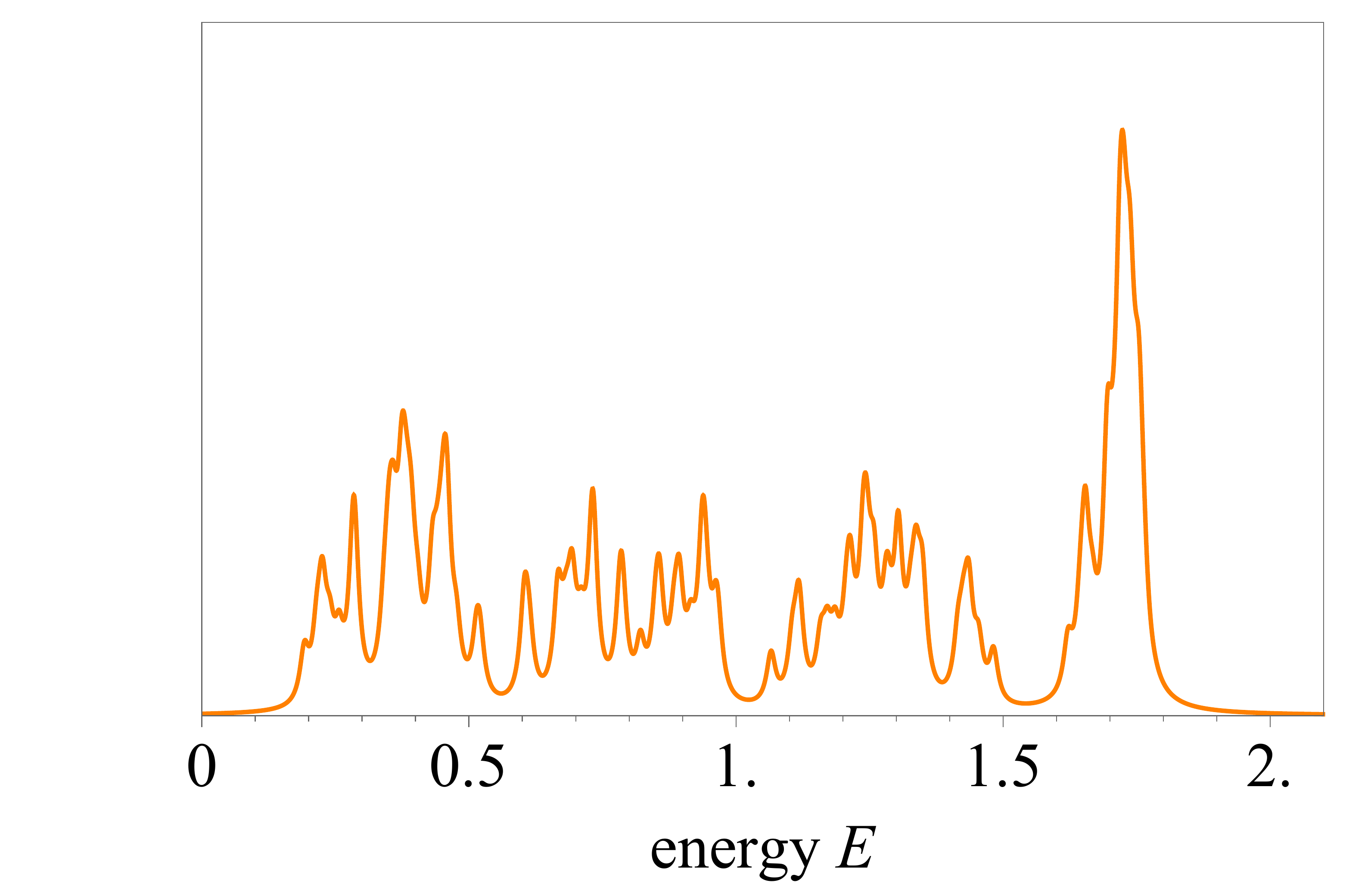} \\ 
   
        \caption{{\bf Low-temperature density of states for the (8,3)$x$ lattice systems.} Data shown is for the linear system size $L=7$, except for (8,3)n  (here: $L = 4$).    
    }
    \label{fig:DOS8}
\end{figure}

The analytical and numerical results for the DOS are given side-by-side in Figs. \ref{fig:DOS10}, \ref{fig:DOS8}, showing distinct features for all lattice geometries. The behavior of the DOS in the region around $E \sim 0$, in particular, is a direct indicator for the topological band structure of the corresponding Majorana (semi-)metal: For systems with a Fermi surface, e.g., lattice geometries (8,3)a and (10,3)a, one finds a {\sl finite} DOS down to the lowest-lying energy levels \cite{Nasu_2015}. The majority of the other lattice systems host Majorana (semi-)metals with Weyl nodes or a nodal line as distinct topological features. For these systems, the DOS approaches zero for $E \rightarrow 0$ (as generally expected). We can further distinguish these topological features by the {\sl shape} of the low-energy DOS: Those systems that possess a nodal line in their ground state -- lattice geometries (10,3)b and (10,3)c -- show a {\sl linear} increase of the DOS close to $E = 0$. This is reminiscent of the 2D Kitaev honeycomb model, which has Dirac cones in the ground state, and also an $\omega$-linear DOS for low energies \cite{Nasu2015thermal}. For lattice geometry (8,3)b -- a system that exhibits Weyl nodes -- the increase of the low-$E$ DOS is instead quadratic. The only lattice geometry not exhibiting such zero-energy (semi-)metallic features is the Kitaev model on lattice (8,3)n which possesses a finite energy gap (even for the isotropic coupling point), clearly visible also in its Majorana DOS. 

Turning an eye on our numerical results we see that the numerical results at low, but finite temperature qualitatively reproduce the key features of the analytical DOS for all lattice geometries. It is only in the region of lowest energies ($E \ll 0.25$) that the numerical results notably deviate from the exact DOS, which is expected due to finite-size and finite-temperature effects in the QMC simulation: For the lowest-lying energy levels, where the DOS is expected to vanish (unless the system possesses a Fermi surface), there is still a certain occupancy probability if the temperature is very low, but still finite. To converge our numerics towards the analytic DOS at this scale, would require simulation times that exceed those of the current calculations by multiple orders of magnitude. But for the broader energy range considered here, our QMC results reproduce the main features of the Majorana DOS in a well-resolved manner.
   

\section{Conclusions}
\label{Conclusions}    

The purpose of this manuscript has been to present a comprehensive overview of the thermodynamics of a family of elementary 3D Kitaev models, based on numerically exact quantum Monte Carlo simulations. We thereby complement an earlier mainly analytical study on the ground state physics of these models. While the latter mainly classified the physics of the emergent {\it Majorana} collective state, we have focused here mainly on the characteristic behavior of the {\it $\mathbb{Z}_2$ gauge field}, whose most interesting physics plays out at finite temperatures and leads to distinct thermodynamic features.

Simulating these 3D Kitaev systems with quantum Monte Carlo techniques, we were able to show that Lieb's theorem, which predicts the energy-minimizing plaquette flux sector for a given lattice geometry, is in fact extendable to lattices beyond the scope of its rigorous applicability. Despite the fact that only one of the systems in this classification, namely lattice geometry (8,3)b, possesses the mirror symmetries that are required to {\it prove} the validity of this theorem, all systems nonetheless exhibit the predicted flux sectors. That is, lattice systems with an even elementary plaquette $|p|$ have a flux-free ground state if $p \mod 4 = 2$, and carry a $\pi$-flux per plaquette if $p \mod 4 = 0$.  Based on this result, the 3D Kitaev systems can be classified in a meaningful way into families of plaquette length 8 / $\pi$-flux ground state and 10 / $0$-flux ground state, respectively. For non-bipartite lattice systems, which possess an odd elementary plaquette length $p$ and flux states $\pm \pi/2$ per plaquette, Lieb's theorem is not applicable, and the ground state flux sector is characterized by the breaking of time-reversal symmetry.

We could further show that the 3D Kitaev systems considered here in general exhibit the thermodynamic behavior established in former QMC studies\cite{Nasu2014vaporization, mishchenko_prb_96_2017}, namely a double peak structure in the specific heat, where the high-temperature peak is a signature of spin fractionalization, while the low-temperature peak indicates a thermal phase transition, at which the $\mathbb{Z}_2$ gauge field assumes its ordered ground state configuration. 
This phase transition is triggered by the creation and proliferation of loop-like (gapped) {\it vison} excitations in the gauge sector, i.e., it separates a low-temperature phase, where no vison loops or only small loops exist in the system, from a high-temperature phase with system-spanning loops. 
Our comprehensive study of this transition across a number of lattice geometries, produces a direct correlation between the critical temperature of this transition and the magnitude of the vison gap for the respective Kitaev model.  
In contrast with the Landau-Ginzburg-Wilson (LGW) paradigm \cite{} for continuous phase transitions, the phase transitions in these 3D Kitaev systems and their associated $\mathbb{Z}_2$ lattice gauge theories generically lack a local order parameter as first pointed out by Wegner \cite{WegnerIsingGaugeTheory}. This makes these 3D Kitaev models one of the most natural habitats to look for continuous phase transitions beyond the conventional LGW paradigm (often discussed in the context of quantum phase transitions \cite{Senthil2004a,Senthil2004b}) in a {\sl thermal} transition \cite{Alet2006,Chen2009}.

We also reviewed the numerical results on two Kitaev systems which are notable exceptions from this generic two-step scenario: For lattice geometry (8,3)c we discussed the unusual phenomenon of a {\sl gauge-frustrated} ground state \cite{2019EschmannGaugeFrustration}. And for the (non-bipartite) (9,3)a system, where the described gauge-ordering mechanism is accompanied by a breaking of time-reversal symmetry and, unexpectedly, a number of lattice symmetries, within a first-order phase transition \cite{2020MishchenkoChiralSpinLiquids}.

On the technical level, we extended the quantum Monte Carlo method that was earlier introduced for Kitaev systems by introducing an approach which relies on the local transformation of spins introduced by Kitaev, instead of a non-local JW transformation. With this approach, we were able to simulate the full gauge sectors in our systems, and could examine systems with periodic boundary conditions.


\acknowledgments
T.E., K. O.,  and S.T. acknowledge partial funding by the Deutsche Forschungsgemeinschaft (DFG, German Research Foundation) -- Projektnummer 277146847 -- SFB 1238 (projects C02 and C03).
 M.H. acknowledges partial funding by the Knut and Alice Wallenberg Foundation and the Swedish Research Council.
P.A.M, T.A.B, Y.K., and Y.M. acknowledge funding by Grant-in-Aid for Scientific Research under Grant No. 15K13533, 16H02206, 18K03447, 19H05825 and 20H00122. Y.M. and Y.K. were also supported by JST CREST (JPMJCR18T2).
The numerical simulations were performed on the JUWELS cluster at the Forschungszentrum J\"ulich.


\bibliography{./Kitaev}


\widetext


\appendix
\section{Lattice definitions}
\label{sec:LatticeDefinitions}

For the different lattices, we used the geometric definitons (unit cells and lattice vectors) which are given in the Tables \ref{TableLatticeDefinitions10x} - \ref{TableLatticeDefinitions9a}.

\begin{table}[h]
\footnotesize
\begin{tabular*}{\columnwidth}{@{\extracolsep{\fill} } l c c c c}
\hline 
\hline
{\bf (10,3) a} & & &\\ 
\hline 
\noalign{\smallskip}
Lattice vectors: & &  ${\bf a}_1 = \left(1, 0, 0 \right)$ & ${\bf a}_2 = \left(\frac{1}{2}, \frac{1}{2}, -\frac{1}{2} \right)$ & ${\bf a}_3 = \left(\frac{1}{2}, \frac{1}{2}, \frac{1}{2} \right)$\\
\noalign{\smallskip}
\hline 
\noalign{\smallskip}
Unit cell: & ${\bf r}_1 = \left(\frac{1}{8}, \frac{1}{8}, \frac{1}{8} \right)$ & ${\bf r}_2 = \left(\frac{5}{8}, \frac{3}{8}, -\frac{1}{8} \right)$ & ${\bf r}_3 = \left(\frac{3}{8}, \frac{1}{8}, -\frac{1}{8} \right)$ & ${\bf r}_4 = \left(\frac{7}{8}, \frac{3}{8}, \frac{1}{8} \right)$ \\
\noalign{\smallskip}
\hline 
\hline

{\bf (10,3) b} & & &\\ 
\hline 
\noalign{\smallskip}
Lattice vectors: & & ${\bf a}_1 = \left(-1, 1, -2 \right)$& ${\bf a}_2 = \left(-1, 1, 2 \right)$ & ${\bf a}_3 = \left(2, 4, 0 \right)$\\
\noalign{\smallskip}
\hline 
\noalign{\smallskip}
Unit cell: & ${\bf r}_1 = \left(0, 0, 0 \right)$ & ${\bf r}_2 = \left(1, 1, 0 \right)$ & ${\bf r}_3 = \left(1, 2, 1 \right)$ & ${\bf r}_4 = \left(0, -1, 1 \right)$\\
\noalign{\smallskip}
\hline 
\hline

{\bf (10,3) c} & & &\\ 
\hline 
\noalign{\smallskip}
Lattice vectors: & & ${\bf a}_1 = \left(1, 0, 0 \right)$ & ${\bf a}_2 = \left(-\frac{1}{2}, \frac{\sqrt{3}}{2}, 0 \right)$ & ${\bf a}_3 = \left(0, 0, \frac{3\sqrt{3}}{2} \right)$\\
\noalign{\smallskip}
\hline 
\noalign{\smallskip}
Unit cell: & ${\bf r}_1 = \left(\frac{1}{4}, \frac{1}{4\sqrt{3}}, \frac{1}{2\sqrt{3}} \right)$  & ${\bf r}_2 = \left(\frac{3}{4}, \frac{1}{4\sqrt{3}}, \frac{2}{\sqrt{3}} \right)$ & ${\bf r}_3 = \left(\frac{1}{2}, \frac{1}{\sqrt{3}}, \frac{7}{2\sqrt{3}} \right)$ & ${\bf r}_4 = \left(\frac{3}{4}, \frac{1}{4\sqrt{3}}, \frac{1}{\sqrt{3}}\right)$ \\
\noalign{\smallskip}
& ${\bf r}_5 = \left(\frac{1}{2}, \frac{1}{\sqrt{3}}, \frac{5}{2\sqrt{3}} \right)$ & ${\bf r}_6 = \left(\frac{1}{4}, \frac{1}{4\sqrt{3}}, \frac{4}{\sqrt{3}}\right)$ & & \\
\noalign{\smallskip}
\hline 
\hline

{\bf (10,3) d} & & &\\ 
\hline 
\noalign{\smallskip}
Lattice vectors: & & ${\bf a}_1 = \left(\frac{1}{2}, -\frac{1}{2}, 0 \right)$ & ${\bf a}_2 = \left(\frac{1}{2}, \frac{1}{2}, 0 \right)$ &  ${\bf a}_3 = \left(0, 0, \frac{1}{2} \right)$\\
\noalign{\smallskip}
\hline 
\noalign{\smallskip}
Unit cell: & & & $a = \frac{1}{4} \left(2 - \sqrt{2} \right)$ & $c = \frac{1}{2}$\\
\noalign{\smallskip}
& ${\bf r}_1 = \left(0, -a, \frac{3}{4}c \right)$ & ${\bf r}_2 = \left(-a, 0, \frac{1}{2}c \right)$ & ${\bf r}_3 = \left(0, a, \frac{1}{4}c \right)$ & ${\bf r}_4 = \left(a, 0, 0 \right)$\\
\noalign{\smallskip}
& ${\bf r}_5 = \left(-a, -\frac{1}{2}, \frac{1}{4}c \right)$ & ${\bf r}_6 = \left(0, a - \frac{1}{2}, \frac{1}{2}c \right)$ & ${\bf r}_7 = \left(a, -\frac{1}{2}, \frac{3}{4}c \right)$ & ${\bf r}_8 = \left(0, -a- \frac{1}{2},0 \right)$\\
\noalign{\smallskip}
\hline 
\hline

\end{tabular*} 
\caption{Lattice definitions of the (10,3)$x$ family.}
\label{TableLatticeDefinitions10x}
\end{table}

\begin{table}[h]
\footnotesize
\begin{tabular*}{\columnwidth}{@{\extracolsep{\fill} } l c c c}
\hline 
\hline
{\bf (8,3) a} & & & \\ 
\hline 
\noalign{\smallskip}
Lattice vectors: & ${\bf a}_1 = \left(1, 0, 0 \right)$  & ${\bf a}_2 = \left(-\frac{1}{2}, \frac{\sqrt{3}}{2}, 0 \right)$ & ${\bf a}_3 = \left(0, 0, \frac{3\sqrt{2}}{5} \right)$  \\ 
\noalign{\smallskip}
\hline 
\noalign{\smallskip}
Unit cell: & ${\bf r}_1 = \left(\frac{1}{2}, \frac{\sqrt{3}}{10}, 0 \right)$  & ${\bf r}_2 = \left(-\frac{3}{5}, \frac{\sqrt{3}}{5}, \frac{2\sqrt{2}}{5} \right)$  & ${\bf r}_3 = \left(\frac{1}{10}, \frac{3\sqrt{3}}{10}, \frac{\sqrt{2}}{5} \right)$ \\ 
\noalign{\smallskip}
 & ${\bf r}_4 = \left(\frac{4}{10}, \frac{\sqrt{3}}{5}, \frac{\sqrt{2}}{5} \right)$  & ${\bf r}_5 = \left(0, \frac{2\sqrt{3}}{5}, 0 \right)$  &  ${\bf r}_6 = \left(-\frac{1}{10}, \frac{3\sqrt{3}}{10}, \frac{2\sqrt{2}}{5} \right)$\\  
\noalign{\smallskip}
\hline 

\hline

{\bf (8,3) b} & & & \\ 
\hline 
\noalign{\smallskip}
Lattice vectors: & ${\bf a}_1 = \left(\frac{1}{2}, \frac{1}{2\sqrt{3}}, \frac{\sqrt{2}}{5\sqrt{3}} \right)$ & ${\bf a}_2 = \left(0, \frac{1}{\sqrt{3}}, \frac{2\sqrt{2}}{5\sqrt{3}} \right)$ &  ${\bf a}_3 = \left(0, 0, \frac{\sqrt{6}}{5} \right)$\\ 
\noalign{\smallskip}
\hline 
\noalign{\smallskip}
Unit cell: & ${\bf r}_1 = \left(\frac{1}{10}, \frac{1}{2\sqrt{3}}, \frac{\sqrt{2}}{5\sqrt{3}} \right)$ & ${\bf r}_2 = \left(\frac{1}{5}, \frac{\sqrt{3}}{5}, \frac{\sqrt{6}}{5} \right)$ & ${\bf r}_3 = \left(\frac{3}{10}, \frac{11}{10\sqrt{3}}, \frac{4\sqrt{2}}{5\sqrt{3}} \right)$ \\ 
\noalign{\smallskip}
 & ${\bf r}_4 = \left(\frac{1}{5}, \frac{2}{5\sqrt{3}}, \frac{2\sqrt{2}}{5\sqrt{3}} \right)$ & ${\bf r}_5 = \left(\frac{3}{10}, \frac{3\sqrt{3}}{10}, \frac{\sqrt{6}}{5} \right)$ & ${\bf r}_6 = \left(\frac{2}{5}, \frac{1}{\sqrt{3}}, \frac{\sqrt{2}}{\sqrt{3}} \right)$ \\  
\noalign{\smallskip}
\hline 

\hline

{\bf (8,3) c} & & & \\ 
\hline 
\noalign{\smallskip}
Lattice vectors: &   ${\bf a}_1 = \left(1, 0, 0 \right)$    &  ${\bf a}_2 = \left(-\frac{1}{2}, \frac{\sqrt{3}}{2}, 0 \right)$   &  ${\bf a}_3 = \left(0, 0, \frac{\sqrt{2}}{5} \right)$   \\ 
\noalign{\smallskip}
\hline 
\noalign{\smallskip}
Unit cell: & ${\bf r}_1 = \left(-\frac{1}{5}, \frac{4}{5\sqrt{3}}, \frac{1}{10} \right)$  & ${\bf r}_2 = \left(0, \frac{7}{5\sqrt{3}}, \frac{1}{10} \right)$  &  ${\bf r}_3 = \left(\frac{1}{5}, \frac{4}{5\sqrt{3}}, \frac{1}{10} \right)$  \\ 
\noalign{\smallskip}
 &  ${\bf r}_4 = \left(\frac{1}{2}, \frac{1}{2\sqrt{3}}, \frac{3}{10} \right)$  &  ${\bf r}_5 = \left(0, \frac{1}{\sqrt{3}}, \frac{1}{10} \right)$  &  ${\bf r}_6 = \left(\frac{3}{10}, \frac{7}{10\sqrt{3}}, \frac{3}{10} \right)$ \\  
\noalign{\smallskip}
 &  ${\bf r}_7 = \left(\frac{1}{2}, \frac{1}{10\sqrt{3}}, \frac{3}{10} \right)$ &  ${\bf r}_8 = \left(\frac{7}{10}, \frac{7}{10\sqrt{3}}, \frac{3}{10} \right)$ & \\
\noalign{\smallskip}

\hline 

\hline

{\bf (8,3) n} & & & \\ 
\hline 
\noalign{\smallskip}
Lattice vectors: &   ${\bf a} = \left(1, 0, 0 \right)$    &   ${\bf b} = \left(0, 1, 0 \right)$  &  ${\bf c} = \left(0, 0, \frac{4}{2\sqrt{3}+\sqrt{2}} \right)$   \\ 
\noalign{\smallskip}
& ${\bf a}_1 = {\bf a}$ & ${\bf a}_2 = {\bf b}$ & ${\bf a}_3 = \frac{1}{2} ({\bf a} + {\bf b} + {\bf c})$ \\
\noalign{\smallskip}
\hline 
\noalign{\smallskip}
Unit cell: & & $x = \frac{\sqrt{3}+\sqrt{2}}{2\left(2\sqrt{3}+\sqrt{2} \right)}$ & $z = \frac{1}{8}$ \\ 
  &${\bf r}_1 = x\cdot{\bf a} + \left(\frac{1}{2} - x \right)\cdot{\bf b} + \frac{1}{4}\cdot{\bf c}$ & ${\bf r}_2 = (1-x)\cdot{\bf a} + \left(\frac{1}{2} - x \right)\cdot{\bf b} + \frac{1}{4}\cdot{\bf c}$ & ${\bf r}_3 = \left(\frac{1}{2} + x \right)\cdot{\bf a} + \frac{1}{2}{\bf b} + \left(\frac{1}{2} - z \right)\cdot{\bf c}$\\  
\noalign{\smallskip}
  & ${\bf r}_4 = (1-x)\cdot{\bf a} + \left(\frac{1}{2} + x \right)\cdot{\bf b} + \frac{1}{4}\cdot{\bf c}$ & ${\bf r}_5 = x\cdot{\bf a} + \left(\frac{1}{2} + x \right)\cdot{\bf b} + \frac{1}{4}\cdot{\bf c}$ & ${\bf r}_6 = \left(\frac{1}{2} - x \right)\cdot{\bf a} + \frac{1}{2}{\bf b} + \left(\frac{1}{2} - z \right)\cdot{\bf c}$\\
\noalign{\smallskip}
& ${\bf r}_7 = (1-x)\cdot{\bf b} + z\cdot{\bf c}$ & ${\bf r}_8 = x\cdot{\bf b} + z\cdot{\bf c}$ & ${\bf r}_9 = \left(\frac{1}{2} - x \right)\cdot{\bf a} + x\cdot{\bf b} + \frac{1}{4}\cdot{\bf c}$\\
& ${\bf r}_{10} =  \frac{1}{2}\cdot{\bf a} + \left(\frac{1}{2} - x \right)\cdot{\bf b} + \left(\frac{1}{2} - z \right)\cdot{\bf c}$ & ${\bf r}_{11} = \left(\frac{1}{2} + x \right)\cdot{\bf a} + x\cdot{\bf b} + \frac{1}{4}\cdot{\bf c}$ & ${\bf r}_{12} = \left(\frac{1}{2} + x \right)\cdot{\bf a} + (1-x)\cdot{\bf b} + \frac{1}{4}\cdot{\bf c}$ \\
\noalign{\smallskip}
& ${\bf r}_{13} = \frac{1}{2}\cdot{\bf a} +  \left(\frac{1}{2} + x \right)\cdot{\bf b} + \left(\frac{1}{2} - z \right)\cdot{\bf c}$ & ${\bf r}_{14} = \left(\frac{1}{2} - x \right)\cdot{\bf a} + (1-x)\cdot{\bf b} + \frac{1}{4}\cdot{\bf c}$ & ${\bf r}_{15} = x\cdot{\bf a} + z\cdot{\bf c}$\\
\noalign{\smallskip}
& ${\bf r}_{16} = (1-x)\cdot{\bf a} + z\cdot{\bf c}$& & \\
\noalign{\smallskip}
\hline 
\hline

\end{tabular*} 
\caption{Lattice definitions of the (8,3)$x$ family.}
\label{TableLatticeDefinitions8x}
\end{table}

\begin{table}[h]
\footnotesize
\begin{tabular*}{\columnwidth}{@{\extracolsep{\fill} } l c c c}
\hline 
\hline
{\bf (9,3) a} & & & \\ 
\hline 
\noalign{\smallskip}
Lattice vectors: &${\bf a}_1 = \left(-\frac{\sqrt{3}}{2}, \frac{1}{2}, \frac{1}{\sqrt{3}} \right)$ &  ${\bf a}_2 = \left(0, -1, \frac{1}{\sqrt{3}} \right)$   &  ${\bf a}_3 = \left(\frac{\sqrt{3}}{2}, \frac{1}{2}, \frac{1}{\sqrt{3}} \right)$ \\ 
\noalign{\smallskip}
\hline 
\noalign{\smallskip}
Unit cell: &${\bf r}_1 = \left(-\frac{2}{\sqrt{3}}, 0, 0 \right)$ &  ${\bf r}_2 = \left(-\frac{7}{4\sqrt{3}}, \frac{1}{4}, 0 \right)$   &  ${\bf r}_3 = \left(-\frac{7}{4\sqrt{3}}, \frac{1}{4}, \frac{1}{\sqrt{3}} \right)$  \\ 
& ${\bf r}_4 = \left(-\frac{\sqrt{3}}{2}, \frac{1}{2}, \frac{1}{\sqrt{3}} \right)$ &  ${\bf r}_5 = \left(-\frac{1}{2\sqrt{3}}, -\frac{1}{2}, 0 \right)$  &  ${\bf r}_6 = \left(-\frac{1}{\sqrt{3}}, -\frac{1}{2}, 0 \right)$  \\
\noalign{\smallskip}
&${\bf r}_7 = \left(-\frac{1}{\sqrt{3}}, -\frac{1}{2}, \frac{1}{\sqrt{3}} \right)$ & ${\bf r}_8 = \left(-\frac{\sqrt{3}}{2}, -\frac{1}{2}, \frac{1}{\sqrt{3}} \right) $ & ${\bf r}_9 = \left(-\frac{1}{2\sqrt{3}}, \frac{1}{2}, 0 \right)$\\
\noalign{\smallskip}
& ${\bf r}_{10} = \left(-\frac{1}{4\sqrt{3}}, \frac{1}{4}, 0 \right) $& ${\bf r}_{11} = \left(-\frac{1}{4\sqrt{3}}, \frac{1}{4}, \frac{1}{\sqrt{3}} \right)$ & ${\bf r}_{12} = \left(0, 0, \frac{1}{\sqrt{3}} \right)$\\
\noalign{\smallskip}
\hline 
\hline
\end{tabular*} 
\caption{Lattice definition of the (9,3)a lattice.}
\label{TableLatticeDefinitions9a}
\end{table}


\section{Lieb's theorem}
\label{sec:LiebTheorem}

The problem of finding the flux ground state for a half-filled band of hopping electrons was intensely studied in mathematical physics in the early 1990's. The interest for this problem rooted in the attention on an intriguing phenomenon that arises in the context of correlated electron systems and superconductivity: It has been noticed that under certain conditions, i.e., in systems with a high electron density, the effect of diamagnetism can be reversed. In these systems, the application of a magnetic field does in fact not raise, but {\it lower} the energy. 

This discovery led to the formulation of the {\it flux phase conjecture}, which states that on a planar square lattice with free hopping electrons, the energy minimizing magnetic flux is $\pi$ per square, if the electron filling factor is $\frac{1}{2}$. More general, it was stated that on planar lattices, the optimum flux choice per plaquette (circuit) is $\pi$ for plaquettes containing 0 (mod 4) sites, and 0 for plaquettes with 2 (mod 4) sites. This conjecture was proved by Lieb and coworkers for several lattice graphs, such as rings, trees of rings, ladders and necklaces \cite{Lieb1993}, which laid the foundation for the later formulation of {\it Lieb's theorem}\cite{Lieb1994}.

\begin{figure}[h]
    \centering
    \includegraphics[width=0.48\columnwidth]{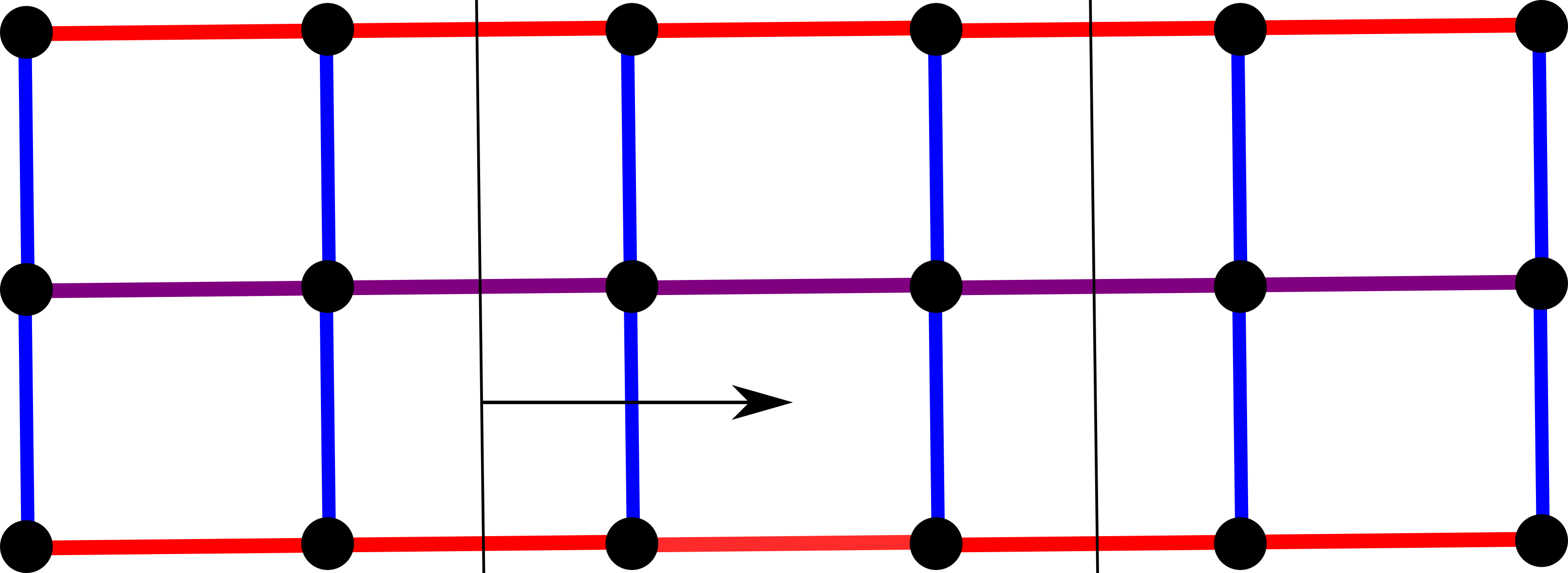}  
      \caption{{\bf Reflection symmetry in Lieb's theorem.} Square lattice with periodic boundary conditions in the horizontal direction (the sites on the left hand side and on the right hand side are the same). The thin black lines are mirror lines. All the bond couplings (indicated in red, purple and blue) are mirror-symmetric. Lieb's theorem states that a flux $\pi$ per square minimizes the energy for the plaquettes that are cut in half by the mirror lines. Since the mirror symmetry condition is also fulfilled if the mirror lines are moved by integer multiples of the horizontal lattice vector (indicated in black), it is proven that the $\pi$ flux is the optimal choice for any square plaquette. The theorem includes the statement that a 0 flux optimizes the mirror-symmetric hexagonal plaquette, and so on. It can be generalized to a $D$-dimensional lattice with ($D-1$)-dimensional hyperplanes that do not intersect any vertices.  }
    \label{fig:MirrorPlaneApp}
\end{figure}

The set up for Lieb's theorem is a finite graph $\Lambda$, consisting of $|\Lambda|$ sites, which are indexed by $x,y$, and hopping amplitudes $t_{xy} = |t_{xy}| e^{i \phi(x,y)}$ (with $\phi(x,y) = -\phi(y,x)$ and $t_{xx} = 0$ for all $x$). The quest is for the numbers $\phi(x,y)$ which minimize the electronic ground state energy of the tight-binding Hamiltonian $K = -\sum_{x,y} t_{xy} c_x^\dagger c_y$ (in fact, different fluxes for up- and down-spins are also allowed, as well as further terms in the Hamiltonian, which introduce longer range density-density or spin-spin interactions). It had been proven before\cite{Lieb1993} that the spectrum of the Hermitian matrix $T = \{t_{xy}\}$ only depends on the numbers $\phi$ trough the {\it fluxes}. The latter are defined on closed loops (circuits), i.e., sequences of connected lattice points $x_1, x_2, ..., x_n, x_1$ (with $t_{x_i, x_{i=1}} \neq 0$ for all i) by
\begin{equation}
\Phi = \sum_i^n \phi(x_i, x_{i+1}) \mod 2\pi.
\end{equation}
Note that in the Kitaev model, the $\mathbb{Z}_2$ gauge field $u_{ij}$ corresponds to the hopping phase factor $e^{i \phi(x,y)}$ in the general setup, and the loop operator eigenvalue $W_p$ to the exponentiated flux term $e^{i\Phi}$. The ground state energy of $K$ is given by the sum over the negative eigenvalues of $T$
\begin{equation}
E_0 = \sum_{\lambda = 1}^{N/2} \epsilon_\lambda(T).
\end{equation}
The flux conjecture was proven for systems with a certain periodicity requirement: The lattice $\Lambda$ has to be (at least) half-periodic in the horizontal direction. Then, it can be cut into two half-cylinders, with the cutting lines intersecting only bonds, such that the two half-cylinders are mirror images of each other in terms of the bond couplings $|t_{xy}|$ (see Fig. \ref{fig:MirrorPlaneApp} for the square lattice). The proof of Lieb's theorem, which we will not reproduce in detail here, now shows that the energy-minimizing flux is $\pi$ for the squares containing the cutting lines. If in addition to the aforementioned geometric requirement, reflection symmetry of the bond couplings is fulfilled for {\it any} choice of cutting lines, it follows from Lieb's theorem that $\pi$ is the optimal flux choice for every square of $\Lambda$. 

The theorem includes the prediction of the respective ground state fluxes of hexagonal, octagonal and further plaquettes, with the same argument. It can be further generalized to D-dimensional hypercubes instead of squares, if reflection symmetry is realized with respect to (D--1)-dimensional hyperplanes. Then, it states that $\pi$ is the optimal flux choice for each 2D square plaquette that is cut by the hyperplane. Also here, it follows that the flux is optimal for every plaquette if periodicity in (D--1) dimensions is fulfilled. Among the 3D Kitaev system, the reflection symmetry condition is completely fulfilled only for (8,3)b, while in (8,3)n, seven out of eight elementary plaquettes per unit cell are mirror-symmetric in the described way.

An alternative proof of Lieb's theorem was later presented by Macris and coworkers\cite{Macris1996}, which makes use of the same symmetry requirements as the proof by Lieb. To summarize, we have seen that the applicability of the mirror symmetry argument is a {\it sufficient}, but not a {\it necessary} condition for the validity of the flux phase conjecture.  


\section{Local transformation}
\label{sec:LocalTransformation}

The major part of the exact solution of the Kitaev model was explained in the main text. By applying the local transformation of spins into Majorana operators, $\sigma_i^\gamma = i b_i^\gamma c_i$, and then recombining the $b$-Majoranas to bond operators via $\hat{u}_{ij}^\gamma = ib_i^\gamma b_j^\gamma$, the Kitaev Hamiltonian takes up the form\cite{Kitaev2006anyons}
\begin{equation}
\mathcal{H} = \frac{i}{4} \sum_{i,j} A_{ij} c_i c_j,
\end{equation}
with the non-zero matrix entries $A_{ij} = 2J^\gamma u_{ij}^\gamma$ for connected bonds $\langle i, j\rangle$. 

After that, the Majoranas are basis-transformed to normal modes according to
\begin{equation}
(b_1', b_1'', ..., b_m', b_m'') = (c_1, c_2, ..., c_{2m-1},c_{2m}) Q,
\end{equation}
where $Q$ is a transformation matrix consisting of the real (imaginary) parts of the eigenvectors of $iA$ in their odd (even) columns. The matrix $A$ and the eigenvalues $\epsilon_i$ of $iA$ are related with the transformation matrix $Q$ by
\begin{equation}
A = Q \left( \begin{array}{ccccc}
0 & \epsilon_1 &  &  &  \\ 
-\epsilon_1 & 0 &  &  &  \\ 
 &  & \ddots &  &  \\ 
 &  &  & 0 & \epsilon_m \\ 
 &  &  & -\epsilon_m & 0
\end{array}  \right) Q^T.
\end{equation}
Finally,  the basis-transformed Majorana operators $b_\lambda', b_\lambda''$ are tranformed into spinless fermion operators via
\begin{align}
a_\lambda^\dagger &= \frac{1}{2} (b_\lambda' - ib_\lambda'')\nonumber\\
a_\lambda &= \frac{1}{2} (b_\lambda' + ib_\lambda'').
\end{align}
Inserting the operators $a^\dagger, a$ into the Hamiltonian gives its diagonal form
\begin{equation}
\mathcal{H} = \sum_{\lambda = 1}^{N/2} \epsilon_\lambda \left(a_\lambda^\dagger a_\lambda - \frac{1}{2}\right). 
\end{equation}


\section{Jordan-Wigner transformation}
\label{sec:JWTransformation}

In contrast with the local transformation ansatz introduced by Kitaev\cite{Kitaev2006anyons}, an alternative approach, which leads to the same exact solution, makes use of a non-local JW transformation  and was applied in earlier QMC studies on Kitaev systems \cite{chen_prb_76_2007, feng_prl_98_2007, chen_j_phys_a_41_2008, Nasu2014vaporization}.
\begin{figure}[h]
   \centering
    \includegraphics[width=0.5 \columnwidth]{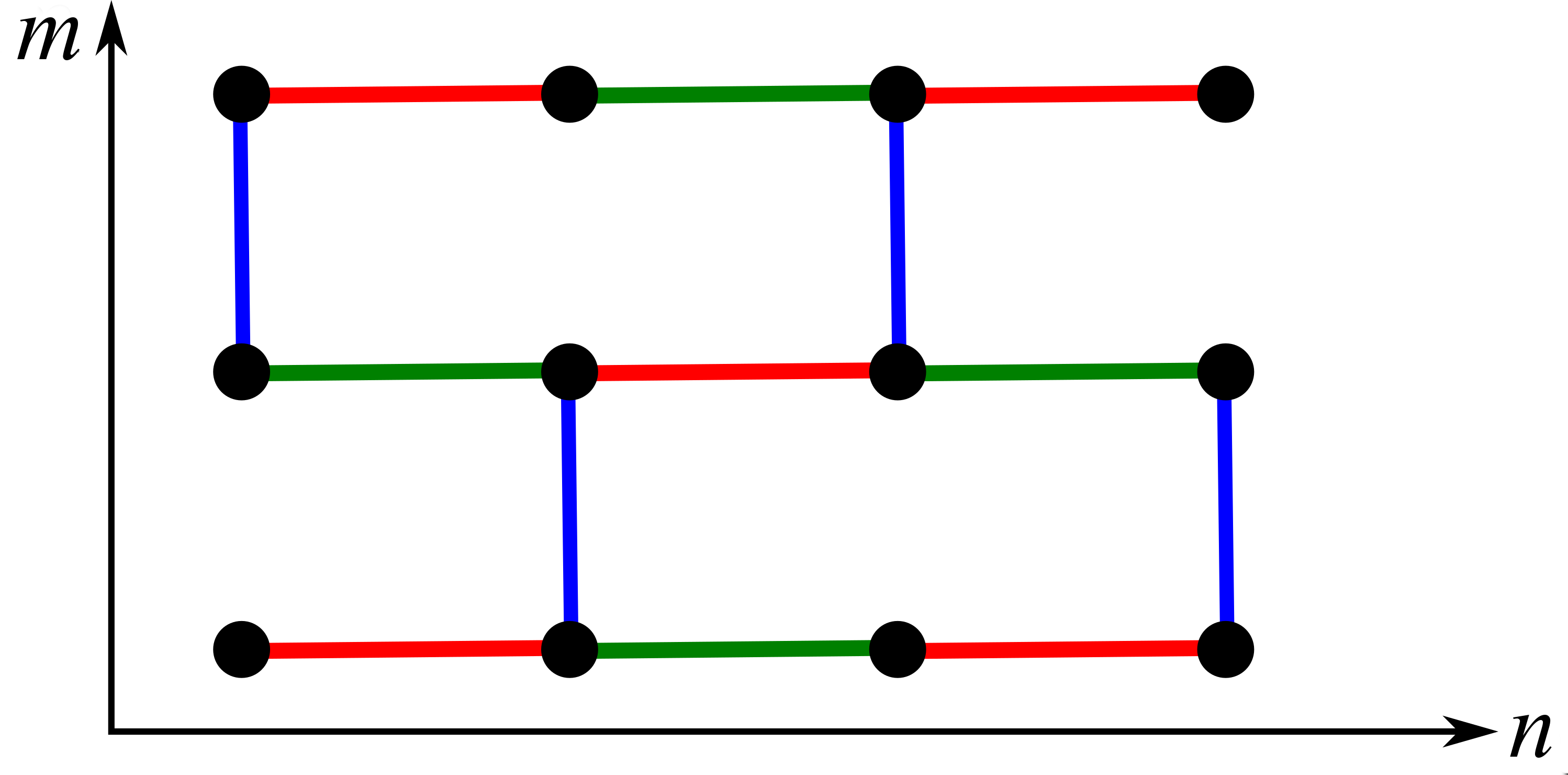}  
     \caption{{\bf Honeycomb lattice in bricklayer geometry.} Rows (columns) of sites are labeled by the coordinates $m$ ($n$). The JW strings are defined along rows of $x$- and $y$-bonds (red and green). The $z$-bonds (blue) host the $\mathbb{Z}_2$ gauge variables $\eta$.}
    \label{fig:JWBricklayer}
\end{figure}
Here, the system is regarded as being composed of one-dimensional strings of bonds, which belong to two of the three subclasses $\gamma$ (e.g., $x$- and $y$-bonds). The strings are connected by bonds of the third subclass. A convenient way to visualize this in 2D systems is to transform the underlying lattice to a bricklayer geometry (see Fig. \ref{fig:JWBricklayer}) for the honeycomb example). In this geometry, where rows and columns of sites can be labeled by the two coordinates $m$ and $n$, the Kitaev Hamiltonian is rewritten as
\begin{equation}
\mathcal{H} = \sum_{m + n even} \left( -J_x \sigma_{m,n}^x \sigma_{m,n+1}^x - J_y \sigma_{m,n-1}^y \sigma_{m,n}^y - J_z \sigma_{m,n}^z \sigma_{m+1,n}^z \right).
\end{equation}
In the following, we choose the  JW strings along $x$- and $y$-bonds. With this choice, the spin operators are replaced with spinless fermion operators $a^\dagger, a$ via
\begin{align}
\sigma_{m,n}^x \sigma_{m,n+1}^x &= -(a_{m,n} - a_{m,n}^\dagger)(a_{m,n+1} + a_{m,n+1}^\dagger)\nonumber\\
\sigma_{m,n}^y \sigma_{m,n+1}^y &= (a_{m,n} + a_{m,n}^\dagger)(a_{m,n+1} - a_{m,n+1}^\dagger)\nonumber\\
\sigma_{m,n}^z \sigma_{m+1,n}^z &= (2 n_{m,n} - 1)(2 n_{m+1,n} - 1).
\end{align}
The Hamiltonian now reads
\begin{align}
\label{TransformedJWHam}
\mathcal{H} = \sum_{m + n even} & \{ J_x (a_{m,n} - a_{m,n}^\dagger)(a_{m,n+1} + a_{m,n+1}^\dagger) + J_y  (a_{m,n-1} + a_{m,n-1}^\dagger)(a_{m,n} - a_{m,n}^\dagger) \nonumber\\ & \left. - J_z  (2 n_{m,n} - 1)(2 n_{m+1,n} - 1) \right \}.
\end{align}
On bipartite lattices, the two sublattices can be distinguished by the labels $A$,$B$. With these labels, different Majorana operators $c_{A/B}$ for the two sublattices can be defined via
\begin{align}
c_A &= -i (a_A - a_A^\dagger)\nonumber\\
c_B &= a_B + a_B^\dagger\nonumber\\
\bar{c}_A &= a_A + a_A^\dagger\nonumber\\
\bar{c}_B &= -i (a_B - a_B^\dagger).
\end{align} 
With these Majorana operators, the Hamiltonian in Eq. \eqref{TransformedJWHam} is finally rewritten as
\begin{equation}
\label{JWFinalHamiltonian}
H = \sum_{m + n even} \left( iJ_x c_{m,n} c_{m,n+1} - i J_y c_{m,n-1}c_{m,n} - iJ_z \eta_{m,m+1} c_{m,n} c_{m+1,n} \right),
\end{equation}
where $\eta_{m,m+1} = i \bar{c}_{m,n}\bar{c}_{m+1,n}$ is the $\mathbb{Z}_2$ gauge variable that is defined on all $z$-bonds.

Note that within this transformation approach, no artificial expansion of the local spin Hilbert space was necessary. Therefore, the JW transformation-based solution is known to be exact without any reprojection. However, this method only gives a Hamiltonian of the easy-to-handle form in Eq. \eqref{JWFinalHamiltonian} for systems with open boundary conditions in the direction of the JW strings. Otherwise, additional nonlocal boundary terms appear in the transformation, which are difficult to deal with.

\begin{figure}[h]
   \centering
    \includegraphics[width=0.48\columnwidth]{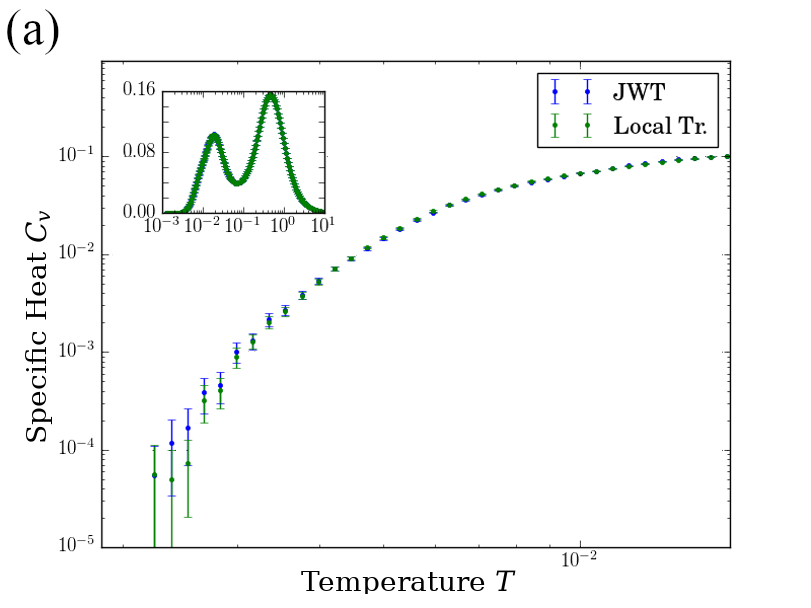}  
    \includegraphics[width=0.48\columnwidth]{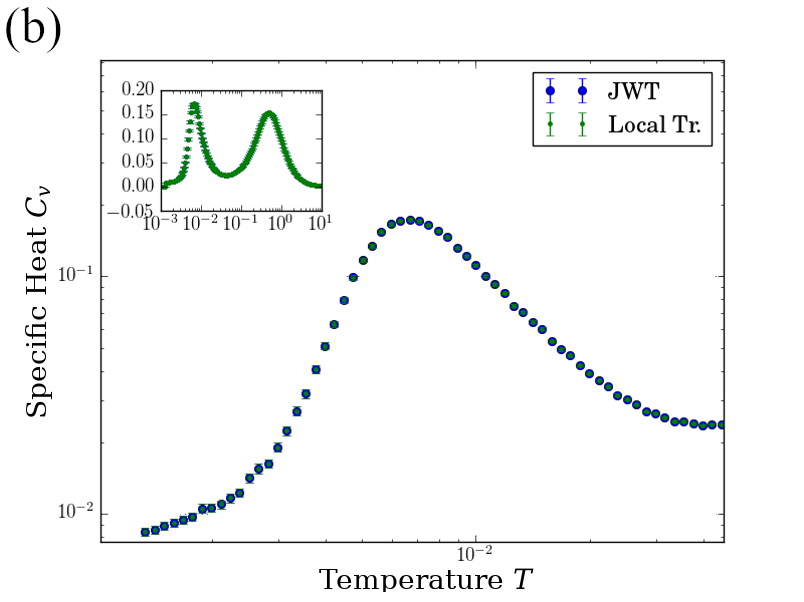}
     \caption{{\bf Jordan-Wigner vs. local transformation.} Benchmark calculations on (10,3)a clusters with 32 sites (a) and 108 sites (b) have shown that the error obtained from sampling $\mathbb{Z}_2$ gauge variables on all lattice bonds (local transformation) is negligible even for small systems (shown here for a double-logarithmic scale).}
    \label{fig:JWBenchmark}
\end{figure}

\begin{figure}[h!]
   \centering
    \includegraphics[width=0.35\columnwidth]{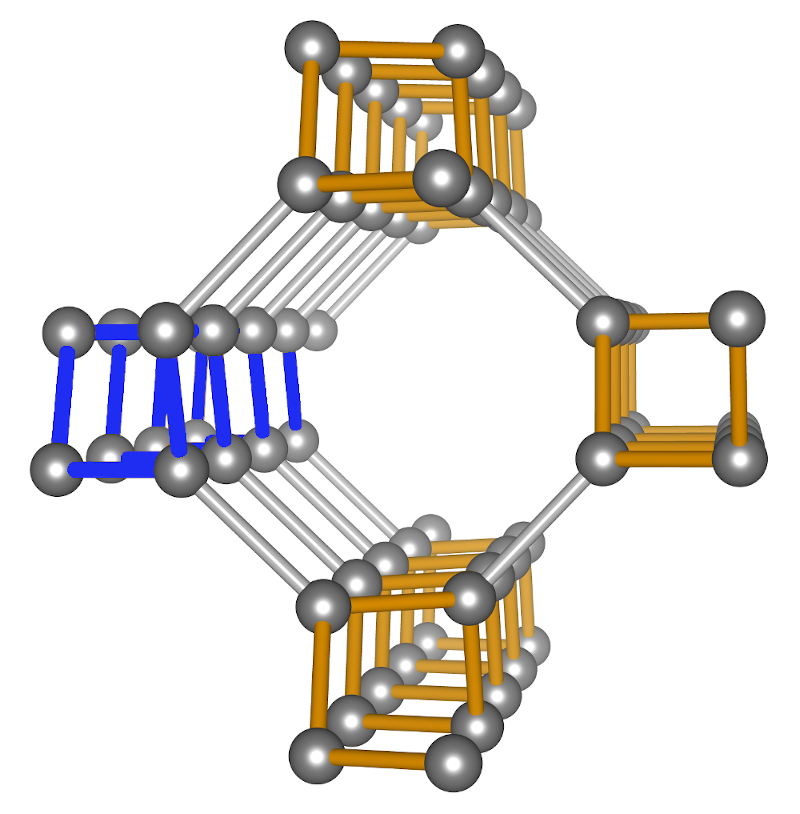}  \hspace{2 cm}
    \includegraphics[width=0.35\columnwidth]{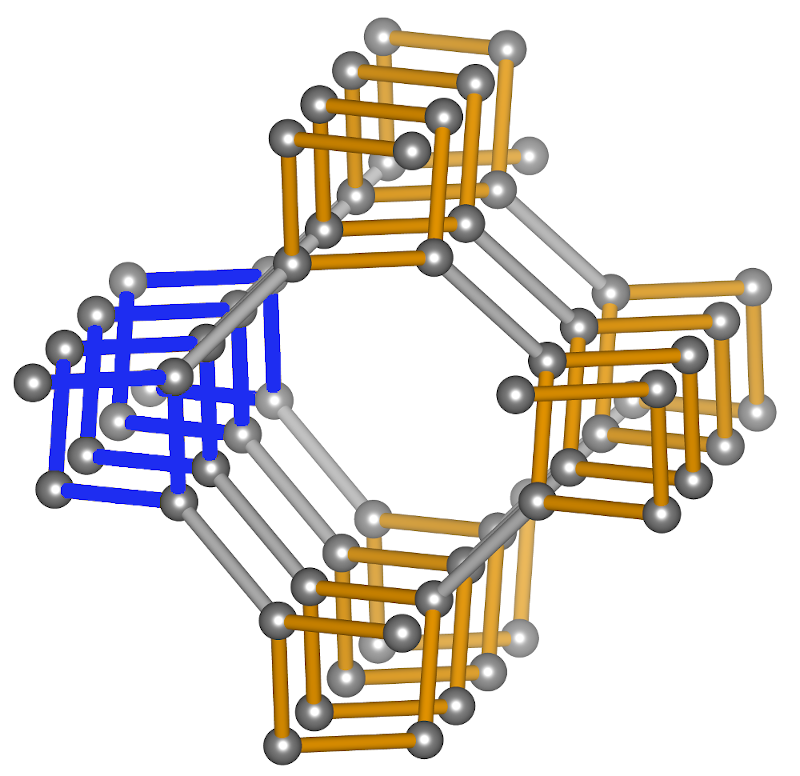}\\
     \caption{{\bf (10,3)a and (10,3)d.} JW strings can be defined along all spirals of $x$- and $y$-bonds (yellow / blue), while the remaining $z$-bonds (grey) host the $\mathbb{Z}_2$ gauge variable $\eta$.}
    \label{fig:JWStrings1}
\end{figure}

\begin{figure}[h!]
   \centering
    \includegraphics[width=0.4\columnwidth]{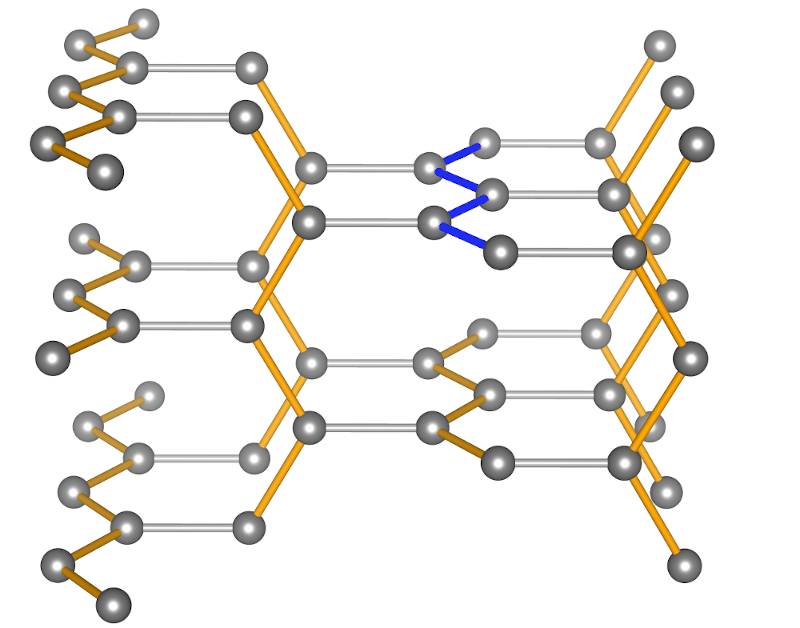} \hspace{1 cm}
    \includegraphics[width=0.4\columnwidth]{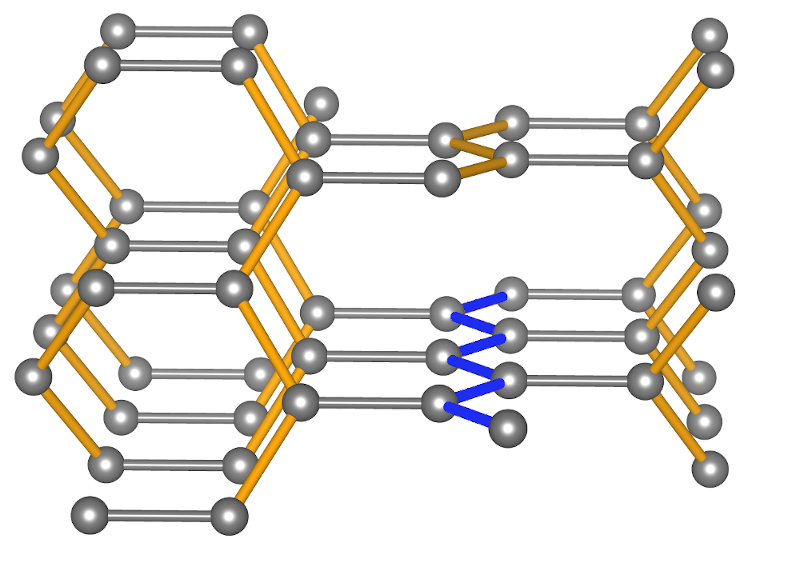}\\  
     \caption{{\bf (10,3)b and (10,3)c.} For both lattices, JW strings are defined along the zigzag chains of $x$- and $y$-bonds (yellow / blue).}
    \label{fig:JWStrings2}
\end{figure}

\begin{figure}[h!]
   \centering
    \includegraphics[width=0.45\columnwidth]{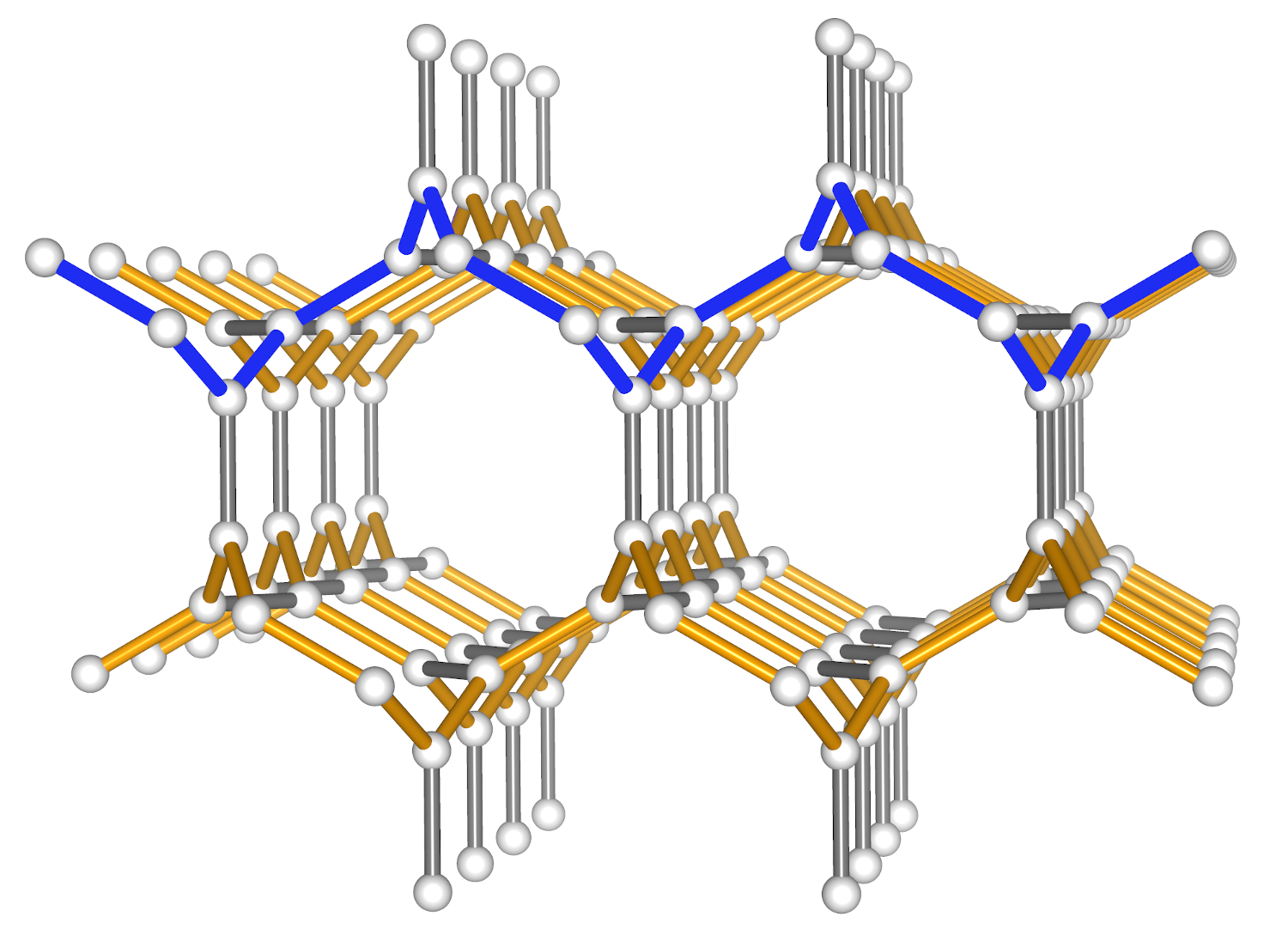}  \hspace{1 cm}
    \includegraphics[width=0.45\columnwidth]{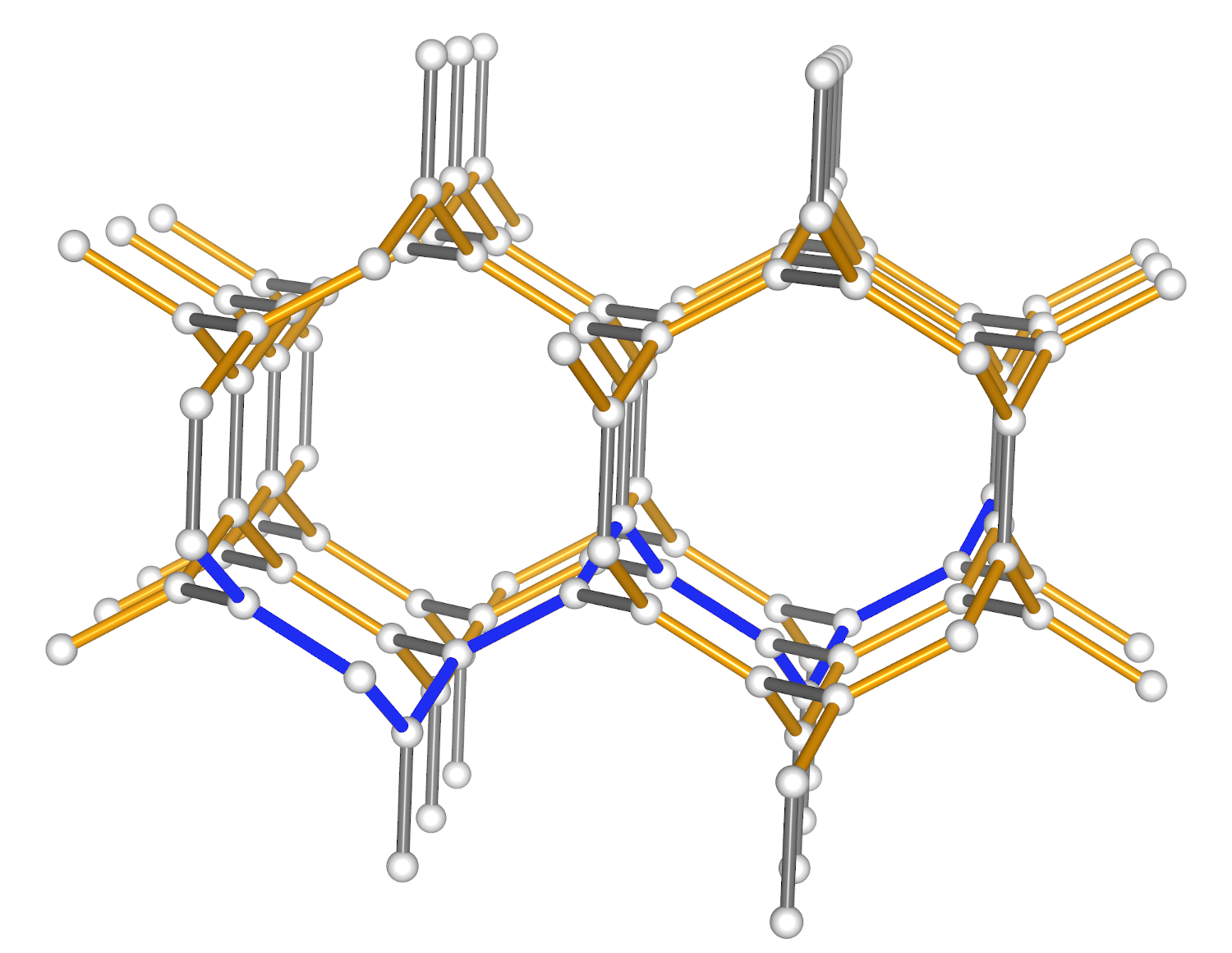}\\
     \caption{{\bf (8,3)a and (8,3)b.} For both lattices, JW strings are defined along chains of $x$- and $y$-bonds (yellow), assigning the $\mathbb{Z}_2$ gauge field $\eta$ to the remaining $z$-bonds (grey). For both lattices, one JW string is highlighted in blue.    
     }
    \label{fig:JWStrings3}
\end{figure}

\begin{figure}[h!]
   \centering
    \includegraphics[width=0.38\columnwidth]{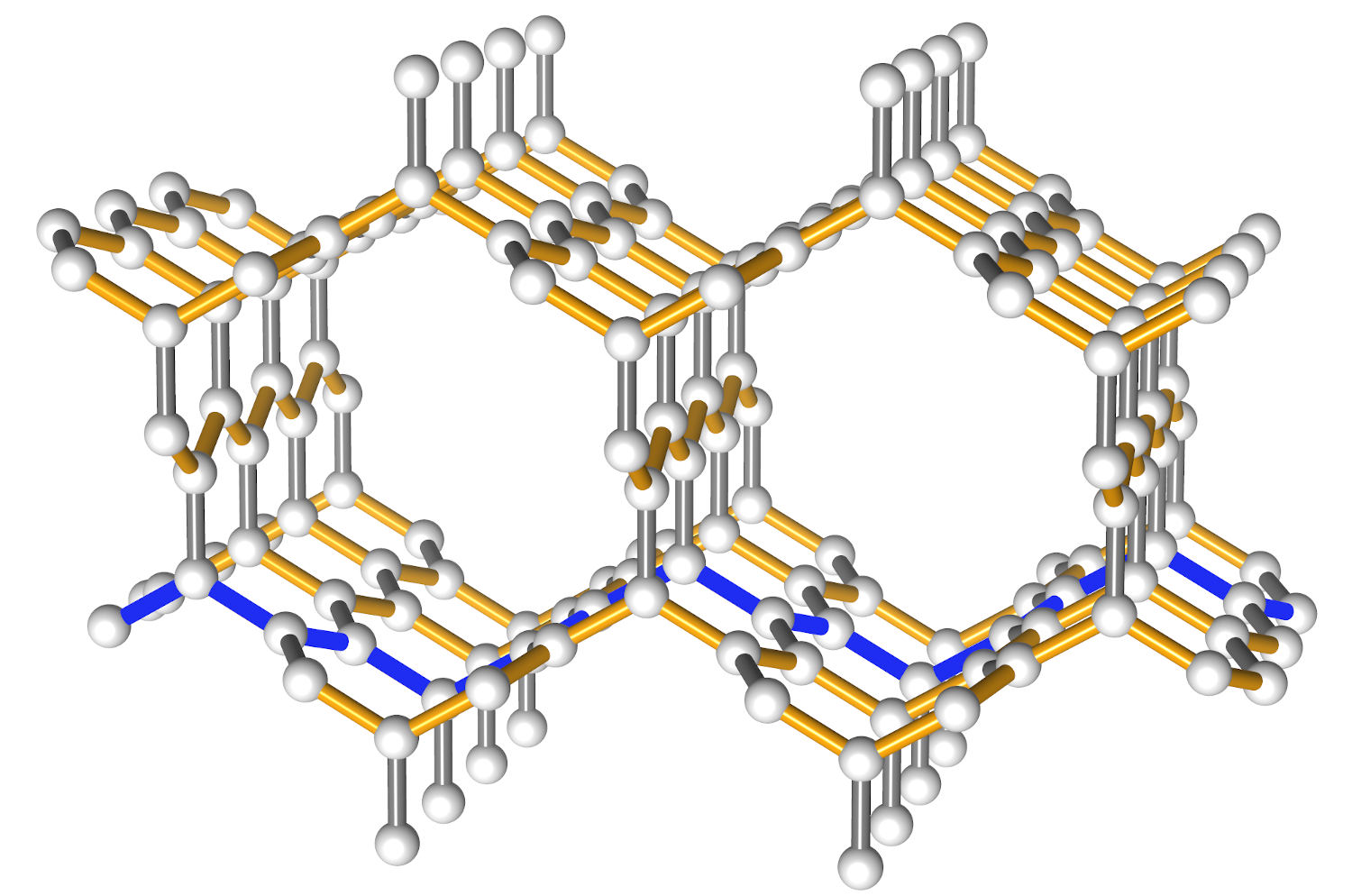}  \hspace{0.5 cm}
    \includegraphics[width=0.55\columnwidth]{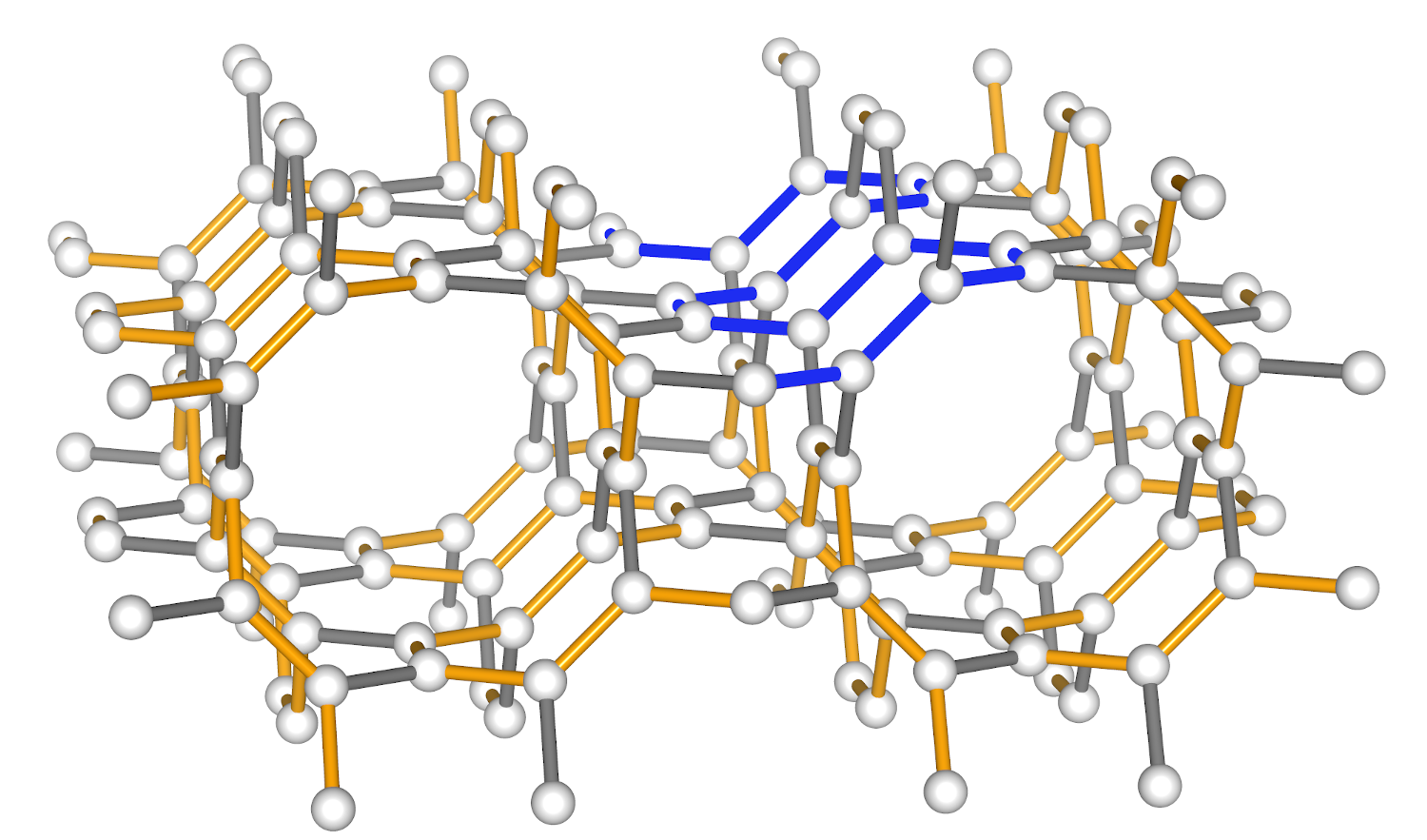}\\
     \caption{{\bf (8,3)c and (8,3)n.} On (8,3)c, JW strings are defined on $x$- and $y$-bonds (yellow). On (8,3)n, it is of more advantage to use $y$- and $z$-bonds. (yellow). For both lattices, individual JW strings are highlighted in blue.}
    \label{fig:JWStrings4}
\end{figure}

\begin{figure}[h!]
   \centering
    \includegraphics[width=0.48\columnwidth]{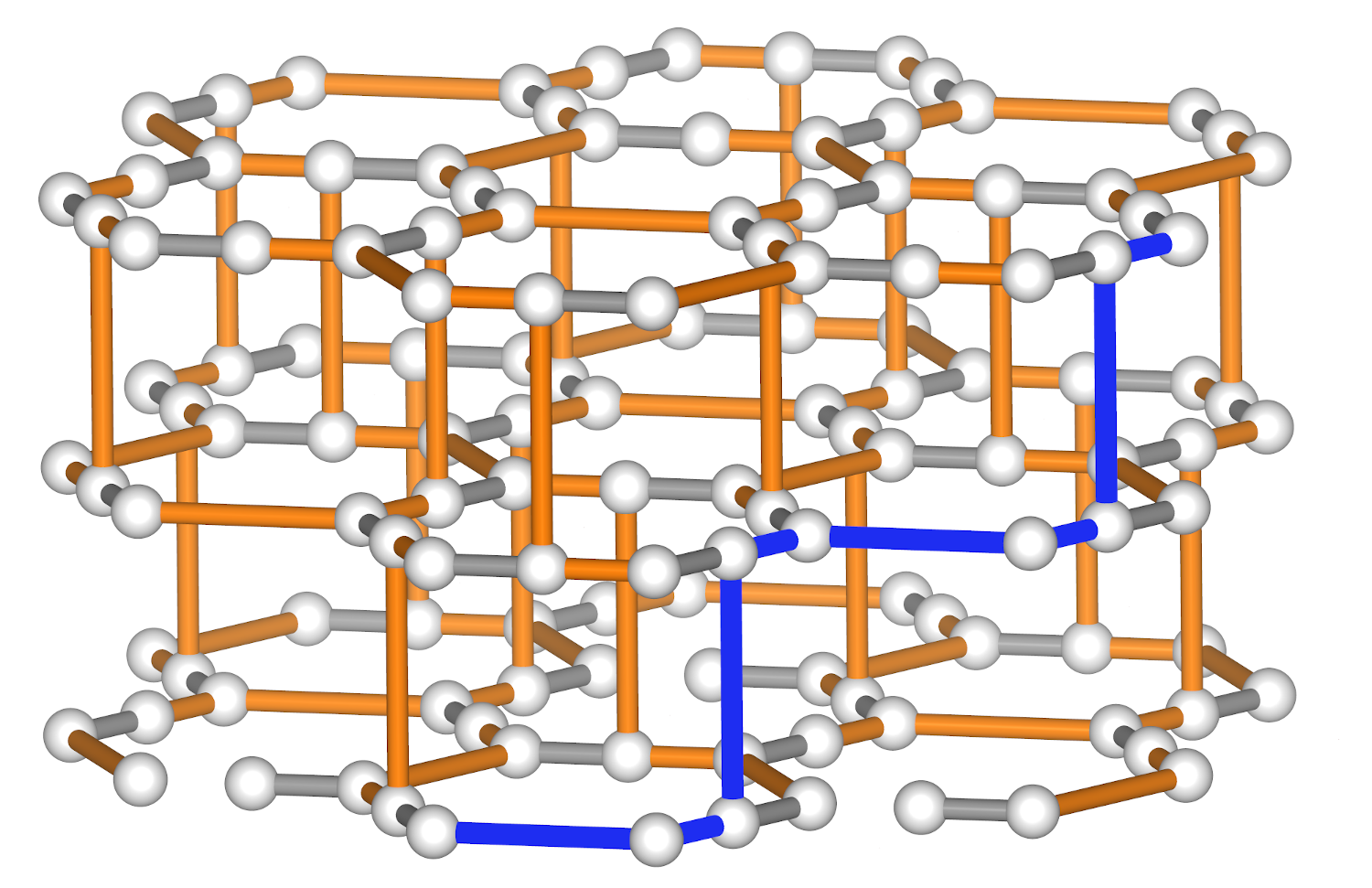}  
     \caption{{\bf (9,3)a.} On the only non-bipartite lattice in this classification, we can define JW strings along the $y$- and $z$-bonds.}
    \label{fig:JWStrings5}
\end{figure}

On the Hamiltonian level, the difference between the local transformation and the JW ansatz is the number of $\mathbb{Z}_2$ gauge variables in the system. The local transformation generates $\mathbb{Z}_2$ gauge variables $u_{ij}$ on {\it all} bonds, while the $\mathbb{Z}_2$ gauge variables $\eta$ in the JW-transformed Hamiltonian only live on {\it one} subclass of bonds (here: the $z$-bonds). For a system with open boundary conditions, both Hamiltonians are equivalent if the gauge field in the local version is fixed on the $x$- and $y$-bonds. However, benchmark calculations on small Kitaev clusters have shown that the QMC simulation based on the local transformation gives results that are within the error bars of the data points that were obtained from a QMC simulation with JW strings (see Fig. \ref{fig:JWBenchmark}), even for very small systems, where the deviations are expected to be the largest. We can therefore conclude that the error arising from the local transformation ansatz, where the Hilbert space is artificially enlarged for each spin, is negligible in the large-scale QMC simulations. The interpretation is that on systems with a well-defined JW-transformed Hamiltonian of the form in Eq. \eqref{JWFinalHamiltonian}, the additional gauge variables $u_{ij}$ of the local transformation only lead to an overcounting of physical states. Thus, the existence of a JW solution on a given Kitaev system ensures us that the results are correct, even if the QMC simulation is based on the local transformation ansatz. Therefore, we checked that for all elementary, tricoordinated 3D lattices considered in this paper, there is a well-defined JW transformation if appropriate bond subsets are chosen for the one-dimensional strings  (see Figs. \ref{fig:JWStrings1} - \ref{fig:JWStrings5}).

\clearpage
\section{Green-function-based kernel polynomial method}
\label{sec:KPM}

In the QMC method used for this article, the $\mathbb{Z}_2$ gauge variables $u_{ij}$ on the lattice bonds $\langle i,j \rangle$ are sampled, which is expressed as rank-2 updates of the matrix $\tilde{A}$ in the Hamiltonian: $\tilde{A} \rightarrow \tilde{A}'$. In the GF-KPM method, it is used that the spectrum of $\tilde{A}'$ is given by the roots of the function$d(E) = \{1 + \Delta_{ij}G_{ji}(E)\}\{1 - \Delta_{ji}G_{ij}(E)\} + \Delta_{ij}\Delta_{ji} G_{jj}G_{ii}$. 

Within this function, it is convenient to express the off-diagonal Green functions $G_{ij}$ ($i \neq j$) in terms of the diagonal Green's functions via
\begin{align}
G_{ab} &= \frac{1}{2} \left\{G_{a+b, a+b} - iG_{a+ib, a+ib} - (1-i) (G_{a,a} + G_{b,b})\right\}\nonumber\\
G_{ba} &= \frac{1}{2} \left\{G_{a+b, a+b} + iG_{a+ib, a+ib} - (1+i) (G_{a,a} + G_{b,b})\right\}
\end{align}
The diagonal Green's functions are then approximated by the expression
\begin{equation}
G_{ii}(E + i \epsilon) = i\frac{\mu_0 + 2 \sum_{m=1}^{M-1}\mu_m \exp \left\{-i m \arccos(E/s) \right\}}{\sqrt{s^2 - E^2}},
\end{equation}
where the key ingredients are the Chebyshev moments 
\begin{equation}
\mu_m = g_m  \bra{i} T_m(H/s) \ket{i}. 
\end{equation}
Here, $g_m$ denotes the Jackson kernel factor
\begin{equation}
g_m = \frac{(M - m + 1) \cos \left(\frac{\pi m}{M + 1} \right) + \sin \left(\frac{\pi m}{M + 1} \right)\cot \left(\frac{\pi}{M + 1} \right)}{M + 1},
\end{equation}
which serves to dampen the Gibbs oscillations that usually occur when a Chebyshev iteration is truncated after a finite amount of steps \cite{Weisse2006}. The expression $\bra{i} T_m(H/s) \ket{i}$ is iterated by the recursion  $T_m(x) = 2x T_{m-1}(x) - T_{m-2}(x)$, which, for the $i$-th element, is realized by successive multiplications with the rescaled Hamiltonian $H/s$ according to
\begin{align}
\ket{u_0} &= \mathbb{I}\ket{u} = T_0(H/s)\ket{u}\nonumber\\
\ket{u_1} &= (H/s)\ket{u_0} = T_1(H/s)\ket{u_0}\nonumber\\
\ket{u_m} &= 2(H/s)\ket{u_{m-1}} - \ket{u_{m-2}}.
\end{align}
Note that the subsquent matrix-vector multiplications used in the calculation of the Chebyshev moments $\mu_m$ are the most time-consuming part in the QMC-KPM method. The necessary calculation steps for the moments can be reduced by a factor of 2 by using the relations
\begin{align}
\mu_{2m} &= 2\braket{u_m, u_m} - \mu_0,\nonumber\\
\mu_{2m+1} &= 2\braket{u_{m+1}, u_m} - \mu_1.
\end{align}
Benchmark calculations for different lattices and system sizes have shown that the GF-KPM method well reproduces the results of the (exact) QMC-ED method (see Fig. \ref{fig:KPMBenchmark} for a (10,3)b system with $L = 6$).


\section{Thermodynamic observables}
\label{sec:Observables}

The major part of the thermodynamic observables are calculated from the Majorana partition function in a fixed $\mathbb{Z}_2$ gauge field configuration $\{u_{ij}\}$. Starting from the full partition function, which we express in terms of the diagonalized spinless fermion Hamiltonian, we obtain the Majorana partition function $\mathcal{Z}_{\rm Maj} \left(\{u_{ij}\}\right)$ by
\begin{align}
\mathcal{Z} &= \mathrm{tr}_{\{u_{ij}\}} \mathrm{tr}_{n_\lambda} e^{-\beta  \mathcal{H}} \nonumber\\
&= \mathrm{tr}_{\{u_{ij}\}} \mathrm{tr}_{n_\lambda} e^{-\beta  \sum_{\lambda = 1}^{N/2} \epsilon_\lambda \left (\hat{n}_\lambda - \frac{1}{2} \right)} \nonumber\\
&= \mathrm{tr}_{\{u_{ij}\}} \underbrace{\prod_{\lambda = 1}^{N/2} \left\{2 \cosh \left( \frac{\beta \epsilon_\lambda}{2}\right) \right\}}_{=: \mathcal{Z}_{\rm Maj}\left(\{u_{ij}\}\right)}\,.
\end{align}
The expressions for the free energy $F\left(\{u_{ij}\}\right)$, the internal energy $E\left(\{u_{ij}\}\right)$ and the specific heat contributions of the Majorana fermions $C_{v, {\rm MF}}(T$ and the $\mathbb{Z}_2$ gauge field $C_{v, {\rm GF}}(T)$ follow as
\begin{align}
F\left(\{u_{ij}\}, T\right) &= -T \sum_{\lambda = 1}^{N/2} \ln \left\{ 2 \cosh\left(\frac{\beta \epsilon_\lambda}{2} \right) \right\}\,, \\
E\left(\{u_{ij}\}, T\right) &= - \sum_{\lambda = 1}^{N/2} \frac{\epsilon_\lambda}{2} \tanh\left(\frac{\beta \epsilon_\lambda}{2} \right)\,, \\
C_{v, {\rm MF}}(T) &= -\frac{1}{T^2} \left \langle \frac{\partial E_f(\{u_{jk}\})}{\partial \beta} \right \rangle_{\rm MC} \,,\\
C_{v,{\rm GF}}(T) &= \frac{1}{T^2} \left ( \langle E_f^2(\{u_{jk}\}) \rangle_{\rm MC} - \langle E_f(\{u_{jk}\}) \rangle_{\rm MC}^2 \right ) \,,\\
C_{v,{\rm total}}(T) &= C_{v,{\rm MF}}(T) + C_{v,{\rm GF}}(T) \,.
\end{align}
Note that the bracket $\langle ... \rangle_{\rm MC}$ indicates the average over the Monte Carlo samples, i.e., averaging over the $\mathbb{Z}_2$ gauge field configurations $\{u_{ij}\}$.

The entropy per site can be calculated from the internal energy by the integration
\begin{equation}
S = \ln(2) + \beta \langle E \rangle_{\rm MC} - \int_0^\beta \langle E \rangle_{\rm MC} d\beta\,.
\end{equation}

\end{document}